\documentclass[12pt,a4paper]{iopart}

\usepackage{iopams}
\usepackage{setstack}
\usepackage{graphicx}
\usepackage{pict2e}
\usepackage{color}
\usepackage{hyperref}

\renewcommand{\P}{\mathbb{P}}
\newcommand{\Li}{\mathrm{Li}}
\renewcommand{\Re}{\mathrm{Re}}

\newcommand{\C}{\mathbb{C}}
\newcommand{\Ch}{\widehat{\mathbb{C}}}
\newcommand{\R}{\mathcal{R}}
\newcommand{\lb}{[\![}
\newcommand{\rb}{]\!]}

\newcommand{\keywords}[1]{\noindent\textbf{Keywords:} #1}
\newcommand{\tfrac}[2]{\mbox{\small$\frac{#1}{#2}$}}
\renewcommand{\text}[1]{\mathrm{#1}}

\begin{document}

\title{Riemann surfaces for integer counting processes}
\author{Sylvain Prolhac}
\address{Laboratoire de Physique Th\'eorique, Universit\'e de Toulouse, UPS, CNRS, France}


\begin{abstract}
Integer counting processes increment of an integer value at transitions between states of an underlying Markov process. The generator of a counting process, which depends on a parameter conjugate to the increments, defines a complex algebraic curve through its characteristic equation, and thus a compact Riemann surface. We show that the probability of a counting process can then be written as a contour integral on that Riemann surface. Several examples are discussed in details.\\


\end{abstract}

\begin{section}{Introduction}
In many instances, non-equilibrium phenomena can be modelled adequately by microscopic dynamics without memory, such that the evolution from time $t$ depends only on the state of the system at time $t$ and not on the evolution prior to $t$. In such cases, Markov processes constitute the natural setting incorporating randomness directly at the level of the microscopic dynamics. The generator $M$ of the Markov process then gives direct access to the statistical properties of the system at time $t$.

Statistics of observables depending on correlations between several times require more work. A prototypical example is counting processes \cite{HS2007.1} $Q_{t}$, which increment only when the underlying Markov process makes a transition between two states, and stay constant otherwise. It turns out that the probability of $Q_{t}$ can be extracted from a deformation $M(\rme^{\gamma})$ of $M$, with $\gamma$ a variable conjugate to the increments of $Q_{t}$.

The deformation parameter $\gamma$ is usually taken real, and the largest eigenvalue of $M(\rme^{\gamma})$ gives access to stationary large deviations of $Q_{t}$, reached in the long time limit. In this paper, we are interested instead in the time evolution of the probability of $Q_{t}$ at \textit{finite time} $t$. In that case, the natural approach consists in an eigenstate expansion of the propagator $\rme^{tM(\rme^{\gamma})}$. All the eigenstates of $M(\rme^{\gamma})$ will then contribute, and not just the one with largest eigenvalue as for stationary large deviations. While this approach works in principle for Markov processes with few states, and can even provide reasonably explicit results after asymptotic analysis in some exactly solvable cases with a large number of states, the lack of manageable expressions for the eigenstates severely limits this approach in general.

In this paper, we turn instead to complex values of the parameter $g=\rme^{\gamma}$, and exploit the well known analytic properties of eigenfunctions of a parameter dependent matrix. An important feature is the unavoidable existence of \textit{exceptional points} \cite{K1995.2} $g_{*}$, where $M(g_{*})$ is not diagonalizable because of the presence of Jordan blocks. Exceptional points can only happen at non-Hermitian $M(g_{*})$, and are associated with exchanges of the eigenstates under analytic continuation along small loops around $g_{*}$. They lead to a rich non-Hermitian physics \cite{H2012.1,EMKMR2018.1,MA2019.1,AGU2020.1,BBK2021.1} induced by the non-trivial topology of the spectrum, in particular in the context of non-Hermitian quantum mechanics \cite{M2011.2}.

For simplicity, we restrict in this paper to integer counting processes, for which the increments of $Q_{t}$ are integers. The characteristic equation of $M(g)$ is then polynomial in $g$, and the eigenvalues and eigenvectors of $M(g)$ live on a compact Riemann surface $\R$. We show in particular the the probability of $Q_{t}$ can be expressed as a contour integral (\ref{Prob[int R]}) on $\R$, and is thus simply equal to a sum of residues.

Since compact Riemann surfaces are not widely used in the study of non-equilibrium statistical mechanics, we provide a reasonably self-contained introduction to the subject in section~\ref{section counting processes}. We stay at a rather elementary level and do not make use of more advanced tools from algebraic geometry: the most essential features used in this paper are the fact that meromorphic functions on a compact Riemann surface have as many zeroes as poles, the Riemann-Hurwitz formula relating the genus $\mathrm{g}$ of $\R$ and the ramification data of any meromorphic function on $\R$, the Newton polygon approach for the genus of $\R$ in terms of an underlying non-singular algebraic curve, and a uniqueness theorem for meromorphic differentials with simple poles.

The Riemann surface approach for integer counting processes discussed in this paper is illustrated in section~\ref{section simple example} on the simplest possible example, where a single transition of a Markov process is monitored. This Riemann surface approach was used earlier by the author in \cite{P2020.2} for the statistics of the current in the totally asymmetric simple exclusion process (TASEP) with periodic boundaries, an exactly solvable model of hard-core particles moving in the same direction on a one-dimensional lattice, for which Bethe ansatz gives a particularly simple representation for $\R$. The Riemann surface for the current of TASEP is described in section~\ref{section single file unidirectional} and compared with the Riemann surface for a more general model where particles move in a single file with generic transition rates. Asymmetric hopping, where particles are allowed to move in both directions, is finally discussed in section~\ref{section single file bidirectional}, both for generic transition rates, and for the exactly solvable case of the asymmetric simple exclusion process (ASEP).
\end{section}

\begin{section}{Probability of a counting process as a contour integral on a Riemann surface}
\label{section counting processes}
In this section, we show that the probability of an integer counting process can be expressed as a contour integral on the compact Riemann surface associated with the generator of the counting process. In order to have a reasonably self-contained presentation, we provide an introduction to the needed aspects of complex algebraic curves and Riemann surfaces in sections~\ref{section algebraic curves} and \ref{section Riemann surfaces}.

\begin{subsection}{Probability of an integer counting process}
\label{section M(g)}
We consider a general Markov process on a finite set of states $\Omega$ of cardinal $|\Omega|$, with transition rates $w_{C\to C'}$ from states $C$ to $C'\neq C$. The Markov process is assumed to be ergodic, i.e. any state $C'$ can be reached from any state $C$ by a finite number of transitions.

The probability $P_{t}(C)$ that the system is in state $C\in\Omega$ at time $t$ evolves by the master equation $\partial_{t}P_{t}(C)=\sum_{C'\neq C}(w_{C'\to C}P_{t}(C')-w_{C\to C'}P_{t}(C))$. In the vector space generated by the basis of configuration vectors $|C\rangle$, $C\in\Omega$, the probability vector $|P_{t}\rangle=\sum_{C\in\Omega}P_{t}(C)|C\rangle$ then evolves as $\partial_{t}|P_{t}\rangle=M|P_{t}\rangle$ with $M$ the Markov matrix. The non-diagonal elements of $M$ are $\langle C'|M|C\rangle=w_{C\to C'}\geq0$, while conservation of probability reads $\langle C|M|C\rangle=-\sum_{C'\neq C}w_{C\to C'}\leq0$, or equivalently $\sum_{C\in\Omega}\langle C|M=0$. The line vector $\sum_{C\in\Omega}\langle C|$ is thus a left eigenvector of $M$ with eigenvalue $0$. The corresponding right eigenvector is the stationary state $|P_{\text{stat}}\rangle=\sum_{C\in\Omega}P_{\text{stat}}(C)|C\rangle$, normalized as $\sum_{C\in\Omega}P_{\text{stat}}(C)=1$. Since the Markov process is ergodic, the stationary state is uniquely defined, has $P_{\text{stat}}(C)>0$ for all $C\in\Omega$, and is reached in the long time limit from any initial condition $|P_{0}\rangle$.

While statistics of observables depending only on the state of the system at time $t$ may be computed directly from the propagator $\rme^{tM}$, we are interested in this paper in \textit{counting processes} \cite{HS2007.1} $Q_{t}$, starting initially at $Q_{0}=0$, and driven by the Markov process above in such a way that $Q_{t}$ is updated only at any transition $C\to C'$ as $Q_{t}\to Q_{t}+\delta Q_{C\to C'}$, for some fixed choice of increments $\delta Q_{C\to C'}$, $C\neq C'$. More precisely, we consider here only \textit{integer valued} counting processes (simply called Markov counting processes in the mathematical literature \cite{J1982.1}, when all the $\delta Q_{C\to C'}$ are non-negative), for which $\delta Q_{C\to C'}\in\mathbb{Z}$, which ensures that the eigenstates of the deformed generator defined below only have algebraic singularities. By convention, we set in the following $\delta Q_{C\to C'}=0$ when $C'=C$ and for forbidden transitions with $w_{C\to C'}=0$. Some increments $\delta Q_{C\to C'}$ for allowed transitions $w_{C\to C'}>0$ may also be chosen equal to zero.

The usual method to obtain informations about the statistics of a counting process at a given time $t$ proceeds by considering the generating function $\langle\rme^{\gamma Q_{t}}\rangle$, where the average is taken over all histories of the Markov process up to time $t$. The logarithm $F(\gamma)=\log\langle\rme^{\gamma Q_{t}}\rangle$ is then the cumulant generating function of $Q_{t}$, and the average and the variance of $Q_{t}$ are in particular given by $\langle Q_{t}\rangle=F'(0)$ and $\langle Q_{t}^{2}\rangle-\langle Q_{t}\rangle^{2}=F''(0)$.

The generator of $\langle\rme^{\gamma Q_{t}}\rangle$ is a deformation of the Markov matrix $M$. In order to obtain in the following an algebraic curve, we work instead in the variable $g=\rme^{\gamma}$. Then, one has
\begin{equation}
\label{GF[M]}
\langle g^{Q_{t}}\rangle=\sum_{C\in\Omega}\langle C|\rme^{tM(g)}|P_{0}\rangle\;,
\end{equation}
see e.g. \cite{HS2007.1}, where the deformed generator $M(g)$ is defined by $\langle C'|M(g)|C\rangle=g^{\delta Q_{C\to C'}}\langle C'|M|C\rangle$. Compared to the Markov matrix $M$, non diagonal elements of $M(g)$ have the additional factor $g^{\delta Q_{C\to C'}}$ while diagonal elements of $M$ and $M(g)$ are identical.

Computing the generating function $\langle g^{Q_{t}}\rangle$ requires in practice to expand the propagator $\rme^{tM(g)}$ over the eigenstates of $M(g)$. The main issue, which eventually leads to the introduction of the Riemann surface $\R$ in the following, is that while the generating function has trivial monodromy in the variable $g$ (i.e. following $\langle g^{Q_{t}}\rangle$ analytically along a closed path for $g$ leads back to the starting value), individual eigenstates may be permuted among themselves along a loop for $g$.

The eigenvalues of $M(g)$ are solutions of an algebraic equation of degree $|\Omega|$ with coefficients depending on the parameter $g$. In general, for any fixed value of $g$, one does not realistically expect these eigenvalues to be so simple that the eigenstate expansion of the generating function has a particularly illuminating expression. In the alternative approach studied in this paper, all the values of the parameter $g$ are instead considered at the same time: labelling the eigenstates of $M(g)$ by an index $r=1,\ldots,|\Omega|$, the couple $(g,r)$ will be interpreted as a point $p$ on a compact Riemann surface $\R$. Then, rather than considering the generating function (\ref{GF[M]}), which is just a sum over $|\Omega|$ points on $\R$ after the eigenstate expansion, we focus on the probability of $Q_{t}$, which is expressed as a contour integral on $\R$, an object with nicer analytic properties.

Since by definition of the mean value, the generating function is written in terms of the probability of $Q_{t}$ as
\begin{equation}
\langle g^{Q_{t}}\rangle=\sum_{Q\in\mathbb{Z}}g^{Q}P(Q_{t}=Q)\;,
\end{equation}
the probability can be extracted with residues, and one has from (\ref{GF[M]})
\begin{equation}
\label{Prob[M]}
\P(Q_{t}=Q)=\oint_{\gamma}\frac{\rmd g}{2\rmi\pi g^{Q+1}}\,\sum_{C\in\Omega}\langle C|\rme^{tM(g)}|P_{0}\rangle\;.
\end{equation}
The matrix elements of $M(g)$ are monomials in $g$ and thus meromorphic functions (i.e. analytic functions whose only singularities are poles) of $g$ with poles at $g=0$ and $g=\infty$. The only singularities of the integrand in (\ref{Prob[M]}) are thus poles at $g=0$ and $g=\infty$, both of infinite order if increments $\delta Q_{C\to C'}$ with either signs exist, and the contour of integration $\gamma$ must have winding number one around $0$.

In section~\ref{section proba integral on R}, we explain that the contour $\gamma$ may be replaced after the eigenstate expansion by a contour $\Gamma$ on the Riemann surface $\R$ mentioned above. Before doing this, we give a short introduction to the aspects of the theory of algebraic curves and compact Riemann surfaces that will be needed in the subsequent sections.
\end{subsection}

\begin{subsection}{Complex algebraic curve for \texorpdfstring{$M(g)$}{M(g)}}
\label{section algebraic curves}
In this section, we summarize known facts about complex algebraic curves \footnote{Complex algebraic \textit{curves} are actually two-dimensional \textit{surfaces} (almost everywhere), i.e. curves over the complex numbers.} such as the one built from the characteristic polynomial of a parameter dependent matrix, see e.g. \cite{E2018.1,T2007.1} for more details.

\begin{table}
	\begin{tabular}{|lcc|}\hline&&\\[-5mm]
		\hspace{1mm}\begin{tabular}{c}Type of point $p_{0}$\\(with respect to $g$)\end{tabular} & Local parameter $z$ & \begin{tabular}{c}Genericness\\of $\mathcal{A}$\end{tabular}\\[3mm]\hline&&\\[-5mm]
		Regular point $P(p_{0})=0$ & \begin{tabular}{cc}$z=g-g_{0}$ & $g_{0}\neq\infty$\\[2mm]$z=g^{-1}$ & $g_{0}=\infty$\end{tabular} & \begin{tabular}{c}any\\[2mm]any\end{tabular}\\[10mm]
		\hspace{-2mm}\begin{tabular}{l}Ramification point\\$\hookrightarrow$ ramified twice\\$P(p_{0})=P^{(1,0)}(p_{0})=0$\end{tabular} & \begin{tabular}{cc}$z=\sqrt{g-g_{0}}$ & $g_{0}\neq\infty$\\[2mm]$z=g^{-1/2}$ & $g_{0}=\infty$\end{tabular} & \begin{tabular}{c}any\\[2mm]non-generic\end{tabular}\\[10mm]
		\hspace{-2mm}\begin{tabular}{l}Ramification point\\$\hookrightarrow$ ramified $m$ times, $m\geq3$\\$P^{(k,0)}(p_{0})=0$, $0\leq k\leq m-1$\end{tabular} & \begin{tabular}{cc}$z=(g-g_{0})^{1/m}$ & $g_{0}\neq\infty$\\[2mm]$z=g^{-1/m}$ & $g_{0}=\infty$\end{tabular} & \begin{tabular}{c}non-generic\\[2mm]non-generic\end{tabular}\\[7mm]\hline&&\\[-5mm]
		\hspace{-2mm}\begin{tabular}{l}Singular point $P(p_{0})=P^{(1,0)}(p_{0})=P^{(0,1)}(p_{0})=0$\\[3mm]$\hookrightarrow$  Nodal $\mathcal{H}\neq0$\\[2mm]$\hookrightarrow$  Non-nodal $\mathcal{H}=0$\end{tabular}\hspace{-35mm} & \hspace{-25mm}\begin{tabular}{l}\\[2mm]$\mathcal{H}=\displaystyle\det\!\Big(\begin{array}{cc}P^{(2,0)}(p_{0}) & P^{(1,1)}(p_{0})\\P^{(1,1)}(p_{0}) & P^{(0,2)}(p_{0})\end{array}\Big)$\end{tabular} & \hspace{-7mm}\begin{tabular}{c}singular, non-generic\\\\\\\\\end{tabular}\\[-5mm]&&\\\hline
	\end{tabular}
	\caption{Summary of the different types of points $p_{0}=(\lambda_{0},g_{0})$ which may appear on an algebraic curve $\mathcal{A}$ defined by a polynomial equation $P(\lambda,g)=0$. The cases in the middle section of the table depend implicitly on the choice of the variable $g$ to parametrize $\mathcal{A}$. The singular points in the bottom part of the table are independent of the choice of parametrization, and only happen for \textit{singular} algebraic curves, which are not generic. Ramification points in the variable $g$ are not seen as singular points, but higher ramification index $m\geq3$ can only happen for non-generic algebraic curves as well.}
	\label{table points algebraic curve}
\end{table}

While all the matrix elements of $M(g)$ are meromorphic functions of $g$, the eigenvalues $\lambda_{r}(g)$ and the corresponding left and right \footnote{The matrix $M(g)$ is not symmetric in general, and its left and right eigenvectors are not transposed of each other. They still verify $\langle\psi_{r}(g)|\psi_{s}(g)\rangle=0$ if $\lambda_{r}(g)\neq\lambda_{s}(g)$ however, and resolution of the identity ${\bf1}=\sum_{r=1}^{|\Omega|}\frac{|\psi_{r}(g)\rangle\langle\psi_{r}(g)|}{\langle\psi_{r}(g)|\psi_{r}(g)\rangle}$ holds, at least if all the eigenvalues are distinct.} eigenvectors $\langle\psi_{r}(g)|$ and $|\psi_{r}(g)\rangle$, $r=1,\ldots,|\Omega|$ are not: branch point singularities appear, associated with non-trivial monodromy around them. This can be understood in terms of the characteristic polynomial
\begin{equation}
\label{P[M]}
P(\lambda,g)=\det(\lambda I-M(g))\;,
\end{equation}
with $I$ the $|\Omega|\times|\Omega|$ identity matrix, which vanishes if $\lambda$ is an eigenvalue of $M(g)$. By construction of $M(g)$, there exists integers $d_{\pm}$, $d_{-}\leq0\leq d_{+}$, such that $P(\lambda,g)=\sum_{k=d_{-}}^{d_{+}}P_{k}(\lambda)g^{k}$, with $P_{k}$ polynomials. Then, $g^{-d_{-}}P(\lambda,g)$ is a polynomial in both variables $\lambda$ and $g$, and
\begin{equation}
P(\lambda,g)=0\;,
\end{equation}
$(\lambda,g)\in\C^{2}$, is the equation of a complex \textit{algebraic curve}, called $\mathcal{A}$ in the following. In the context of classical integrable systems, where $M(g)$ is a Lax matrix depending on a spectral parameter, see e.g. \cite{E2018.1,BBEIM1994.1,BBT2003.1}, $\mathcal{A}$ is called the spectral curve of $M(g)$.

We summarize in the rest of this section some useful facts about the local shape of an algebraic curve $\mathcal{A}$ near a point $p_{0}=(\lambda_{0},g_{0})\in\mathcal{A}$, see e.g. \cite{E2018.1} for more details. Both $\lambda_{0}$ and $g_{0}$ are assumed to be finite in this section, unless explicitly stated otherwise. The classification of the various possible cases, discussed below in decreasing order of genericness (see table~\ref{table points algebraic curve} for a summary), depends implicitly on the variable used to parametrize the algebraic curve, and which we take as the variable $g$, the most natural choice for us since an integral over $g$ appears in (\ref{Prob[M]}). We emphasize that even highly non-generic cases do appear in practice, as illustrated in sections~\ref{section TASEP} and \ref{section ASEP} with TASEP and ASEP.

We begin with the most generic case (with respect to the parametrization with the variable $g$) of finite $\lambda_{0}$ and $g_{0}$ with a non-zero partial derivative $P^{(1,0)}(p_{0})$. In a neighbourhood of $p_{0}$, the equation $P(\lambda,g)=0$ has a unique solution for $\lambda$, which is analytic in $g$, with $\lambda-\lambda_{0}=-(g-g_{0})P^{(0,1)}(p_{0})/P^{(1,0)}(p_{0})+\mathcal{O}(g-g_{0})^{2}$. Around such a point $p_{0}$, the algebraic curve is then a surface (two-dimensional real manifold). We say that $z=g-g_{0}$, which vanishes at $p_{0}$, is a \textit{local parameter} for the surface around $p_{0}$: any small disk centred at the origin in the complex plane is mapped bijectively to a neighbourhood of $p_{0}$ under $z\mapsto(\lambda,g)$, and both $\lambda$ and $g$ are locally analytic functions of $z$. This is the most generic situation for a point on an algebraic curve, and we call $p_{0}$ a regular point with respect to the variable $g$. We emphasize that the condition $P^{(1,0)}(p_{0})\neq0$ does depend on our choice to parametrize $\mathcal{A}$ in terms of the variable $g$.

\begin{figure}
	\begin{center}
	\begin{tabular}{lll}
	\begin{tabular}{l}\includegraphics[width=35mm]{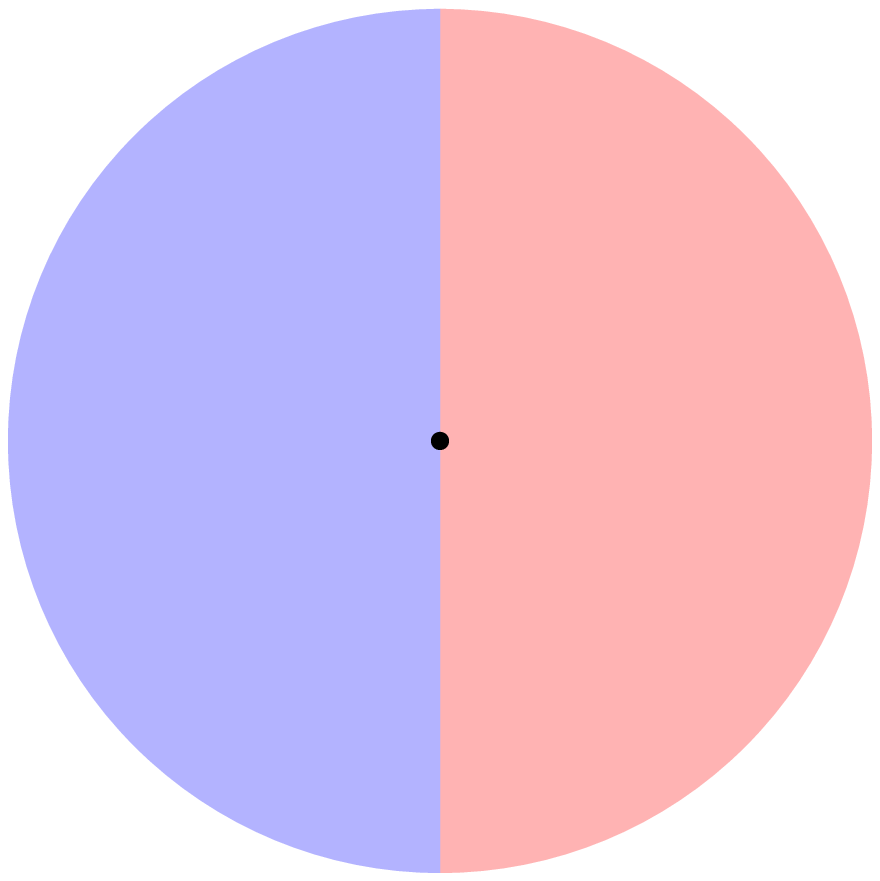}\end{tabular}
	&\hspace*{5mm}&
	\begin{tabular}{l}\includegraphics[width=35mm]{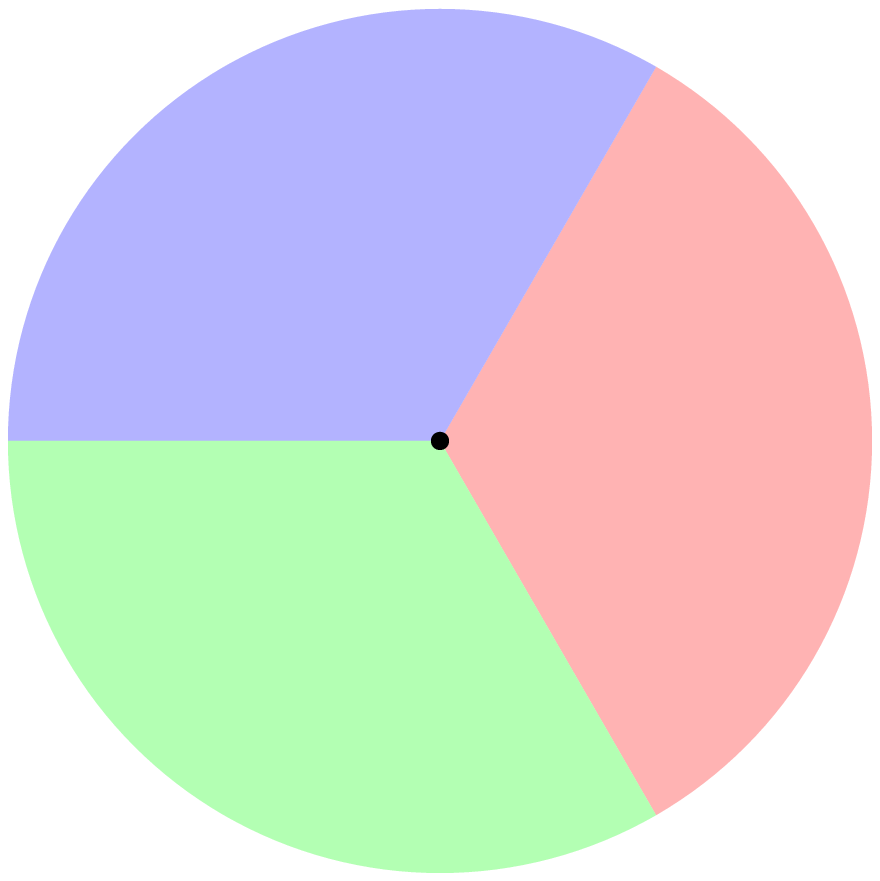}\end{tabular}
	\end{tabular}
	\end{center}
	\caption{Neighbourhood in an algebraic curve $\mathcal{A}$ of a ramification point $p_{0}=(\lambda_{0},g_{0})$ in the variable $g$, with ramification index $m=e_{p_{0}}$ for $m=2$ (left) and $m=3$ (right). Both disks represent values $|z|<\epsilon$ of the local parameter $z$, such that $g-g_{0}=z^{m}$. Different colors represent different determinations of the $m$-th root function, such that a same value of $g$ corresponds to two (left) or three (right) points of the neighbourhood. The choice of cuts between the different portions of the circle are largely arbitrary and depend on the choice of branch cut for the $m$-th root function.}
	\label{fig half disks}
\end{figure}

If $P^{(1,0)}(p_{0})=0$, on the other hand, $\lambda-\lambda_{0}$ is no longer analytic in $g$ in a neighbourhood of $g_{0}$. Indeed, assuming first that $P^{(2,0)}(p_{0})\neq0$, we observe that it is now $(\lambda-\lambda_{0})^{2}$ which is an analytic function of $g$, such that $(\lambda-\lambda_{0})^{2}=-2(g-g_{0})P^{(0,1)}(p_{0})/P^{(2,0)}(p_{0})+\mathcal{O}(g-g_{0})^{2}$. Assuming further that $P^{(0,1)}(p_{0})\neq0$ and taking the square root (with for definiteness the usual choice of branch cut $\mathbb{R}^{-}$ for the square root, corresponding to $\sqrt{r\rme^{\rmi\theta}}=\sqrt{r}\,\rme^{\rmi\theta/2}$, $r\geq0$, $-\pi<\theta\leq\pi$), one finds $\lambda-\lambda_{0}=\pm c\sqrt{g-g_{0}}\,(1+\mathcal{O}(g-g_{0}))$, with $c=\sqrt{-2P^{(0,1)}(p_{0})/P^{(2,0)}(p_{0})}\neq0$, an overall ambiguity for the choice of the sign, and a non-analyticity in $g$ coming from $\sqrt{g-g_{0}}$. The parameter $z$, defined as $z=\sqrt{g-g_{0}}$ in the sector where $\lambda-\lambda_{0}\simeq c\sqrt{g-g_{0}}$ and as $z=-\sqrt{g-g_{0}}$ in the sector where $\lambda-\lambda_{0}\simeq-c\sqrt{g-g_{0}}$, is then a local parameter for $\mathcal{A}$ around $p_{0}$, such that both $g=g_{0}+z^{2}$ and $\lambda\simeq\lambda_{0}+cz+\mathcal{O}(z^{3})$ are locally analytic functions of $z$. A small disk $|z|<\epsilon$ is then in bijection with a neighbourhood of $p_{0}$ in $\mathcal{A}$, and $\mathcal{A}$ is thus still a surface locally. In terms of the variable $g$, this neighbourhood comes from the union of two disjoint half-disks $z=\sqrt{g-g_{0}}$, $|g-g_{0}|<\epsilon^{2}$, for which $\Re\,z\geq0$, and $z=-\sqrt{g-g_{0}}$, $|g-g_{0}|<\epsilon^{2}$, for which $\Re\,z\leq0$, see figure~\ref{fig half disks}. In the situation described in this paragraph, we note that we could instead parametrize $\mathcal{A}$ with the variable $\lambda$ and not $g$, in which case we are back to the generic situation of the previous paragraph; in practice, however, it is usually more convenient to always use the same base variable everywhere.

The case described in the previous paragraph requires both $P(\lambda,g)=0$ and $P^{(1,0)}(\lambda,g)=0$, and thus happen only at a finite number of points $p_{0}\in\mathcal{A}$. Such a point $p_{0}$ is called a \textit{ramification point} for the variable $g$, or equivalently for the map $(\lambda,g)\mapsto g$ from $\mathcal{A}$ to $\C$, with \textit{ramification index} $e_{p_{0}}=2$. The value $g_{0}$ is then called a \textit{branch point} for the variable $g$, a denomination justified by the fact that a generic function analytic in the local parameter, $f(z)=\sum_{n=0}^{\infty}a_{n}z^{n}$, has an algebraic branch point at $g=g_{0}$ when expressed as a function of the variable $g$. Such a branch point is characterized by the fact that a small loop winding once around $g_{0}$ lifts to an open path on $\mathcal{A}$ in a neighbourhood of $p_{0}$, while a loop on $\mathcal{A}$ with winding number $2$ around $g_{0}$ lifts to a loop winding once around $p_{0}$. More generally, a ramification point $p_{0}$ with ramification index $e_{p_{0}}=m\geq2$ only lifts small loops around $g_{0}$ with winding number proportional to $m$ into loops around $p_{0}$. Such a point $p_{0}$ requires that all $P^{(k,0)}$, $k=0,\ldots,m-1$ vanish at $p_{0}$ but neither $P^{(m,0)}$ nor $P^{(0,1)}$, and thus only happens for non-generic algebraic curves if $m\geq3$. The algebraic curve is still locally a surface around such a point, and a local parameter is $z$ such that $g-g_{0}=z^{m}$, see figure~\ref{fig half disks}.

\begin{figure}
	\begin{center}
	\begin{tabular}{lll}
	\begin{tabular}{l}
	\begin{picture}(15,32)
	\put(0,0){\line(1,2){15}}\put(0,30){\line(1,-2){15}}
	\qbezier(0,0)(7.5,-5)(15,0)
	{\color[rgb]{0.9,0.9,0.9}\qbezier(0,0)(7.5,5)(15,0)}
	\qbezier(0,30)(7.5,25)(15,30)\qbezier(0,30)(7.5,35)(15,30)
	{\color{red}\put(7.5,15){\circle*{1}}}
	\end{picture}
	\end{tabular}
	&\hspace*{10mm}&
	\begin{tabular}{l}
	\begin{picture}(15,32)
	\qbezier(0,0)(7.5,20)(15,0)\qbezier(0,30)(7.5,10)(15,30)
	\qbezier(0,0)(7.5,-5)(15,0)
	{\color[rgb]{0.9,0.9,0.9}\qbezier(0,0)(7.5,5)(15,0)}
	\qbezier(0,30)(7.5,25)(15,30)\qbezier(0,30)(7.5,35)(15,30)
	{\color{blue}\put(7.5,10){\circle*{1}}\put(7.5,20){\circle*{1}}}
	\end{picture}
	\end{tabular}	
	\end{tabular}
	\end{center}
	\caption{Singular algebraic curve $\mathcal{A}$ near a nodal point (left) and corresponding desingularized Riemann surface $\R$ (right) obtained after splitting $\mathcal{A}$ at the nodal point. The nodal point of $\mathcal{A}$ (red dot located at the apex of the double cone) then corresponds to two points (the blue dots) on $\R$.}
	\label{fig cone}
\end{figure}
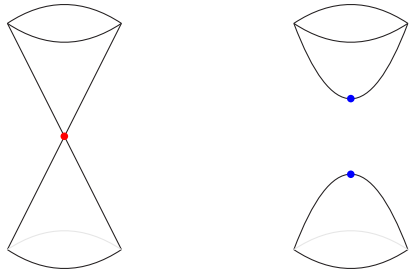

In all the situations described above, the algebraic curve $\mathcal{A}$ is locally a surface at $p_{0}$. When this is true for any $p_{0}\in\mathcal{A}$, we say that $\mathcal{A}$ is \textit{non-singular}. Singular algebraic curves $\mathcal{A}$, on the other hand, have points $p_{0}$ around which $\mathcal{A}$ is not a surface locally. This requires that both $P^{(1,0)}$ and $P^{(0,1)}$ vanish at $p_{0}$, which we refer to as a \textit{singular point} of the algebraic curve, and which is independent of the way we choose to parametrize $\mathcal{A}$. The most generic singular points are \textit{nodal points} (sometimes also called diabolical points \cite{BW1984.1} or conical intersections \cite{Y2001.1} in the context of quantum Hamiltonians), for which the Hessian determinant of $P$ does not vanish at $p_{0}$, and the algebraic curve then looks like the neighbourhood of the apex $p_{0}$ of a double cone, see figure~\ref{fig cone}. The situation is more complicated in the presence of non-nodal, higher singular points, at which the Hessian determinant vanishes too. Since the presence of singular points only happens for non-generic algebraic curves, one could be tempted to simply ignore them, at least in a first approach. We do not want to do that here since the algebraic curves for prominent examples of counting processes, such as the current for TASEP and ASEP studied in sections~\ref{section TASEP} and \ref{section ASEP}, have a high number of singular points, both nodal and non-nodal.

We have restricted so far to points $p_{0}$ with both $\lambda_{0}$ and $g_{0}$ finite. A nice feature of algebraic curves is however that points $p_{0}$ at infinity may be treated on an equal footing with other points, by adding to $\C$ a point $\infty$ representing complex infinity reached from any direction, so that algebraic curves may be thought of as \textit{compact} objects. The classification discussed above into regular, ramified and singular points still applies for points $p_{0}$ at infinity. In particular, a local parameter for the neighbourhood of $p_{0}$ in $\mathcal{A}$ is through the variable $z=g^{-1}$ for points $p_{0}$ regular with respect to the variable $1/g$, and $z=g^{-1/m}$ for ramification points with ramification index $e_{p_{0}}=m$. Since ramification points are non-generic points of an algebraic curve, a generic algebraic curve does not have ramification points at infinity. The situation is however different for the algebraic curves considered in this paper, which are built from the characteristic equation of a non-diagonal deformation of a matrix independent of $g$: depending on the choice of increments $\delta Q_{C\to C'}$, branch points may appear at $g_{0}=\infty$ and $g_{0}=0$ even for generic transition rates $w_{C\to C'}$, see sections~\ref{section simple example}, \ref{section single file unidirectional} and \ref{section single file bidirectional} for specific examples.
\end{subsection}

\begin{subsection}{Riemann surfaces}
\label{section Riemann surfaces}
In this section, we consider the compact Riemann surface associated with the algebraic curve $\mathcal{A}$, and summarize some known properties about meromorphic functions and meromorphic differentials. We refer to \cite{B2013.3,E2018.1,S1957.1} for additional material on the subject, and detailed derivations of some properties that are stated here without proofs.

\begin{subsubsection}{Algebraic curves and Riemann surfaces}\hfill\\
As we have seen in the previous section, the neighbourhood of any point $p_{0}=(\lambda_{0},g_{0})$ of a non-singular algebraic curve $\mathcal{A}$ is a two-dimensional surface, which can always be parametrized \textit{locally} by a complex number $z$ in such a way that both $\lambda$ and $g$ are holomorphic (respectively meromorphic) functions of $z$ for finite (resp. infinite) $\lambda_{0}$, $g_{0}$. We say that the functions $\lambda(p)$ and $g(p)$ are then globally meromorphic on $\mathcal{A}$. We emphasize that because of ramification, there does not exist in general (except for genus zero, see below) a \textit{global} parametrization $z$ for $\mathcal{A}$ such that $\lambda$ and $g$ are meromorphic functions of $z$ everywhere.

Non-singular algebraic curves, which look locally like the complex plane, are the natural setting to extend complex analysis to the compact setting. It should be noted, however, that many algebraic curves can accommodate exactly the same meromorphic functions up to changes of variables. Equivalence classes are then called \textit{compact Riemann surfaces}, and may alternatively be defined in a more abstract way without referring to an underlying algebraic curve, by looking at how a local parameter $z$ transforms from a neighbourhood to another, see e.g. \cite{B2013.3,E2018.1}.

The discussion above can be extended to singular algebraic curves $\mathcal{A}$, for which a procedure called desingularization associates to $\mathcal{A}$ a (non-singular) compact Riemann surface $\R$, which is locally a surface everywhere. In the presence of nodal points, the algebraic curve is in particular cut as in figure~\ref{fig cone}, so that the nodal point of $\mathcal{A}$ gives two distinct points $p_{1},p_{2}\in\R$, which no longer have a special nature with respect to a generic parametrization of $\R$ (but are of course still special for the variables $\lambda$ and $g$, as $(\lambda,g)$ takes the same value at both points).
\end{subsubsection}

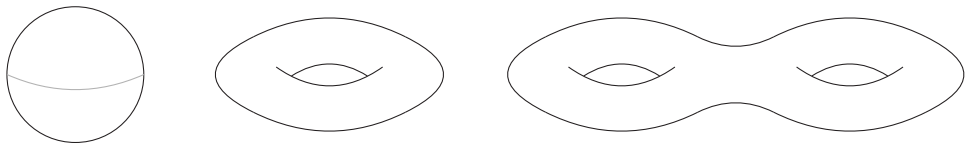
\begin{figure}
	\begin{center}
	\begin{tabular}{lll}
		\begin{tabular}{l}
			\begin{picture}(20,20)
				\put(10,10){\circle{18}}
				{\color[rgb]{0.7,0.7,0.7}\qbezier(1,10)(10,6)(19,10)}
			\end{picture}
		\end{tabular}
		&
		\begin{tabular}{l}
			\begin{picture}(30,20)
				\qbezier(5,5)(-5,10)(5,15)
				\qbezier(5,15)(15,20)(25,15)
				\qbezier(5,5)(15,0)(25,5)
				\qbezier(8,11)(15,6)(22,11)\qbezier(10,9.8)(15,13)(20,9.8)
				\qbezier(25,5)(35,10)(25,15)
			\end{picture}
		\end{tabular}
		&
		\begin{tabular}{l}
			\begin{picture}(60,20)
				\qbezier(5,5)(-5,10)(5,15)
				\qbezier(5,15)(15,20)(25,15)
				\qbezier(5,5)(15,0)(25,5)
				\qbezier(8,11)(15,6)(22,11)\qbezier(10,9.8)(15,13)(20,9.8)
				\qbezier(25,15)(30,12.5)(35,15)
				\qbezier(25,5)(30,7.5)(35,5)
				\qbezier(35,15)(45,20)(55,15)
				\qbezier(35,5)(45,0)(55,5)
				\qbezier(38,11)(45,6)(52,11)\qbezier(40,9.8)(45,13)(50,9.8)
				\qbezier(55,5)(65,10)(55,15)
			\end{picture}
		\end{tabular}
	\end{tabular}
	\end{center}
	\caption{Surfaces of genus $0$, $1$, $2$, from left to right.}
	\label{fig genus}
\end{figure}

\begin{subsubsection}{Connected components, genus}\hfill\\
\label{section genus}
Topologically, a compact Riemann surface $\R$ is an \textit{orientable} two dimensional manifold: continuous deformations of any simply connected domain on $\R$ (i.e. path connected domain inside which any closed curve can be contracted to a point by continuous deformations within the domain), whose boundary $\Gamma$ is a simple closed curve on $\R$ (i.e. a closed curve without self intersection), preserves the orientation of $\Gamma$.

The Riemann surface $\R$ associated to an algebraic curve $\mathcal{A}$ generically has a single connected component. Multiple connected components correspond to the characteristic polynomial $P$ from which $\mathcal{A}$ is defined factorizing as a product of polynomials, $P(\lambda,g)=P_{1}(\lambda,g)P_{2}(\lambda,g)$, and thus to a singular algebraic curve (solutions of $P_{1}(\lambda,g)=P_{2}(\lambda,g)=0$ are indeed singular points of $\mathcal{A}$).

The genus $\mathrm{g}\in\mathbb{N}$ of a connected, orientable surface counts its number of holes (or its number of handles), see figure~\ref{fig genus}, with in particular $\mathrm{g}=0$ for a sphere and $\mathrm{g}=1$ for a torus. The geometric \footnote{When $\R$ has $K>1$ connected components, it is sometimes useful to consider instead the \textit{arithmetic genus} $\mathrm{g}-K+1$, for which the Riemann-Hurwitz formula (\ref{Riemann-Hurwitz}) below is then independent of $K$. In this paper, $\mathrm{g}$ always refers to the geometric genus.} genus of a Riemann surface $\R$ is then the sum over all its connected components of the genus of each component. Varying the parameters of the algebraic curve $\mathcal{A}$ (i.e. the coefficients of the polynomial $P$) while keeping it non-singular preserves the genus. Up to appropriate isomorphisms between Riemann surfaces accommodating the same meromorphic functions, there exists a single connected Riemann surface of genus $0$, the Riemann sphere $\Ch$ obtained by adding the point at infinity to $\C$. On the other hand, the space of Riemann surfaces of genus $\mathrm{g}=1$ (respectively $\mathrm{g}\geq2$) is of complex dimension one (resp. $3\mathrm{g}-3$).

\begin{figure}
	\begin{center}
	\begin{tabular}{lll}
		\begin{tabular}{l}
		\begin{picture}(60,20)
		{\color[rgb]{1,0.5,0.5}\qbezier(15,2.5)(16,6)(15,8.5)\qbezier(30,6.2)(32,10)(30,13.8)}
		\qbezier(5,5)(-5,10)(5,15)
		\qbezier(5,15)(15,20)(25,15)
		\qbezier(5,5)(15,0)(25,5)
		\qbezier(8,11)(15,6)(22,11)\qbezier(10,9.8)(15,13)(20,9.8)
		\qbezier(25,15)(30,12.5)(35,15)
		\qbezier(25,5)(30,7.5)(35,5)
		\qbezier(35,15)(45,20)(55,15)
		\qbezier(35,5)(45,0)(55,5)
		\qbezier(38,11)(45,6)(52,11)\qbezier(40,9.8)(45,13)(50,9.8)
		\qbezier(55,5)(65,10)(55,15)
		\end{picture}
		\end{tabular}
		&&
		\begin{tabular}{l}
		\begin{picture}(60,20)
		\qbezier(5,5)(-5,10)(5,15)
		\qbezier(5,15)(15,20)(25,15)
		\qbezier(5,5)(10,2.5)(15,7)\qbezier(15,7)(20,2.5)(25,5)
		\qbezier(8,11)(15,3)(22,11)\qbezier(10,9)(15,13)(20,9)
		\qbezier(25,15)(27.5,13.75)(30,10)\qbezier(30,10)(32.5,13.75)(35,15)
		\qbezier(25,5)(27.5,6.25)(30,10)\qbezier(30,10)(32.5,6.25)(35,5)
		\qbezier(35,15)(45,20)(55,15)
		\qbezier(35,5)(45,0)(55,5)
		\qbezier(38,11)(45,6)(52,11)\qbezier(40,9.8)(45,13)(50,9.8)
		\qbezier(55,5)(65,10)(55,15)
		{\color{red}\put(15,7){\circle*{1}}\put(30,10){\circle*{1}}}
		\end{picture}
		\end{tabular}
	\end{tabular}
	\end{center}
	\caption{Schematic representations of a non-singular (left) and singular (right) complex algebraic curve. The non-singular curve $P(\lambda,g)=0$ on the left corresponds to a Riemann surface of genus $2$. Varying the parameters of the polynomial $P$, the $2$ red circles may be pinched into $2$ nodal points. This leads to the singular curve on the right, which corresponds after desingularization to a Riemann surface split into $2$ connected components (one sphere and one torus).}
	\label{fig 2-torus}
\end{figure}
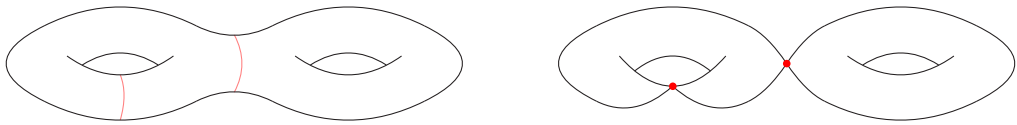

There exists a simple way to compute the genus from the knowledge of the coefficients of the polynomial $P(\lambda,g)$ using the Newton polygon, defined as the convex hull in $\mathbb{R}^{2}$ of the points $\{(j,k)\in\mathbb{Z}^{2},P^{(j,k)}(0,0)\neq0\}$. It can be shown that the genus is simply equal to the number of points with integer coordinates $(j,k)$ (independently on whether $P^{(j,k)}(0,0)$ is equal to zero or not) in the interior of the Newton polygon (i.e. excluding points on the boundary of the polygon). We emphasize however that since the desingularization procedure may reduce the genus for a singular algebraic curve compared to a non-singular perturbation, see figure~\ref{fig 2-torus}, this method only works as formulated above for non-singular algebraic curves: each nodal point then either decreases the genus by one or increases the number of connected components by one compared to the number given by the Newton polygon, and non-nodal singular points further decrease the genus or increase the number of connected components by some amount. The Newton polygon approach for the genus comes from an explicit construction using monomials $\lambda^{j}g^{k}$ from the interior of the Newton polygon, see e.g. \cite{E2018.1}, of a basis of the space of meromorphic differentials on $\R$ without poles, whose dimension is known to be equal to the genus of $\R$.

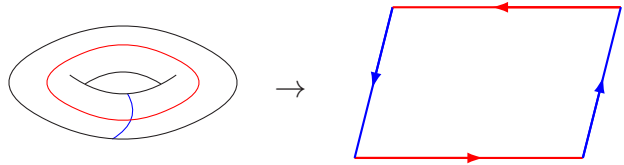
\begin{figure}
	\begin{center}
	\begin{tabular}{lll}
		\begin{tabular}{l}
		\begin{picture}(30,20)
		{\color{blue}\qbezier(15,2.5)(19,5)(17,8.5)}
		{\color{red}\qbezier(10,6)(0,10)(10,14)\qbezier(10,6)(15,4)(20,6)\qbezier(20,6)(30,10)(20,14)\qbezier(10,14)(15,16)(20,14)}
		\qbezier(5,5)(-5,10)(5,15)
		\qbezier(5,15)(15,20)(25,15)
		\qbezier(5,5)(15,0)(25,5)
		\qbezier(8,11)(15,6)(22,11)\qbezier(10,9.8)(15,13)(20,9.8)
		\qbezier(25,5)(35,10)(25,15)
		\end{picture}
		\end{tabular}
		&$\to$&
		\begin{tabular}{l}
		\begin{picture}(35,20)
		\put(0,0){\thicklines\color{red}\line(1,0){30}}\put(0,0){\thicklines\color{red}\vector(1,0){17}}
		\put(0,0){\thicklines\color{blue}\line(0.25,1){5}}\put(5,20){\thicklines\color{blue}\vector(-0.25,-1){2.7}}
		\put(5,20){\thicklines\color{red}\line(1,0){30}}\put(35,20){\thicklines\color{red}\vector(-1,0){17}}
		\put(30,0){\thicklines\color{blue}\line(0.25,1){5}}\put(30,0){\thicklines\color{blue}\vector(0.25,1){2.7}}
		\end{picture}
		\end{tabular}
	\end{tabular}
	\end{center}
	\caption{Planar representation of a torus obtained by cutting the torus along two curves intersecting at one point. Opposite edges in the resulting parallelogram on the right are identified in the torus and correspond to the two curves drawn on the torus on the left. The oriented contour made by the edges of the parallelogram corresponds for the torus to a closed path passing through both curves drawn on the torus twice, in both directions, and the integral of any meromorphic differential on this contour is necessarily equal to zero.}
	\label{fig cut torus}
\end{figure}

A useful planar representation of a connected orientable surface of genus $\mathrm{g}$ comes by cutting the surface along $2\mathrm{g}$ loops intersecting at the same point on the surface, see e.g. \cite{B2013.3}. For a torus, this gives the usual representation as a parallelogram with opposite edges identified, see figure~\ref{fig cut torus}. More generally, for genus $\mathrm{g}\geq1$, this leads to a polygon with $4\mathrm{g}$ edges identified two by two. The cutting path is then a graph on $\R$ with $F=1$ face, $E=2\mathrm{g}$ edges and $V=1$ vertex, which does correspond to an Euler characteristic $\chi=V-E+F=2-2\mathrm{g}$. We emphasize that the oriented contour made by the edges of the polygon corresponds for the surface to a closed path passing through all $2\mathrm{g}$ curves along which the surface has been cut twice, in both directions, see figure~\ref{fig cut torus} for the example of the torus. The integral of any meromorphic differential (see the next section) on this contour is then necessarily equal to zero.
\end{subsubsection}

\begin{subsubsection}{Meromorphic functions and meromorphic differentials}\hfill\\
When the algebraic curve $\mathcal{A}$ defined from $P(\lambda,g)=0$ is non-singular, it is possible to show that any function meromorphic on the corresponding Riemann surface $\R$ can be written as a rational function of $\lambda(p)$ and $g(p)$. If $\mathcal{A}$ has nodal points, however, any rational function of $\lambda(p)$ and $g(p)$ necessarily takes the same value at the points $p_{1}$ and $p_{2}$ on $\R$ corresponding to the same nodal point, see figure~\ref{fig cone}, and there exists additional meromorphic functions on $\R$ taking distinct values at $p_{1}$ and $p_{2}$, and which can not be expressed as rational functions of $\lambda(p)$ and $g(p)$.

Meromorphic functions on the Riemann sphere $\Ch$ are simply rational functions of some variable. For genus $\mathrm{g}\geq1$, on the other hand, there exists meromorphic functions that can not be expressed globally as rational functions. In particular, for genus one, meromorphic functions are elliptic functions, which can either be seen as meromorphic functions on $\C$ that are periodic in two directions, or as functions with periodic boundary conditions on a parallelogram that can be folded into a torus by identifying opposite edges, see figure~\ref{fig cut torus}.

Given a non-constant meromorphic function $f$ on $\R$, the antecedents of $a\in\Ch$ by $f$, i.e. the solutions $p\in\R$ of $f(p)=a$, are locally analytic functions of $a$ away from a finite number of values $a=f_{*}$ called the branch points of $f$ (and coinciding for $f$ equal to the function $g$ with the branch points of the algebraic curve discussed in section~\ref{section algebraic curves}). The number of antecedents $d$ away from branch points, called the degree of $f$, is constant, and the function $f$ is then also called a \textit{ramified covering} from $\R$ to $\Ch$. For any branch point $f_{*}\in\Ch$, there exists at least one ramification point $p_{*}\in\R$, antecedent of $f_{*}$ by $f$, such that in a neighbourhood of $p_{*}$, a local parameter for $\R$ is $z$ with $f(p)-f_{*}=z^{m}$, $m\geq2$. The ramification index $m=e_{p_{*}}$ corresponds to the multiplicity for $f$ of the value $f_{*}$ at $p_{*}$, and any value in $\Ch$ including branch points has then the same number $d$ of antecedents by $f$ \textit{counting multiplicity}. In particular, any non-constant meromorphic function on a Riemann surface has the same number of poles and zeroes, again counting multiplicity, and only constant functions have no poles and are holomorphic everywhere. Poles of meromorphic functions should then not be considered as particularly special points, but merely as the antecedents of the point $\infty\in\Ch$.

On the other hand, poles have a special place in complex analysis because of Cauchy's integral formula, but in the setting of a compact Riemann surface, one has to make the distinction between poles of meromorphic functions, for which the concept of a residue makes no sense, and poles of \textit{meromorphic differentials} and their residues, which are the object of Cauchy's integral formula. Working on a compact Riemann surface indeed forces to consider more closely meromorphic differentials, which are the extension to $\R$ of e.g. the integrand $\rmd g\,f(g)$ appearing in (\ref{Prob[M]}). A major difference with complex analysis on $\C$ is that it is not possible to use the same $\rmd g$ everywhere on $\R$ because $\rmd g$ is singular at the ramification points of $\R$ with respect to the parametrization $g$. Indeed, at such a point $p_{0}$ an analytic description requires switching to a local parameter $z$, such that $g(p)-g(p_{0})=z^{m}$, and the usual change of variable formula gives $\rmd g=m z^{m-1}\rmd z$, which is interpreted as the presence of a zero of order $m-1$ for the meromorphic differential $\rmd g$. More generally, a meromorphic differential $\omega$ written in a neighbourhood of a point $p_{0}\in\R$ as $\omega=\varphi(z)\rmd z$ with $z$ a local parameter vanishing at $p_{0}$ is said to have a pole (respectively a zero) of order $n$ if $\varphi$ has a pole (resp. a zero) of order $n$ at $0$. The degrees of poles and zeroes are independent from the choice of local parameter $z$, and so is the coefficient of $z^{-1}$ in the expansion of $\varphi(z)$ near $z=0$, which is called the residue of $\omega$ at $p_{0}$.

Given a connected Riemann surface $\R$ and a simple closed contour $\Gamma$ on $\R$ which is \textit{contractible} (i.e. $\Gamma$ can be deformed continuously on $\R$ into a point), we consider the planar representation of $\R$ as a polygon with edges identified two by two mentioned at the end of section~\ref{section genus}. Since $\Gamma$ is contractible, we can choose a cutting path on $\R$ that does not intersect $\Gamma$, and $\Gamma$ then splits the polygon into an inside domain, which is simply connected, and an outside domain. Cauchy's integral formula then states that $\oint_{\Gamma}\omega$ (defined as usual by taking local coordinates) is equal to $2\rmi\pi$ (respectively $-2\rmi\pi$) times the sum of the residues of the poles of $\omega$ in the inside domain if the curve $\Gamma$ has positive (resp. negative) orientation with respect to the inside domain. Since the integral of $\omega$ over the polygon is necessarily equal to zero, see figure~\ref{fig cut torus}, we observe in particular that the sum of all the residues of a meromorphic differential is necessarily equal to zero.

Compared with complex analysis on $\C$, another type of contour integral of meromorphic differentials has to be considered for compact Riemann surfaces, namely integrals over non-contractible closed contours. Such integrals are called \textit{periods} of the meromorphic differential. An important uniqueness theorem used in the following states that given a connected compact Riemann surface $\R$, $n$ distinct points $p_{1},\ldots,p_{n}\in\R$ and $n$ real numbers $\alpha_{1},\ldots,\alpha_{n}$ such that $\alpha_{1}+\ldots+\alpha_{n}=0$, there exists a unique meromorphic differential $\omega$ on $\R$ whose only poles are simple poles with residues $\alpha_{i}$ at the points $p_{i}$, $i=1,\ldots,n$ and with purely imaginary periods, see e.g. \cite{GW1987.1} for a detailed proof. In particular, any contour integral of $\omega$ over a closed curve on $\R$ is purely imaginary.
\end{subsubsection}

\begin{subsubsection}{Riemann-Hurwitz formula for the genus}\hfill\\
The genus $\mathrm{g}$ of a Riemann surface $\R$ with $K$ connected components can be computed directly from the ramification data of any non-constant meromorphic function $f$ on $\R$. The Riemann-Hurwitz formula states that
\begin{equation}
\label{Riemann-Hurwitz}
\mathrm{g}=-d+K+\frac{1}{2}\sum_{p\in\R}(e_{p}-1)\;,
\end{equation}
where $d$ is the degree of $f$ and $e_{p}$ the ramification indices for $f$ (with $e_{p}=1$ for $p$ not a ramification point of $f$). For $\R$ with a single connected component, considering a graph $\mathcal{G}$ on $\R$ obtained by lifting with $f^{-1}$ a graph on $\Ch$ whose vertices are the branch points of $f$, the Riemann-Hurwitz formula is a simple consequence of the expression $\chi=V-E+F$ for the Euler characteristic $\chi=2-2\mathrm{g}$ of $\R$ in terms of the number $V$ of vertices, the number $E$ of edges and the number $F$ of faces of $\mathcal{G}$. Summing over all the connected components of $\R$ then leads to (\ref{Riemann-Hurwitz}).

The Riemann-Hurwitz formula (\ref{Riemann-Hurwitz}) is especially useful to compute the genus of $\R$ when one can not work with the algebraic curve $\mathcal{A}$ and the Newton polygon, in particular when the algebraic curve has many singular points, see the examples of TASEP and ASEP in sections~\ref{section TASEP} and \ref{section ASEP}.

A consequence of the Riemann-Hurwitz formula is that the number of zeroes minus the number of poles of a meromorphic differential, counted with multiplicity, is equal to $2\mathrm{g}-2$. In order to show this, let us consider a meromorphic function $f$ of degree $d$, assumed for simplicity to have only simple poles, and the corresponding exact differential $\rmd f$. The ramification points $p_{*}$ of $f$ are the zeroes of $\rmd f$: indeed, in terms of a local parameter $z$, $f(p)=f_{*}+z^{m}$ with $m\geq2$ implies $\rmd f=mz^{m-1}\rmd z$, and $p_{*}$ is a zero of $\rmd f$ of order $m-1$. Additionally, the poles of $\rmd f$ are the poles of $f$: $f(p)=z^{-1}$ implies $\rmd f=-\rmd z/z^{2}$, which is a pole of order $2$ for $\rmd f$. The number of zeroes minus the number of poles of $\rmd f$, counted with multiplicity, is thus equal to $\sum_{p\in\R}(e_{p}-1)-2d$ which, using the Riemann-Hurwitz formula (\ref{Riemann-Hurwitz}), does reduce to $2\mathrm{g}-2$. If $f$ has multiple poles, these poles are also ramification points for $f$ and the counting is slightly modified, but the conclusion still holds. Finally, since the ratio of two meromorphic differential is a meromorphic function, which has as many zeroes as poles, the result is also true for differentials that are not exact.
\end{subsubsection}

\end{subsection}

\begin{subsection}{Riemann surface \texorpdfstring{$\R$}{R} and eigenstates of \texorpdfstring{$M(g)$}{M(g)}}
\label{section eigenvectors on R}
We consider in this section the eigenstates of $M(g)$ from the point of view of the Riemann surface $\R$ introduced in the previous section. This perspective is standard, and is used for instance in the theory of classical integrable systems for the eigenstates of Lax matrices depending on a spectral parameter, see e.g. \cite{E2018.1,BBEIM1994.1,BBT2003.1}.

Let us consider a point $p$ on the algebraic curve $\mathcal{A}$ defined above. If $p$ is neither a singular point of $\mathcal{A}$ nor a ramification point for the variable $g$, then $M(g(p))$ has a single eigenstate with eigenvalue $\lambda(p)$, corresponding to single (up to normalization) left and right eigenvectors $\langle\psi(p)|$ and $|\psi(p)\rangle$. If $p$ is a singular point of $\mathcal{A}$, then several eigenvalues of $M(g(p))$ coincide with $\lambda(p)$. At the level of the corresponding Riemann surface $\R$, however, the neighbourhoods of the points $p_{i}\in\R$ corresponding to the same singular point $p\in\mathcal{A}$ are separated, see figure~\ref{fig cone} for the case of nodal points, and a unique pair of left and right eigenvectors can still be defined at each $p_{i}\in\R$ by continuity. Finally, if $p$ is a ramification point for the variable $g$, with ramification index $m$, then $m$ eigenstates of $M(g(p))$ coincide, and eigenvectors no longer form a complete basis of the vector space of dimension $|\Omega|$ on which $M(g)$ acts. In that case, $M(g(p))$ is not diagonalizable, but has a representation in terms of Jordan blocks.

We emphasize that unlike singular points, which are the result of an accidental degeneracy in $\mathcal{A}$ for $(\lambda,g)$, the existence of ramification points is generic, and simply come from the fact that representing a surface by a covering map of dimension $d\geq2$ necessarily comes with ramification points, as can be seen from the Riemann-Hurwitz formula (\ref{Riemann-Hurwitz}). Ramification may happen at any value of the parameter $g\in\Ch$, see for instance \cite{AM2010.1} for an example where $g=1$ is a branch point, and the non-deformed Markov matrix $M$ itself has Jordan blocks.

We have seen that there exists a correspondence between points on the Riemann surface $\R$ and eigenstates: each point $p\in\R$ is associated to a single eigenstate of $M(g(p))$, with eigenvalue $\lambda(p)$ and left and right eigenvectors $\langle\psi(p)|$ and $|\psi(p)\rangle$. Both $\lambda(p)$ and $g(p)$ are meromorphic functions on $\R$, see the previous section. The same is true for all the coordinates of (properly normalized) eigenvectors: choosing for instance the normalization $\langle\psi(p)|C_{0}\rangle=1$ and $\langle C_{0}|\psi(p)\rangle=1$ for some state $C_{0}$, the other coefficients of the eigenvectors are solution of the linear equations $\langle\psi(p)|M(g(p))=\lambda(p)\langle\psi(p)|$ and $M(g(p))|\psi(p)\rangle=\lambda(p)|\psi(p)\rangle$, whose coefficients are the matrix elements of $M(g(p))$, which are monomials in $g(p)$. Solving these linear equations using e.g. Cramer's rule, all the entries of both eigenvectors can then be expressed as rational functions of $\lambda(p)$ and $g(p)$, and are thus meromorphic functions on $\R$.
\end{subsection}

\begin{subsection}{Probability as a contour integral on a Riemann surface}
\label{section proba integral on R}
We finally come back to the probability (\ref{Prob[M]}) of the integer counting process $Q_{t}$. Calling $\tilde{g}$ the integration variable to avoid confusion with the function $g:\R\to\Ch$, the eigenstate expansion of $M(\tilde{g})$ gives
\begin{equation}
\label{Prob[N]}
\P(Q_{t}=Q)=\oint_{\gamma}\frac{\rmd\tilde{g}}{2\rmi\pi\tilde{g}^{Q+1}}\,\sum_{p\in g^{-1}(\tilde{g})}\mathcal{N}(p)\,\rme^{t\lambda(p)}\;,
\end{equation}
where $\mathcal{N}$ is the meromorphic function on $\R$ defined by
\begin{equation}
\label{N(p)}
\mathcal{N}(p)=\frac{\sum_{C\in\Omega}\langle C|\psi(p)\rangle\langle\psi(p)|P_{0}\rangle}{\langle\psi(p)|\psi(p)\rangle}\;,
\end{equation}
and which is independent of the choice of normalization for the eigenvectors. The function $g$ on $\R$ has degree $|\Omega|$, and the sum $\sum_{p\in g^{-1}(\tilde{g})}$ in (\ref{Prob[N]}) over all the points $p\in\R$ such that $g(p)=\tilde{g}$ represents the sum over all $|\Omega|$ eigenstates of $M(\tilde{g})$.

\begin{figure}
	\begin{tabular}{lll}
	\begin{tabular}{l}
	\begin{picture}(50,25)
		\put(0,0){\line(1,0){45}}\put(0,0){\line(1,1){20}}
		\put(32.5,11){\scalebox{1}[0.5]{\circle{25}}}
		\put(19,11){\line(1,0){2}}\put(21.7,10.5){$g_{0}$}
		\put(32.5,11){\circle*{1}}\put(34,9.7){$0$}
		\put(32.5,4.75){\thicklines\vector(1,0){2}}
		\put(42,17){$\gamma$}
		\put(7,1.5){$\Ch$}
	\end{picture}
	\end{tabular}
	&\begin{tabular}{c}$g^{-1}$\\[-2mm]$\longrightarrow$\end{tabular}&
	\hspace{-5mm}
	\begin{tabular}{l}
	\begin{picture}(80,35)
		\put(0,0){\line(1,0){80}}\put(0,0){\line(1,1){35}}
		\put(15,2){\includegraphics[width=30mm]{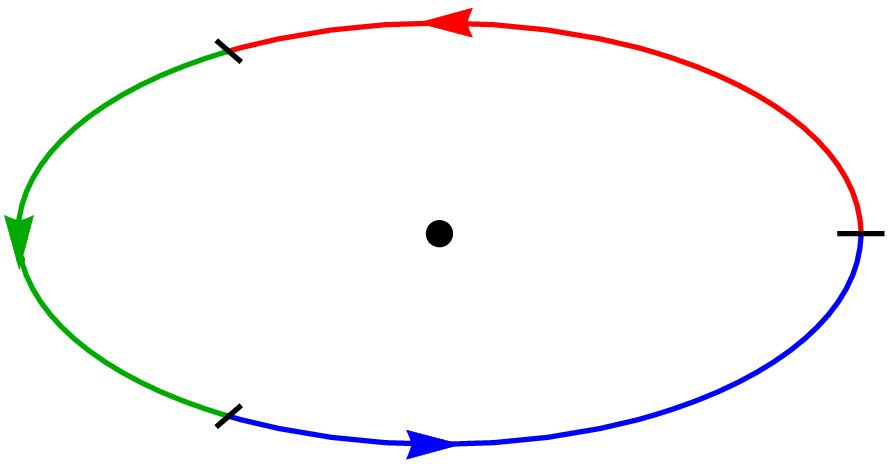}}
		\put(32,18){\includegraphics[width=30mm]{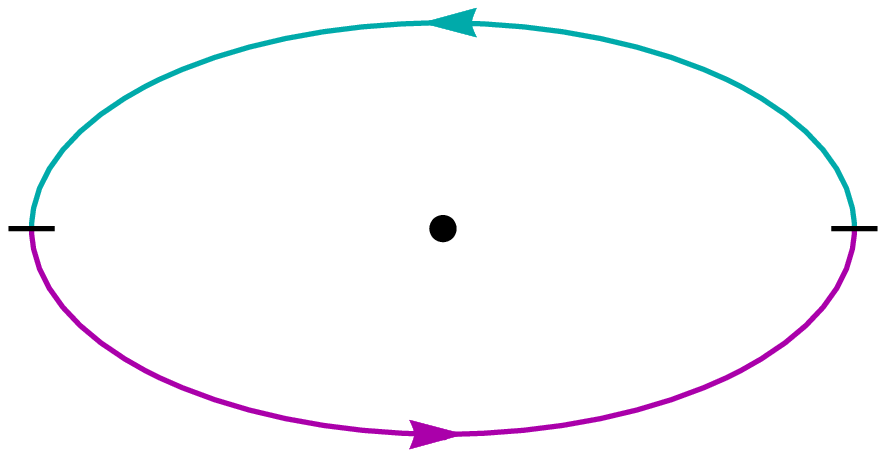}}
		\put(50,2){\includegraphics[width=30mm]{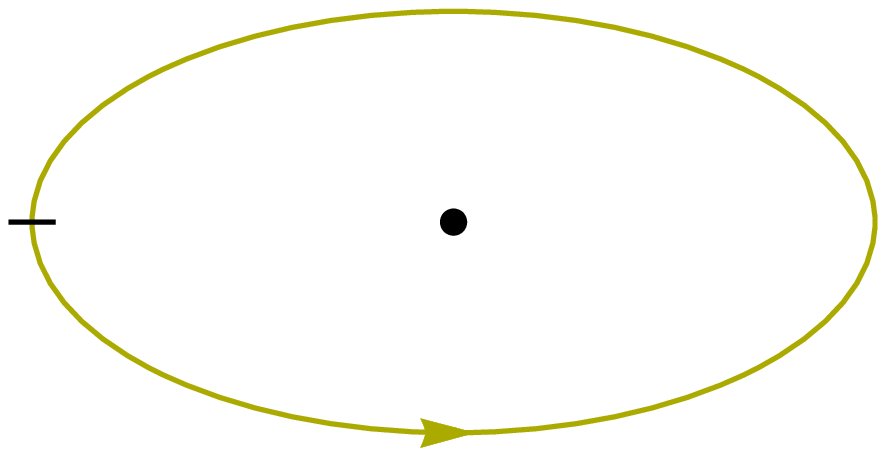}}
		\put(46,12.5){$\Gamma$}
		\put(5.5,1.5){$\R$}
	\end{picture}
	\end{tabular}
	\end{tabular}
	\caption{Lifts of a small loop $\gamma\subset\Ch$ around $0$ (left) to the antecedents on $\R$ of $g_{0}\in\gamma$ (right) by a ramified covering $g$ of degree $6$. The $6$ elements of $g^{-1}(g_{0})$ are represented on the right by the small segments intersecting the ovals. The point $0\in\Ch$ (left) has three antecedents (black dots on the right) by $g$: two ramification points, with ramification indices $2$ (top) and $3$ (lower left), around which $\gamma$ lifts to open paths on $\R$ represented with distinct colors, and a regular point (lower right), around which $\gamma$ lifts to a loop on $\R$. The union $\Gamma$ of the $6$ lifts forms three disjoint loops on $\R$.}
	\label{fig lift path}
\end{figure}

\begin{subsubsection}{Contour of integration \texorpdfstring{$\Gamma$}{Gamma} on \texorpdfstring{$\R$}{R}}\hfill\\
Choosing a loop $\gamma$ which avoids all the branch points for $g$ and an origin $g_{0}\in\gamma$, any point $p_{0}\in g^{-1}(g_{0})$ can be followed unambiguously when $\tilde{g}$ goes from $g_{0}$ to $g_{0}$ along $\gamma$. The final point of the lifted path $p\in g^{-1}(\tilde{g})$ on $\R$ still belongs to $g^{-1}(g_{0})$ but may be distinct from $p_{0}$ if branch points for $g$ are enclosed by $\gamma$, and the lifted path is then not a loop. However, the union $\Gamma$ of the lifted paths starting from any point of $g^{-1}(g_{0})$ does form a union of closed contours on $\R$. For a small loop $\gamma$ around a branch point $g_{*}$ of $g$, the closed loop around $p_{*}\in g^{-1}(g_{*})$ obtained by lifting comes in particular as the union of $m$ open paths, with $m$ the ramification index of $p_{*}$ for $g$, see figure~\ref{fig lift path}.

We deduce from the discussion in the previous paragraph that for any meromorphic function $f$ on $\R$, one has
\begin{equation}
\label{intgamma->intGamma}
\oint_{\gamma}\rmd\tilde{g}\sum_{p\in g^{-1}(\tilde{g})}f(p)=\oint_{\Gamma}\rmd g\,f(p)\;,
\end{equation}
where $\Gamma=g^{-1}(\gamma)=\bigcup_{\tilde{g}\in\gamma}g^{-1}(\tilde{g})$ is a union of closed loops on $\R$. In the case of (\ref{Prob[N]}), the integrand is actually not meromorphic since the factor $\rme^{t\lambda(p)}$ has essential singularities at the poles of $\lambda$, but (\ref{intgamma->intGamma}) can still be used since $\rme^{t\lambda(p)}$ is not more ramified than $\lambda(p)$.

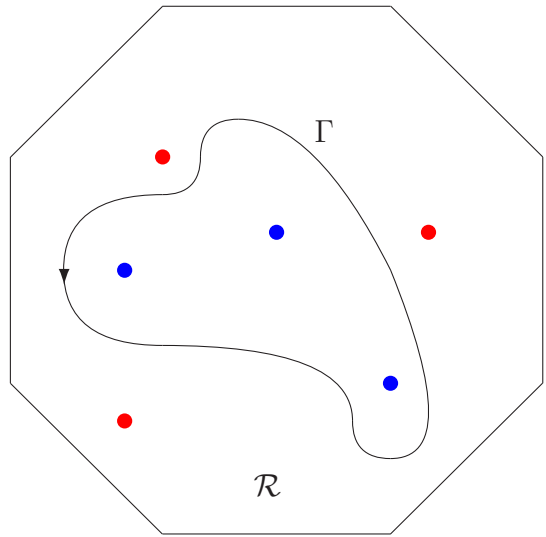
\begin{figure}
	\begin{center}
	\begin{picture}(70,70)
		\put(20,0){\line(1,0){30}}
		\put(50,0){\line(1,1){20}}
		\put(70,20){\line(0,1){30}}
		\put(70,50){\line(-1,1){20}}
		\put(50,70){\line(-1,0){30}}
		\put(20,70){\line(-1,-1){20}}
		\put(0,50){\line(0,-1){30}}
		\put(0,20){\line(1,-1){20}}
		\put(15,15){\color{red}\circle*{2}}
		\put(55,40){\color{red}\circle*{2}}
		\put(20,50){\color{red}\circle*{2}}
		\put(50,20){\color{blue}\circle*{2}}
		\put(15,35){\color{blue}\circle*{2}}
		\put(35,40){\color{blue}\circle*{2}}
		\qbezier(7,35)(7,25)(20,25)
		\qbezier(20,25)(45,25)(45,15)
		\qbezier(45,15)(45,10)(50,10)
		\qbezier(50,10)(60,10)(50,35)
		\qbezier(50,35)(40,55)(30,55)
		\qbezier(30,55)(25,55)(25,50)
		\qbezier(25,50)(25,45)(20,45)
		\qbezier(20,45)(7,45)(7,35)
		\put(7.05,33){\thicklines\vector(0,-1){0}}
		\put(40,52){$\Gamma$}
		\put(32,5){$\R$}
	\end{picture}
	\end{center}
	\caption{Simple contour $\Gamma$ with a single component for (\ref{Prob[int R]}), pictured on a planar representation of a Riemann surface $\R$ with one connected component and genus $2$ obtained after cutting $\R$ along four closed curves intersecting at a single point on $\R$. The blue points, inside $\Gamma$, are the elements of $g^{-1}(0)$. The red points, outside $\Gamma$, are the elements of $g^{-1}(\infty)$.}
	\label{fig Gamma}
\end{figure}

In terms of the function $\mathcal{N}$ defined in (\ref{N(p)}), of the eigenvalue $\lambda$ and of the contour $\Gamma\subset\R$ discussed above, we finally obtain the probability of the counting process $Q_{t}$ as
\begin{equation}
\label{Prob[int R]}
\P(Q_{t}=Q)=\oint_{\Gamma}\frac{\rmd g}{2\rmi\pi}\,\frac{\mathcal{N}(p)\,\rme^{t\lambda(p)}}{g(p)^{Q+1}}\;,
\end{equation}
where $p$ is the current point on $\Gamma$. The meromorphic differential $\rmd g$ is evaluated at $p$. From the planar representation of $\R$ as a polygon discussed at the end of section~\ref{section genus}, we observe that the contour $\Gamma$ can always be deformed into a single simple closed contour on each connected component of $\R$, see figure~\ref{fig Gamma}, as long as the contour does not cross the poles of the integrand, studied in the next section.
\end{subsubsection}

\begin{subsubsection}{Pole structure}\hfill\\
\label{section pole structure}
Beside poles of infinite order in $\lambda^{-1}(\infty)\subset g^{-1}(0)\cup g^{-1}(\infty)$ (eigenvalues can only become infinite when some matrix elements of $M(g)$ are infinite) coming from the function $\rme^{t\lambda(p)}$, the integrand of (\ref{Prob[int R]}) also has poles of finite order coming from the differential $\rmd g\,\mathcal{N}(p)/g(p)^{Q+1}$. The factor $g(p)^{-Q-1}$ only contributes poles in $g^{-1}(0)$ or $g^{-1}(\infty)$ depending on the sign of $Q+1$.

We now focus on the remaining factor $\rmd g\,\mathcal{N}(p)$, and assume that $M(g)$ is generic to avoid pathological cases. In a neighbourhood of $p\in\R$, we choose a normalization of the left and right eigenvectors $\langle\psi(p)|$ and $|\psi(p)\rangle$ so that all their entries are finite. Then, the numerator of $\mathcal{N}(p)$ in (\ref{N(p)}) may not diverge. Additionally, the denominator $\langle\psi(p)|\psi(p)\rangle$ can not vanish if $M(g(p))$ is diagonalizable. At ramification points $p_{*}$ for $g$ with ramification index $m\geq2$, however, $M(g(p_{*}))$ is not diagonalizable and $\langle\psi(p)|\psi(p)\rangle$ has a zero of order $m-1$ (i.e. eigenstates are self-orthogonal \cite{M2011.2}) since in a neighbourhood of $p_{*}$ the eigenvectors at the $m$ points $p$ converging to $p_{*}$ with the same value of $g(p)$ are orthogonal to each other. The poles of $\mathcal{N}(p)$ outside $g^{-1}(0)\cup g^{-1}(\infty)$ are thus necessarily ramification points $p$ for $g$, where $M(g(p))$ is not diagonalizable, and their number counted with multiplicity is equal from the Riemann-Hurwitz formula (\ref{Riemann-Hurwitz}) for the function $g$ to $2\mathrm{g}-2+2|\Omega|$ with $\mathrm{g}$ the genus of $\R$ (the number of connected components of $\R$ is $K=1$ since the algebraic curve is generic). These poles of $\mathcal{N}(p)$ however cancel in (\ref{Prob[int R]}) with zeroes of the meromorphic differential $\rmd g=mz^{m-1}\rmd z$, where $z$ is a local parameter at $p$.

The differential in the integrand of (\ref{Prob[int R]}) thus only has poles in $g^{-1}(0)\cup g^{-1}(\infty)$. Since $\gamma$ in (\ref{Prob[N]}) must have winding number one around $0$, the elements of $g^{-1}(0)$ are contained inside a simply connected domain on $\R$ with boundary $\Gamma$, see figure~\ref{fig Gamma}, while the elements of $g^{-1}(\infty)$ are outside that domain. From Cauchy's integral formula, the probability $\P(Q_{t}=Q)$ is then equal to the sum of the residues of the differential $\rmd g\,\mathcal{N}(p)\,\rme^{t\lambda(p)}/g(p)^{Q+1}$ at the points of $g^{-1}(0)$, or minus the sum of the residues at the points of $g^{-1}(\infty)$.
\end{subsubsection}

\begin{subsubsection}{Exponential representation}\hfill\\
\label{section exponential form}
We discuss now an exponential representation for the integrand of (\ref{Prob[int R]}) in the case where $\R$ has a single connected component, and which appears naturally for the simple example treated in section~\ref{section simple example} and for TASEP in section~\ref{section TASEP}.

Any meromorphic function $f$ on $\R$ can be expressed as $f(p)=f(p_{0})\,\rme^{\int_{p_{0}}^{p}\rmd\log f}$, independent of the choice of integration path between $p_{0}$ and $p$. The meromorphic differential $\rmd\log f$ has only simple poles, otherwise $f$ would have essential singularities. Additionally, the poles of $\rmd\log f$ are the zeroes and the poles of $f$. The residues of the poles of $\rmd\log f$ are necessarily integers, equal to the orders of the zeroes and minus the orders of the poles of $f$.

The initial point $p_{0}$ may be chosen arbitrarily. The best choice for the integrand in (\ref{Prob[int R]}) is however the \textit{stationary point} $o\in\R$, corresponding to the stationary eigenstate of the non-deformed Markov matrix $M=M(1)$, and characterized uniquely by $g(o)=1$ and $\lambda(o)=0$ from the Perron-Frobenius theorem. Indeed, one has additionally $\mathcal{N}(o)=1$ since $\langle\psi(o)|$ is proportional to $\sum_{C\in\Omega}\langle C|$ and the initial probabilities are normalized as $\sum_{C\in\Omega}\langle C|P_{0}\rangle=1$. The factor in front of the exponential for $f(p)=\mathcal{N}(p)\,\rme^{t\lambda(p)}/g(p)^{U+1}$ is thus simply equal to $1$.

The meromorphic differential $\rmd\log f$ is uniquely determined by the knowledge of its poles and residues. Indeed, any contour integral of $\rmd\log f$ along a closed loop $\Gamma\subset\R$ must be equal to an integer multiple of $2\rmi\pi$, otherwise analytic continuation of $\rme^{\int_{p_{0}}^{p}\rmd\log f}$ for $p$ along $\Gamma$ would not leave the function unchanged. The periods of $\rmd\log f$ are in particular purely imaginary, and the uniqueness property for meromorphic differentials with specified simple poles and their real residues discussed in section~\ref{section Riemann surfaces} applies. In the simplest cases, this is sufficient to find an explicit expression for $\rmd\log\mathcal{N}$.

In particular, with stationary initial condition, the function $\mathcal{N}$ has double zeroes at any $p\in g^{-1}(1)\setminus\{o\}$ by orthogonality of the eigenstates since $\langle\psi(o)|\propto\sum_{C\in\Omega}\langle C|$ and $|\psi(o)\rangle\propto|P_{0}\rangle$, which accounts for $2|\Omega|-2$ zeroes of $\mathcal{N}$. If $M(g)$ is generic, $\mathcal{N}$ has $2\mathrm{g}-2+2|\Omega|$ poles with $\mathrm{g}$ the genus of $\R$, see the previous section, and the number of extra zeroes of $\mathcal{N}$ with $g\neq1$ is then equal to $2\mathrm{g}$. For $\mathrm{g}=0$, the function $\mathcal{N}$ has in particular no extra zeroes, see section~\ref{section simple example} for an example where $\rmd\log\mathcal{N}$ can be guessed from such considerations. It may also happen that all the extra zeroes of $\mathcal{N}$ for some initial conditions have simple values for $g$, see sections~\ref{section TASEP} and \ref{section ASEP} for TASEP and ASEP.

Writing $\rmd\log\mathcal{N}=\rmd g\,\partial_{g}\log\mathcal{N}$ allows to express $\rmd\log\mathcal{N}$ explicitly in terms of the derivative of the eigenvectors with respect to $g$. These derivatives can be computed explicitly from the eigenvalue equation. In terms of $M'(g)=\partial_{g}M(g)$, one finds $\partial_{g}\lambda(p)=\langle\psi(p)|M'(g)|\psi(p)\rangle/\langle\psi(p)|\psi(p)\rangle$ (sometimes called the Hellmann-Feynman theorem) and $\langle\psi(p)|\partial_{g}|\psi(q)\rangle=-(\partial_{g}\langle\psi(p)|)|\psi(q)\rangle=\langle\psi(q)|M'(g)|\psi(p)\rangle/(\lambda(p)-\lambda(q))$ for $p\neq q$ with $g=g(p)=g(q)$. This leads to
\begin{eqnarray}
&&\fl\hspace{15mm} \rmd\log\mathcal{N}(p)=\rmd g\sum_{q\neq p\atop g(q)=g(p)}\frac{1}{\lambda(p)-\lambda(q)}\Bigg(\frac{\langle\psi(q)|M'(g)|\psi(p)\rangle}{\langle\psi(q)|\psi(q)\rangle}\frac{\sum_{C\in\Omega}\langle C|\psi(q)\rangle}{\sum_{C\in\Omega}\langle C|\psi(p)\rangle}\\
&&\hspace{65mm} +\frac{\langle\psi(p)|M'(g)|\psi(q)\rangle}{\langle\psi(q)|\psi(q)\rangle}\frac{\langle\psi(q)|P_{0}\rangle}{\langle\psi(p)|P_{0}\rangle}\Bigg)\;,\nonumber
\end{eqnarray}
where the sum is over all the point $q\in\R$ distinct from $p$ and with the same value of $g$ as $p$. This expression is convenient numerically since it can be computed directly from the eigenstates of $M(g)$ at a given value of $g$, without needing to follow them under changes of $g$.

An important issue with the exponential representation is the choice of the function that is written in exponential form, and thus the choice of the meromorphic differential which is left out. Indeed, the seemingly natural choice $f(p)=\mathcal{N}(p)\,\rme^{t\lambda(p)}/g(p)^{U+1}$ is in fact rather arbitrary at the level of the Riemann surface since the remaining differential $\rmd g$ in (\ref{Prob[int R]}) is not particularly special. What happens for TASEP, see equation (\ref{Prob TASEP}) in section~\ref{section TASEP} below, is that one should rather single out in the integrand of (\ref{Prob[int R]}) the differential $\frac{\rmd B}{B}=\kappa(p)^{-1}\frac{\rmd g}{g}$, where $B$ is a meromorphic function on $\R$ appearing naturally in the Bethe ansatz formulation. It is currently an open question whether there exists a natural choice of function $B$ for more general classes of integer counting processes, coming with some of the special properties that the one for TASEP displays.
\end{subsubsection}

\begin{subsubsection}{Multiple time statistics}\hfill\\
The Riemann surface approach extends easily to joint statistics of $Q_{t}$ at multiple times. Considering $n$ times ordered as $0<t_{1}<\ldots<t_{n}$ and using repeatedly the definition of conditional probabilities and the Markov property, one finds that the joint probability $\P(Q_{t_{1}}=Q_{1},\ldots,Q_{t_{n}}=Q_{n},C_{t_{1}}=C_{1},\ldots,C_{t_{n}}=C_{n})$ with $C_{t_{\ell}}$ the state of the system at time $t_{\ell}$ is equal to
\begin{equation}
\sum_{C_{0}\in\Omega}P_{0}(C_{0})\prod_{\ell=1}^{n}\P(Q_{t_{\ell}}=Q_{\ell},C_{t_{\ell}}=C_{\ell}|Q_{t_{\ell-1}}=Q_{\ell-1},C_{t_{\ell-1}}=C_{\ell-1})\;.
\end{equation}
The generating function
\begin{equation}
\label{GF[P] multi-time}
\langle h_{1}^{Q_{t_{1}}}\ldots\,h_{n}^{Q_{t_{n}}}\rangle=\sum_{Q_{1},\ldots,Q_{n}\in\mathbb{Z}}h_{1}^{Q_{1}}\ldots\,h_{n}^{Q_{n}}\,\P(Q_{t_{1}}=Q_{1},\ldots,Q_{t_{n}}=Q_{n})
\end{equation}
is then equal to
\begin{eqnarray}
&&\fl\hspace{5mm}
\langle h_{1}^{Q_{t_{1}}}\ldots\,h_{n}^{Q_{t_{n}}}\rangle=
\sum_{Q_{1},\ldots,Q_{n}\in\mathbb{Z}}
h_{1}^{Q_{1}}\ldots\,h_{n}^{Q_{n}}
\sum_{C_{0},\ldots,C_{n}\in\Omega}
P_{0}(C_{0})\\
&&\hspace{30mm} \times\prod_{\ell=1}^{n}\P(Q_{t_{\ell}}=Q_{\ell},C_{t_{\ell}}=C_{\ell}|Q_{t_{\ell-1}}=Q_{\ell-1},C_{t_{\ell-1}}=C_{\ell-1})\nonumber
\end{eqnarray}
Using translation invariance in time and $\langle h^{Q_{t}}\rangle_{C_{0}\to C_{1}}=\langle C_{1}|\rme^{tM(h)}|C_{0}\rangle$ for the average over all histories starting from state $C_{0}$ at time $0$ and ending in state $C_{1}$ at time $t$, we finally obtain
\begin{equation}
\label{GF[M] multiple-time}
\langle h_{1}^{Q_{t_{1}}}\ldots\,h_{n}^{Q_{t_{n}}}\rangle=\sum_{C}\langle C|\prod_{\ell=n}^{1}\rme^{(t_{\ell}-t_{\ell-1})M(\prod_{m=\ell}^{n}h_{m})}|P_{0}\rangle\;,
\end{equation}
where $t_{0}=0$. From (\ref{GF[P] multi-time}), the joint probability can be extracted with residues as
\begin{equation}
\P(Q_{t_{1}}=Q_{1},\ldots,Q_{t_{n}}=Q_{n})=\oint\frac{\rmd h_{1}\ldots\rmd h_{n}}{(2\rmi\pi)^{n}}\,\frac{\langle h_{1}^{Q_{t_{1}}}\ldots\,h_{n}^{Q_{t_{n}}}\rangle}{h_{1}^{Q_{1}+1}\ldots h_{n}^{Q_{n}+1}}\;,
\end{equation}
where the contours of integration encircle $0$ once. Using (\ref{GF[M] multiple-time}) and the change of variables $g_{\ell}=\prod_{m=\ell}^{n}h_{m}$, with Jacobian $\prod_{\ell=2}^{n}g_{\ell}^{-1}$, leads to
\begin{equation}
\fl\hspace{5mm}
\P(Q_{t_{1}}=Q_{1},\ldots,Q_{t_{n}}=Q_{n})=\oint\frac{\rmd g_{1}\ldots\rmd g_{n}}{(2\rmi\pi)^{n}}\,\frac{\sum_{C}\langle C|\prod_{\ell=n}^{1}\rme^{(t_{\ell}-t_{\ell-1})M(g_{\ell})}|P_{0}\rangle}{\prod_{\ell=1}^{n}g_{\ell}^{1+Q_{\ell}-Q_{\ell-1}}}\;,
\end{equation}
where $Q_{0}=0$. As for the statistics at a single time $t$, the integrals over the $g_{\ell}$ can finally be replaced by $n$ contour integrals on $\R$ after the expansion over the eigenstates of the $M(g_{\ell})$. One has
\begin{eqnarray}
\label{Prob[int R] multiple-time}
&&\fl\hspace{5mm}
\P(Q_{t_{1}}=Q_{1},\ldots,Q_{t_{n}}=Q_{n})=\oint_{\Gamma}\frac{\rmd g_{1}\ldots\rmd g_{n}}{(2\rmi\pi)^{n}}\,\Big(\prod_{\ell=1}^{n}\frac{\rme^{(t_{\ell}-t_{\ell-1})\lambda(p_{\ell})}}{g(p_{\ell})^{1+Q_{\ell}-Q_{\ell-1}}}\Big)\\
&&\hspace{35mm}
\times\frac{\sum_{C}\langle C|\psi(p_{n})\rangle(\prod_{\ell=1}^{n-1}\langle\psi(p_{l+1})|\psi(p_{\ell})\rangle)\langle\psi(p_{1})|P_{0}\rangle}{\prod_{\ell=1}^{n}\langle\psi(p_{\ell})|\psi(p_{\ell})\rangle}\;,\nonumber
\end{eqnarray}
with $\rmd g_{\ell}$ evaluated at $p_{\ell}$, and the contour $\Gamma$ as in (\ref{Prob[int R]}).

One can consider instead the cumulative probability $\P(Q_{t_{1}}\leq Q_{1},\ldots,Q_{t_{n}}\leq Q_{n})$, obtained from (\ref{Prob[int R] multiple-time}) by summing $Q_{\ell}$ from $-\infty$. The geometric series give the constraints $|g(p_{1})|<\ldots<|g(p_{n})|<1$ for the contours of integration, and the integrand has the additional factor $(1-g(p_{n}))^{-1}\prod_{\ell=1}^{n-1}(1-g(p_{\ell})/g(p_{\ell+1}))^{-1}$ compared to (\ref{Prob[int R] multiple-time}), corresponding to apparent poles when $g(p_{n})=1$ and $g(p_{\ell})=g(p_{\ell+1})$. Because of orthogonality of the eigenstates, however, the poles at $g(p_{n})=1$ actually cancel with the factor $\sum_{C}\langle C|\psi(p_{n})\rangle$ in the integrand except if $p_{n}$ is the stationary point $o$, and the poles at $g(p_{\ell})=g(p_{\ell+1})$ cancel with the factor $\langle\psi(p_{\ell+1})|\psi(p_{\ell})\rangle$ except if $p_{\ell}=p_{\ell+1}$. The contours of integration for the $p_{\ell}$ must then be nested, and enclose both $o$ and the points with $g=0$.
\end{subsubsection}

\end{subsection}

\end{section}

\begin{section}{A simple example: monitoring a single transition \texorpdfstring{$C_{1}\to C_{2}$}{C1->C2}}
\label{section simple example}
In this section, we show how the formalism of the previous section can be applied to the case where a single transition of a Markov process is monitored. This example is particularly simple since the corresponding Riemann surface is of genus zero, which allows us to simply guess the exact form of the meromorphic differential $\rmd\log\mathcal{N}$ for stationary initial condition.

\begin{subsection}{Definition of the model}
We consider in this section a general Markov process on a finite number of states $C\in\Omega$, ergodic, with Markov matrix $M$. Additional genericness requirements will be needed as several points of the calculation, but can eventually be lifted by continuity on the final expression for the probability.

Given two distinct states $C_{1}$ and $C_{2}$ in $\Omega$ with non-zero transition rate $w_{C_{1}\to C_{2}}$, the deformation $M(g)$, defined by
\begin{equation}
\label{M(g) single transition}
\langle C'|M(g)|C\rangle=\Bigg\{\!\!
\begin{array}{cc}
g\,w_{C_{1}\to C_{2}} & \text{if}\;C=C_{1}\;\text{and}\;C'=C_{2}\\
\langle C'|M|C\rangle & \text{otherwise}
\end{array}\;,
\end{equation}
counts the number of times $Q_{t}\in\mathbb{N}$ that the system transitions from $C_{1}$ to $C_{2}$ (in this direction only) between time $0$ and time $t$. In the following, the system is initially prepared in its unique stationary state.

We observe from (\ref{M(g) single transition}) that the characteristic polynomial $P$ of $M(g)$ has the form $P(\lambda,g)=P_{0}(\lambda)+g P_{1}(\lambda)$, where $P_{0}$ (respectively $P_{1}$) is a polynomial of degree $|\Omega|$ (resp. $|\Omega|-2$, if $w_{C_{2}\to C_{1}}\neq0$, which we assume in the following unless explicitly stated otherwise). Having $P$ of degree one in $g$ only happens for very special choices of deformation $M(g)$, and leads to a Riemann surface of genus $0$, see below. This is crucial for the following since meromorphic functions are then simply rational functions, which allows for explicit expressions. For other cases of interest with $P$ of degree one in $g$, corresponding to rank one deformations $M(g)=M+(g-1)|U\rangle\langle V|$, one can mention monitoring instead all the transitions to or from a single state $C_{1}$, for which the final results (\ref{Prob[contour lambda] simple example}) and (\ref{Prob[Fourier] simple example}) below for the probability of $Q_{t}$ still hold with the corresponding polynomials $P_{0}$ and $P_{1}$.

In the long time limit, the generating function $\langle g^{Q_{t}}\rangle$ with $g>0$ has from (\ref{GF[M]}) the asymptotics $\log\langle g^{Q_{t}}\rangle\simeq t\lambda_{0}(g)$, where $\lambda_{0}(g)\in\mathbb{R}$ is the eigenvalue of $M(g)$ with largest real part, equal at $g=1$ to the stationary eigenvalue $\lambda_{0}(1)=0$ of the non-deformed Markov matrix $M$. All the cumulants of $Q_{t}$ are then proportional to $t$ at long times, in particular $\langle Q_{t}\rangle\simeq Jt$ with $J=\lambda_{0}'(1)$. The eigenvalue $\lambda_{0}(g)$ for $g>0$ is solution of the characteristic equation $P(\lambda_{0}(g),g)=0$, and expanding at first order around $g=1$ gives
\begin{equation}
\label{P0(0)+P1(0)=0}
P_{0}(0)+P_{1}(0)=0
\end{equation}
and
\begin{equation}
\label{J[P0,P1]}
J=\frac{P_{0}(0)}{P_{0}'(0)+P_{1}'(0)}\;.
\end{equation}
\end{subsection}

\begin{subsection}{Algebraic curve \texorpdfstring{$\mathcal{A}$}{A} and Riemann surface \texorpdfstring{$\R\sim\Ch$}{R~Chat}}
The formalism of section~\ref{section counting processes} applies to the deformed Markov matrix (\ref{M(g) single transition}), and the characteristic equation $P(\lambda,g)=0$ defines a complex algebraic curve $\mathcal{A}$. We assume in the following that $\mathcal{A}$ is non-singular, which is true for generic Markov matrix $M$. In particular, $P_{0}$ and $P_{1}$ do not have a common zero, and the Riemann surface $\R$ corresponding to $\mathcal{A}$ has a single connected component. The Newton polygon of $\mathcal{A}$, represented in figure~\ref{fig Newton polygon single transition}, has no point with integer coordinates in its interior, and the genus $\mathrm{g}$ of $\R$ is then equal to zero, which means that $\R$ is the Riemann sphere $\Ch$.

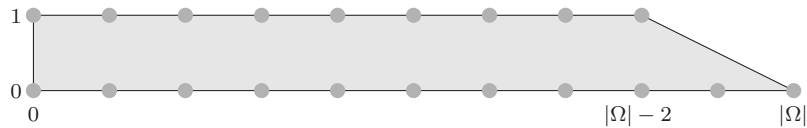
\begin{figure}
	\begin{center}
		\begin{picture}(100,10)
			{\color[rgb]{0.9,0.9,0.9}\polygon*(0,0)(100,0)(80,10)(0,10)}
			\put(0,0){\line(1,0){100}}
			\put(0,0){\line(0,1){10}}
			\put(0,10){\line(1,0){80}}
			\put(80,10){\line(2,-1){20}}
			\multiput(0,0)(10,0){11}{\color[rgb]{0.7,0.7,0.7}\circle*{2}}
			\multiput(0,10)(10,0){9}{\color[rgb]{0.7,0.7,0.7}\circle*{2}}
			\put(-3,-1){\scriptsize$0$}
			\put(-3,9){\scriptsize$1$}
			\put(-0.7,-4){\scriptsize$0$}
			\put(75,-4){\scriptsize$|\Omega|-2$}
			\put(98,-4){\scriptsize$|\Omega|$}
		\end{picture}
	\end{center}
	\caption{Newton polygon for the algebraic curve $P_{0}(\lambda)+g P_{1}(\lambda)=0$ of section~\ref{section simple example} with generic transition rates, such that $P_{0}$ and $P_{1}$ are of respective degrees $|\Omega|$ and $|\Omega|-2$. Powers of $\lambda$ are represented horizontally and powers of $g$ vertically. All the points with integer coordinates in the polygon lie on its boundary, and the corresponding Riemann surface $\R$ has then genus zero.}
	\label{fig Newton polygon single transition}
\end{figure}

The meromorphic function $g$ on $\R$ has degree $|\Omega|$ since there are $|\Omega|$ distinct eigenstates corresponding to a generic value of $g$. On the other hand the meromorphic function $\lambda$ on $\R$ has degree $1$, since setting $\lambda\in\Ch$ fixes uniquely $g$ from $P(\lambda,g)=0$ as
\begin{equation}
\label{g(lambda)}
g=-P_{0}(\lambda)/P_{1}(\lambda)\;,
\end{equation}
and thus also the point $(\lambda,g)$ on the algebraic curve. The function $\lambda$ is then an analytic bijection on $\Ch$, and can thus be used as a global parametrization for $\R$. Therefore, we identify points $p\in\R$ with the value $\lambda=\lambda(p)$ in the following. Additionally, any meromorphic function on $\R$ may be written as a rational function of $\lambda$ and $g$, and thus also as a rational function of $\lambda$ alone from (\ref{g(lambda)}). This is consistent with genus $\mathrm{g}=0$ since meromorphic functions being rational functions of some parameter characterizes the Riemann sphere.
\end{subsection}

\begin{subsection}{Ramification for the variable \texorpdfstring{$g$}{g}}
From the degrees of $P_{0}$ and $P_{1}$, we observe that the point $\lambda=\infty$ of $\R$ is a ramification point for $g$ with ramification index $2$ generically. Additional ramification points $\lambda_{*}\in\Lambda_{*}$, which also have ramification index $2$ generically, are solutions of the system $P(\lambda_{*},g_{*})=P^{(1,0)}(\lambda_{*},g_{*})=0$. Eliminating $g_{*}$ using (\ref{g(lambda)}) gives the polynomial equation
\begin{equation}
\label{eq lambda*}
P_{0}'(\lambda_{*})P_{1}(\lambda_{*})-P_{0}(\lambda_{*})P_{1}'(\lambda_{*})=0
\end{equation}
of degree $2|\Omega|-3$ generically for $\lambda_{*}$, which implies that $\Lambda_{*}$ has $|\Lambda_{*}|=2|\Omega|-3$ elements. This is consistent with the Riemann-Hurwitz formula (\ref{Riemann-Hurwitz}) applied to the function $g$ since $-|\Omega|+1+\frac{1}{2}+\frac{2|\Omega|-3}{2}=0$. In the following, $\Lambda_{*}$ is identified with the corresponding set of $2|\Omega|-3$ points on $\R$.

\begin{table}
	\begin{center}
		\begin{tabular}{|c|c|c|cccccc|}\hline
			Point $p_{0}\in\R$ & \begin{tabular}{c}Number\\of points\end{tabular} & Local parameter & $\lambda$ & $g$ & $\rmd\lambda$ & $\rmd g$ & $\mathcal{N}_{\text{stat}}$ & $\omega_{\text{stat}}$\\\hline
			$\lambda_{0}=0$, $g_{0}=1$ & $1$ & $\lambda\sim g-1$ & $1$ & $\cdot$ & $\cdot$ & $\cdot$ & $\cdot$ & $\cdot$\\
			$g_{0}=1$, $\lambda_{0}\neq0$ & $|\Omega|-1$ & $\lambda-\lambda_{0}\sim g-1$ & $\cdot$ & $\cdot$ & $\cdot$ & $\cdot$ & $2$ & $-1$\\
			$g_{0}=0$ & $|\Omega|$ & $\lambda-\lambda_{0}\sim g$ & $\cdot$ & $1$ & $\cdot$ & $\cdot$ & $\cdot$ & $\cdot$\\
			$\lambda_{0}=g_{0}=\infty$ & $1$ & $\lambda^{-1}\sim g^{-1/2}$ & $-1$ & $-2$ & $-2$ & $-3$ & $-1$ & $-1$\\
			$\lambda_{0}\neq g_{0}=\infty$ & $|\Omega|-2$ & $\lambda-\lambda_{0}\sim g^{-1}$ & $\cdot$ & $-1$ & $\cdot$ & $-2$ & $\cdot$ & $\cdot$\\
			$\lambda_{0}\in\Lambda_{*}$ & $2|\Omega|-3$ & $\lambda-\lambda_{0}\sim \sqrt{g-g_{0}}$ & $\cdot$ & $\cdot$ & $\cdot$ & $1$ & $-1$ & $-1$\\\hline
		\end{tabular}
	\end{center}
	\caption{Poles and zeroes $p_{0}\in\R$ (with $\lambda(p_{0})=\lambda_{0}$ and $g(p_{0})=g_{0}$) of various functions and differentials for the model studied in section~\ref{section simple example} with generic transition rates. The positive numbers are the orders of the zeroes, the negative numbers are minus the orders of the poles, and $\cdot$ indicates that the point is neither a pole nor a zero. The zeroes of the differential $\omega_{\text{stat}}=\rmd\log\mathcal{N}_{\text{stat}}$ are not shown.}
	\label{table poles zeroes single transition}
\end{table}

The variable $g$ is a local parameter for $\R$, except in the neighbourhood of $p_{*}\in\Lambda_{*}$ where a local parameter is $z=\sqrt{g-g(p_{*})}$, or in the neighbourhood of the point $p\in\R$ with $g(p)=\infty$ and $\lambda(p)=\infty$ (respectively the $|\Omega|-2$ points with $g(p)=\infty$ and $\lambda(p)$ finite), where a local parameter is $z=g^{-1/2}$ (resp. $z=g^{-1}$). These points correspond to the poles and zeroes of the meromorphic differential $\rmd g$, and are summarized in table~\ref{table poles zeroes single transition}.
\end{subsection}

\begin{subsection}{Meromorphic function \texorpdfstring{$\mathcal{N}_{\text{stat}}(p)$}{Nstat(p)} and meromorphic differential \texorpdfstring{$\omega=\rmd\log\mathcal{N}_{\text{stat}}$}{omega=dlogNstat}}
The expression (\ref{Prob[int R]}) for the probability of $Q_{t}$ with stationary initial condition involves the meromorphic function
\begin{equation}
\mathcal{N}_{\text{stat}}(p)=\frac{\sum_{C\in\Omega}\langle C|\psi(p)\rangle\,\langle\psi(p)|P_{\text{stat}}\rangle}{\langle\psi(p)|\psi(p)\rangle}\;.
\end{equation}
As discussed in section~\ref{section pole structure}, $\mathcal{N}_{\text{stat}}$ only has poles at the ramification points for $g$, the orders of the poles being equal to the ramification indices minus one. Generically, the poles of $\mathcal{N}_{\text{stat}}$ are then the $2|\Omega|-3$ elements of $\Lambda_{*}$ plus the point $\lambda=\infty$, which are all ramified twice for $g$, and are thus simple poles. Additionally, the function $\mathcal{N}_{\text{stat}}(p)$ has zeroes of order $2$ at the points with $g=1$, $\lambda\neq0$ (i.e. $p\in g^{-1}(1)\setminus\{o\}$ with $o$ the stationary point characterized by $\lambda(o)=0$, $g(o)=1$) because of orthogonality of the eigenstates, see section~\ref{section exponential form}. This gives $2|\Omega|-2$ zeroes (counted with multiplicity) for $\mathcal{N}_{\text{stat}}$, which matches with its number of poles, and all the zeroes of $\mathcal{N}_{\text{stat}}$ are thus accounted for. The locations of the poles and zeroes of $\mathcal{N}_{\text{stat}}$ are summarized in table~\ref{table poles zeroes single transition}.

We consider now the meromorphic differential
\begin{equation}
\label{omega[N] simple example}
\omega_{\text{stat}}=\rmd\log\mathcal{N}_{\text{stat}}\;,
\end{equation}
which by construction has only simple poles, located at the poles and zeroes of $\mathcal{N}_{\text{stat}}$. The poles of $\omega_{\text{stat}}$ have integer residues, equal to the orders of the zeroes and minus the orders of the poles of $\mathcal{N}_{\text{stat}}$. From the discussion above, we conclude that the poles of $\omega_{\text{stat}}$ are generically the points $g=1$, $\lambda\neq0$ (with residue $2$), $\lambda=\infty$ (with residue $-1$) and $\lambda\in\Lambda_{*}$ (with residue $-1$). From the uniqueness property for meromorphic differentials with simple poles discussed above (\ref{Riemann-Hurwitz}) in section~\ref{section Riemann surfaces}, there exists a single meromorphic differential on $\R$ with those poles and residues (since $\R$ has genus zero, there is no constraint about integrals of $\omega_{\text{stat}}$ over non-contractible loops on $\R$ here). Defining $\Lambda_{1}=\{\lambda(p),p\in g^{-1}(1)\setminus\{o\}\}$ (i.e. $\Lambda_{1}$ is the set of $|\Omega|-1$ non-zero eigenvalues of $M=M(1)$, which are distinct if the algebraic curve is non-singular and $\Lambda_{1}\cap\Lambda_{*}=\emptyset$), we observe that the meromorphic differential
\begin{equation}
\label{omega stat simple example}
\omega_{\text{stat}}=\rmd\lambda\,\Big(2\sum_{\lambda_{1}\in\Lambda_{1}}\frac{1}{\lambda-\lambda_{1}}-\sum_{\lambda_{*}\in\Lambda_{*}}\frac{1}{\lambda-\lambda_{*}}\Big)
\end{equation}
does have the correct poles and residues, and must then be equal to $\rmd\log\mathcal{N}_{\text{stat}}$. In the non-generic case $w_{C_{2}\to C_{1}}=0$, where the degree of $P_{1}$ is $|\Omega|-2-k$ with $k\geq1$, $\Lambda_{*}$ has $k$ less elements while $\lambda=\infty$ is ramified $k+2$ times, so that $\mathcal{N}_{\text{stat}}$ has a pole of order $k+1$ and $\omega_{\text{stat}}$ a simple pole with residue $-k-1$ at $\lambda=\infty$, and (\ref{omega stat simple example}) still holds.

Since $\mathcal{N}_{\text{stat}}(o)=1$, one can write $\mathcal{N}_{\text{stat}}$ in terms of the differential (\ref{omega[N] simple example}) as
\begin{equation}
\mathcal{N}_{\text{stat}}(p)=\rme^{\int_{o}^{p}\omega_{\text{stat}}}\;,
\end{equation}
as long as $\R$ has a single connected component, which is generically true. The integral in the exponential can be computed explicitly in terms of logarithms, and gives after exponentiation the rational function of $\lambda(p)$
\begin{equation}
\mathcal{N}_{\text{stat}}(p)=\frac{\prod_{\lambda_{1}\in\Lambda_{1}}(1-\lambda(p)/\lambda_{1})^{2}}{\prod_{\lambda_{*}\in\Lambda_{*}}(1-\lambda(p)/\lambda_{*})}\;,
\end{equation}
normalized such that $\mathcal{N}_{\text{stat}}(p)=1$ when $\lambda(p)=0$. The identities
\begin{equation}
\prod_{\lambda_{1}\in\Lambda_{1}}(1-\lambda/\lambda_{1})=\frac{P_{0}(\lambda)+P_{1}(\lambda)}{\lambda(P_{0}'(0)+P_{1}'(0))}
\end{equation}
and
\begin{equation}
\prod_{\lambda_{*}\in\Lambda_{*}}(1-\lambda/\lambda_{*})=\frac{P_{0}(\lambda)P_{1}'(\lambda)-P_{0}'(\lambda)P_{1}(\lambda)}{P_{0}(0)P_{1}'(0)-P_{0}'(0)P_{1}(0)}
\end{equation}
finally lead to
\begin{equation}
\label{N simple example}
\mathcal{N}_{\text{stat}}(p)=\frac{J}{\lambda(p)^{2}}\,\frac{(P_{0}(\lambda(p))+P_{1}(\lambda(p)))^{2}}{P_{0}(\lambda(p))P_{1}'(\lambda(p))-P_{0}'(\lambda(p))P_{1}(\lambda(p))}\;,
\end{equation}
where we used (\ref{P0(0)+P1(0)=0}) and (\ref{J[P0,P1]}) to make some simplifications.

For general initial condition, the zeroes of $\mathcal{N}$ are not known a priori, and one can not guess $\omega=\rmd\log\mathcal{N}$ in the same way as above with stationary initial condition. The function $\mathcal{N}(p)$ can however still be computed in principle, by solving the left and right eigenvalue equations for given eigenvalue $\lambda(p)$: choosing for normalization e.g. $\langle C_{2}|\psi(p)\rangle=\langle\psi(p)|C_{1}\rangle=1$, the other components of the eigenvectors are then rational functions of $\lambda(p)$, expressed as explicit ratios of determinants by Cramer's rule.
\end{subsection}

\begin{subsection}{Probability of \texorpdfstring{$Q_{t}$}{Qt}}
The probability of the integer counting process $Q_{t}$ is given by (\ref{Prob[int R]}). Using the explicit formula (\ref{N simple example}) for the function $\mathcal{N}_{\text{stat}}$ with stationary initial condition, one has for $Q\in\mathbb{N}$
\begin{equation}
\fl\hspace{2mm}
\P(Q_{t}=Q)=J\oint_{\Gamma}\frac{\rmd g}{2\rmi\pi g^{Q+1}}\,\frac{1}{\lambda^{2}}\,\frac{(P_{0}(\lambda)+P_{1}(\lambda))^{2}}{P_{0}(\lambda)P_{1}'(\lambda)-P_{0}'(\lambda)P_{1}(\lambda)}\,\rme^{t\lambda}\;,
\end{equation}
where all functions and differentials are evaluated at the same point $p\in\Gamma$ with $\Gamma\subset\R$ a simple closed contour splitting $\R$ into a domain containing all the points of $\R$ with $g=0$, around which $\Gamma$ has winding number one, and a domain containing all the points of $\R$ with $g=\infty$, see figure~\ref{fig Gamma}.

\begin{figure}
	\begin{tabular}{ll}
		\hspace{-3mm}
		\begin{tabular}{l}\includegraphics[width=72mm]{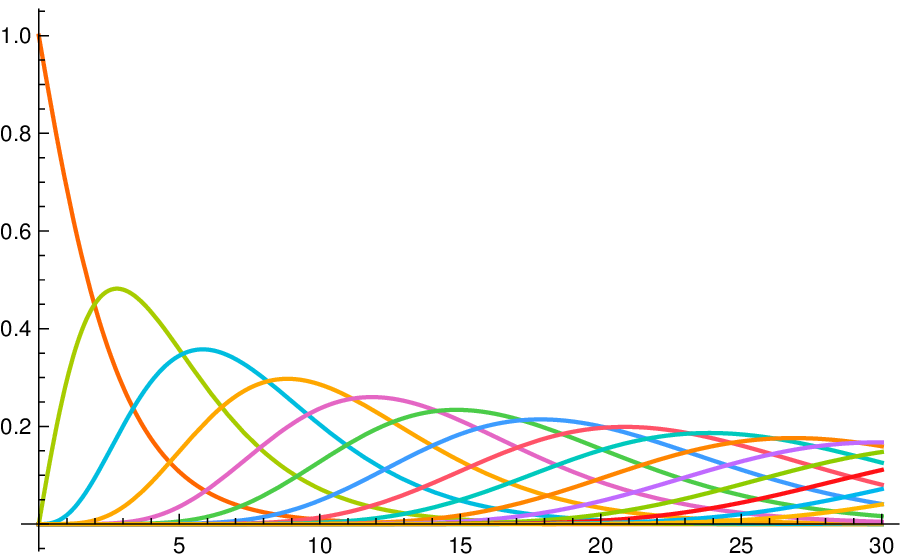}\end{tabular}
		\begin{picture}(0,0)
			\put(-70,19.5){$\P(Q_{t}=Q)$}
			\put(-3.5,-16.5){$t$}
		\end{picture}
		&
		\begin{tabular}{l}\includegraphics[width=72mm]{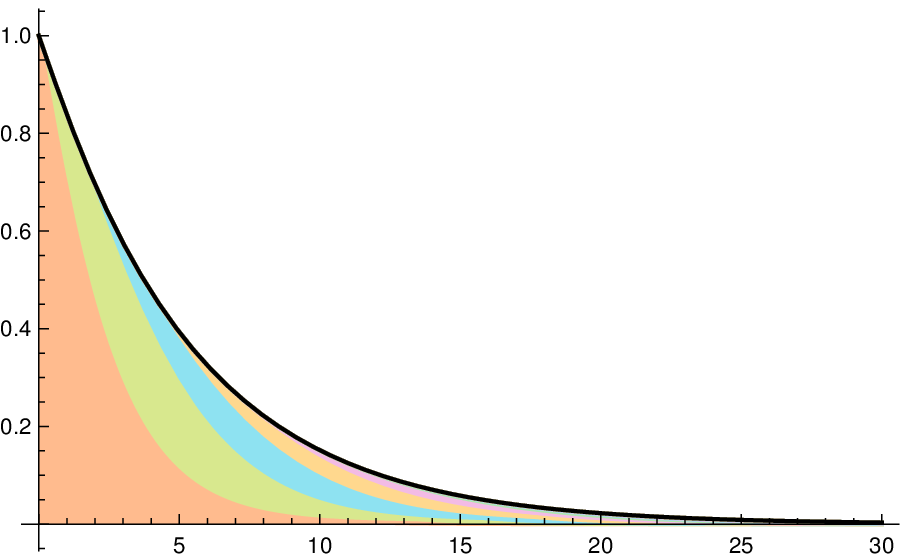}\end{tabular}
		\begin{picture}(0,0)
			\put(-70,19.5){$\langle g^{Q_{t}}\rangle$}
			\put(-6,-15){$t$}
		\end{picture}
	\end{tabular}
	\caption{Probability (left) and generating function (right) of $Q_{t}$ plotted as a function of time for the system with $|\Omega|=2$ states, transition rates $w_{C_{1}\to C_{2}}=1$, $w_{C_{2}\to C_{1}}=1/2$, and stationary initial condition. The curves on the left correspond to $\P(Q_{t}=Q)$ evaluated from (\ref{Prob[contour lambda] simple example}) with $P_{0}(\lambda)=(\lambda+1)(\lambda+1/2)$, $P_{1}(\lambda)=-1/2$, and $Q$ increasing from $0$ to $14$ from left to right. The height of the coloured domains on the right correspond to $g^{Q}\,\P(Q_{t}=Q)$ with $g=1/2$, $Q$ increasing from $0$ from left to right, and probabilities evaluated from (\ref{Prob[contour lambda] simple example}). The black curve on the right is the exact generating function $\langle g^{Q_{t}}\rangle$ given by (\ref{GF[M]}).}
	\label{fig plots GF simple example}
\end{figure}

The contour integral on the Riemann surface $\R$ can be understood as a contour integral for $\lambda=\lambda(p)$ in the complex plane, encircling the zeroes of $P_{0}$ but not the zeroes of $P_{1}$. The differential $\rmd\lambda$ can be expressed in terms of $\rmd g$ as
\begin{equation}
\rmd\lambda=\frac{P_{1}(\lambda)^{2}}{P_{0}(\lambda)P_{1}'(\lambda)-P_{0}'(\lambda)P_{1}(\lambda)}\,\rmd g\;.
\end{equation}
Using (\ref{g(lambda)}), this gives the expression
\begin{equation}
\label{Prob[contour lambda] simple example}
\P(Q_{t}=Q)=(-1)^{Q+1}J\oint_{P_{0}^{-1}(0)}\frac{\rmd\lambda}{2\rmi\pi}\,\frac{(P_{0}(\lambda)+P_{1}(\lambda))^{2}}{\lambda^{2}}\,\frac{P_{1}(\lambda)^{Q-1}}{P_{0}(\lambda)^{Q+1}}\,\rme^{t\lambda}\;,
\end{equation}
where the contour of integration is a union of small circles around the zeroes of $P_{0}$.

The expression above for the probability of $Q_{t}$ is plotted in figure~\ref{fig plots GF simple example} as a function of time for an example with $|\Omega|=2$ states. The expression (\ref{Prob[contour lambda] simple example}) is also checked in figure~\ref{fig plots GF simple example} by plotting the generating function $\langle g^{Q_{t}}\rangle$ against $\sum_{Q=0}^{Q_{\text{max}}}g^{Q}\,\P(Q_{t}=Q)$ as a function of time for small values of $Q_{\text{max}}$.

We observe that the zeroes of $P_{0}$, which are by definition the eigenvalues of $M(0)$, generically have a strictly negative real part. In order to show that, we consider the matrix $A=I+\epsilon M(0)$, whose coefficients are non-negative for small enough $\epsilon>0$. A consequence of the Perron-Frobenius theorem, see e.g. \cite{M2000.1}, states that the eigenvalues $a$ of $A$ verify $|a|\leq\max_{j}\sum_{i}A_{i,j}$, and hence $|a|\leq1$ since $M(0)$ is a Markov matrix except for a single missing non-diagonal element. The eigenvalues $\lambda=(a-1)/\epsilon$ of $M(0)$ then verify $|1+\epsilon\lambda|^{2}\leq1$, which implies $\Re\,\lambda\leq-\epsilon^{2}|\lambda|^{2}/2$. Thus, either $\lambda=0$, which does not happen generically, or $\Re\,\lambda<0$.

The portion of the contour of integration with $\Re\,\lambda<0$ in (\ref{Prob[contour lambda] simple example}) does not contribute for $t>0$ when pushed to infinity because of the factor $\rme^{t\lambda}$. For $Q\geq1$, the contour of integration can then be replaced by the imaginary axis $\rmi\mathbb{R}$ since the only poles of the integrand are the zeroes of $P_{0}$ ($\lambda=0$ is not a pole since $P_{0}(0)+P_{1}(0)=0$). For $Q=0$, a subtraction of some terms in the integrand is needed in order to remove the poles at the zeroes of $P_{1}$, which also have a negative real part. This leads an alternative representation of the probability of $Q_{t}$ as the Fourier transform of a rational function,
\begin{equation}
\label{Prob[Fourier] simple example}
\fl\hspace{5mm}
\P(Q_{t}=Q)=(-1)^{Q}J\int_{-\infty}^{\infty}\frac{\rmd\lambda}{2\pi}\,\frac{(P_{0}(\rmi\lambda)+P_{1}(\rmi\lambda))^{2}-\delta_{U,0}R(\lambda)}{\lambda^{2}}\,\frac{P_{1}(\rmi\lambda)^{Q-1}}{P_{0}(\rmi\lambda)^{Q+1}}\,\rme^{\rmi t\lambda}\;,
\end{equation}
where $R(\lambda)=P_{0}(\rmi\lambda)^{2}+(1+\frac{\rmi\lambda}{J})P_{0}(\rmi\lambda)P_{1}(\rmi\lambda)$.
\end{subsection}

\end{section}

\begin{section}{Current for unidirectional simple exclusion process in one dimension}
\label{section single file unidirectional}
In this section we consider the integer counting process equal to the local time-integrated current of particles for a simple exclusion process on a one-dimensional periodic lattice, where particles of a single species move in a single file by local hops from any site $i$ to the next site $i+1$ if the latter is empty. In a first part, we focus on the model with generic transition rates $w_{C\to C'}$ between allowed states. In a second part, we consider the totally asymmetric simple exclusion process (TASEP), where all the transition rates are equal.

\begin{subsection}{Process with generic transition rates}
\label{section single file unidirectional generic}
We study in this section the model with generic transition rates, whose algebraic curve is non-singular.

\begin{subsubsection}{Definition of the model}\hfill\\
We consider a periodic one-dimensional lattice with $L$ sites labelled $i=1,\ldots,L$ with $i\equiv i+L$. The set $\Omega$ of all possible states with $N$ particles allocated at the sites of the lattice, with the exclusion constraint that there is at most one particle per site, has cardinal $|\Omega|={{L}\choose{N}}$. We restrict to $L\geq2$ and $1\leq N\leq L-1$ in order to avoid cases with a single state. The set $\Omega$ is supplemented with the following Markovian dynamics in continuous time, with Markov matrix $M$: a particle at arbitrary site $i$ may hop to the next site $i+1$, if the latter is empty, with generic rate $w_{C\to C'}>0$ depending on the states $C$ and $C'$ of the system before and after the particle has moved, and not just on the site $i$.

We are interested in the local time-integrated current of particles, and focus without loss of generality to $Q_{t}$ counting the number of times particles have hopped from site $L$ to site $1$ up to time $t$. Following section~\ref{section M(g)}, this integer counting process is associated with a deformation $M(g)$ of the Markov matrix $M$.

It is useful to consider also the total current $Q_{t}^{\text{tot}}$, counting all the particle hops wherever they happen in the system, and associated with a deformation $M_{\text{tot}}(g)$ of $M$. Then, we observe that $M_{\text{tot}}(g^{1/L})$ and $M(g)$ are related by the similarity transformation
\begin{equation}
\label{M[Mtot]}
M(g)=g^{-S/L}M_{\text{tot}}(g^{1/L})g^{S/L}\;,
\end{equation}
where $S$ is the diagonal matrix such that $\langle C|S|C\rangle$ is equal to the sum of the labels $i$, $1\leq i\leq L$ of the sites occupied by particles in the state $C$. Indeed, $\langle C'|M_{\text{tot}}(g^{1/L})|C\rangle=g^{1/L}\,w_{C\to C'}$ for allowed transitions $C\to C'$. For transitions $C\to C'$ corresponding to a particle hopping from site $i\neq L$, one has furthermore $\langle C'|S|C'\rangle=\langle C|S|C\rangle+1$, which implies $\langle C'|g^{-S/L}M_{\text{tot}}(g^{1/L})g^{S/L}|C\rangle=w_{C\to C'}=\langle C'|M(g)|C\rangle$. On the other hand, for transitions $C\to C'$ corresponding to a particle hopping from site $L$ to site $1$, one has $\langle C'|S|C'\rangle=\langle C|S|C\rangle+1-L$, which implies $\langle C'|g^{-S/L}M_{\text{tot}}(g^{1/L})g^{S/L}|C\rangle=g\,w_{C\to C'}=\langle C'|M(g)|C\rangle$.

At the level of characteristic polynomials, (\ref{M[Mtot]}) leads to
\begin{equation}
\label{P[M,Mtot]}
P(\lambda,g)=\det(\lambda I-M(g))=\det(\lambda I-M_{\text{tot}}(g^{1/L}))\;,
\end{equation}
which is a polynomial in both $\lambda$ and $g$. Both local current $Q_{t}$ and total current $Q_{t}^{\text{tot}}/L$ thus lead to the same algebraic curve $\mathcal{A}$, which is also independent from the local bond $L\to1$ chosen for $Q_{t}$.
\end{subsubsection}

\begin{subsubsection}{Degree of the characteristic polynomial}\hfill\\
The choice of generic transition rates for $M$ ensures that the algebraic curve $\mathcal{A}$ is non-singular. The corresponding Riemann surface $\R$ has then a single connected component, whose genus $\mathrm{g}$ can be obtained from the Newton polygon built from the structure of the characteristic polynomial $P$, see section~\ref{section Riemann surfaces}.

\begin{figure}
	\begin{center}
		\begin{picture}(65,102.5)
			\put(0,90){
				\multiput(0,0)(0,5){2}{\line(1,0){25}}
				\multiput(0,0)(5,0){6}{\line(0,1){5}}
				\put(2.5,2.5){\circle*{2}}
				\put(12.5,2.5){\circle*{2}}
			}
			\put(40,90){
				\multiput(0,0)(0,5){2}{\line(1,0){25}}
				\multiput(0,0)(5,0){6}{\line(0,1){5}}
				\put(17.5,2.5){\circle*{2}}
				\put(22.5,2.5){\circle*{2}}
			}
			\put(0,70){
				\multiput(0,0)(0,5){2}{\line(1,0){25}}
				\multiput(0,0)(5,0){6}{\line(0,1){5}}
				\put(7.5,2.5){\circle*{2}}
				\put(12.5,2.5){\circle*{2}}
			}
			\put(40,70){
				\multiput(0,0)(0,5){2}{\line(1,0){25}}
				\multiput(0,0)(5,0){6}{\line(0,1){5}}
				\put(2.5,2.5){\circle*{2}}
				\put(17.5,2.5){\circle*{2}}
			}
			\put(0,50){
				\multiput(0,0)(0,5){2}{\line(1,0){25}}
				\multiput(0,0)(5,0){6}{\line(0,1){5}}
				\put(7.5,2.5){\circle*{2}}
				\put(17.5,2.5){\circle*{2}}
			}
			\put(40,50){
				\multiput(0,0)(0,5){2}{\line(1,0){25}}
				\multiput(0,0)(5,0){6}{\line(0,1){5}}
				\put(2.5,2.5){\circle*{2}}
				\put(22.5,2.5){\circle*{2}}
			}
			\put(0,30){
				\multiput(0,0)(0,5){2}{\line(1,0){25}}
				\multiput(0,0)(5,0){6}{\line(0,1){5}}
				\put(12.5,2.5){\circle*{2}}
				\put(17.5,2.5){\circle*{2}}
			}
			\put(40,30){
				\multiput(0,0)(0,5){2}{\line(1,0){25}}
				\multiput(0,0)(5,0){6}{\line(0,1){5}}
				\put(7.5,2.5){\circle*{2}}
				\put(22.5,2.5){\circle*{2}}
			}
			\put(0,10){
				\multiput(0,0)(0,5){2}{\line(1,0){25}}
				\multiput(0,0)(5,0){6}{\line(0,1){5}}
				\put(12.5,2.5){\circle*{2}}
				\put(22.5,2.5){\circle*{2}}
			}
			\put(40,10){
				\multiput(0,0)(0,5){2}{\line(1,0){25}}
				\multiput(0,0)(5,0){6}{\line(0,1){5}}
				\put(2.5,2.5){\circle*{2}}
				\put(7.5,2.5){\circle*{2}}
			}
			\put(12.5,101.5){\vector(0,-1){4}}\put(33.5,101.5){\thicklines\color{red}\vector(1,-1){4}}\put(31.5,101.5){\thicklines\color{red}\vector(-1,-1){4}}
			\put(12.5,87.5){\thicklines\color{red}\vector(0,-1){10}}\put(52.5,87.5){\thicklines\color{red}\vector(0,-1){10}}\put(27.5,87.5){\vector(1,-1){10}}
			\put(12.5,67.5){\thicklines\color{red}\vector(0,-1){10}}\put(52.5,67.5){\thicklines\color{red}\vector(0,-1){10}}\put(37.5,67.5){\vector(-1,-1){10}}
			\put(12.5,47.5){\thicklines\color{red}\vector(0,-1){10}}\put(52.5,47.5){\thicklines\color{red}\vector(0,-1){10}}\put(27.5,47.5){\vector(1,-1){10}}
			\put(12.5,27.5){\thicklines\color{red}\vector(0,-1){10}}\put(52.5,27.5){\thicklines\color{red}\vector(0,-1){10}}\put(37.5,27.5){\vector(-1,-1){10}}
			\put(12.5,7.5){\vector(0,-1){10}}\put(27.5,7.5){\thicklines\color{red}\vector(1,-1){10}}\put(37.5,7.5){\thicklines\color{red}\vector(-1,-1){10}}
		\end{picture}
	\end{center}
	\caption{Graph of the dynamics for the unidirectional simple exclusion process with $N=2$ particles on $L=5$ sites. The arrows at the bottom point to the states at the top. A Hamiltonian cycle is represented with red, thicker arrows.}
	\label{fig graph TASEP L=5 N=2}
\end{figure}
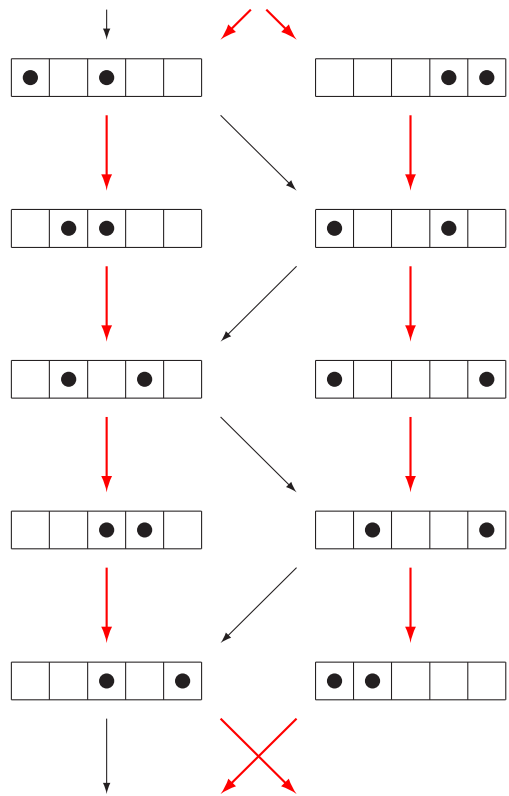

Since all the non-diagonal elements of $M_{\text{tot}}(g^{1/L})$ are equal to $g^{1/L}$ and all the diagonal elements of $M_{\text{tot}}(g^{1/L})$ are non-zero constants, the expansion of the determinant in (\ref{P[M,Mtot]}) as a sum over permutations of $\Omega$ gives
\begin{equation}
\label{P[sigma]}
\fl\hspace{5mm}
P(\lambda,g)=\sum_{\sigma\in S_{\Omega}}(-1)^{\sigma}
\Bigg(\prod_{{C\in\Omega}\atop{\sigma(C)=C}}\!\!\Big(\lambda+\sum_{C'\neq C}w_{C\to C'}\Big)\Bigg)
\Bigg(\prod_{{C\in\Omega}\atop{\sigma(C)\neq C}}\!\!\Big(-g^{1/L}\,w_{C\to\sigma(C)}\Big)\Bigg)\;.
\end{equation}
Writing
\begin{equation}
\label{P[Pk,TASEP]}
P(\lambda,g)=\sum_{k=0}^{d_{+}}P_{k}(\lambda)g^{k}\;,
\end{equation}
the degree of the polynomial $P_{k}$ is then exactly equal to ${{L}\choose{N}}-kL$, since no cancellation is expected for generic transition rates.

One could then naively expect that the exponent $d_{+}$ for the variable $g$ is equal to $\lfloor\frac{1}{L}{{L}\choose{N}}\rfloor$. This is not the case in general due to the sparse nature of the Markov matrix of an exclusion process, and $d_{+}$ depends on the precise structure of the graph of the dynamics (i.e. the graph of all allowed transitions). Indeed, barring cancellations, which are not expected for generic transition rates, $L\,d_{+}$ is equal from (\ref{P[sigma]}) to the maximal number of non-fixed points of a permutation $\sigma$ on $\Omega$ such that all $C\to\sigma(C)$ with $\sigma(C)\neq C$ are allowed transitions for the dynamics. Decomposing $\sigma$ as a product of cycles, $L\,d_{+}$ is then equal to the length of the longest cycle (or product of cycles) which can be drawn on the graph of the dynamics and passes at most once through any state $C$. In some cases, see figure~\ref{fig graph TASEP L=5 N=2}, there exists a Hamiltonian cycle on the graph (i.e. a cycle passing through each state once), and $d_{+}$ is equal to $\frac{1}{L}{{L}\choose{N}}$. Conversely, there is no Hamiltonian cycle when $\frac{1}{L}{{L}\choose{N}}$ is not an integer, see figure~\ref{fig graph TASEP L=4 N=2} for an example. Checks up to $L=7$ seem to indicate that a Hamiltonian cycle exists if and only if $L$ and $N$ are co-prime, in which case $\frac{1}{L}{{L}\choose{N}}$ is indeed an integer.

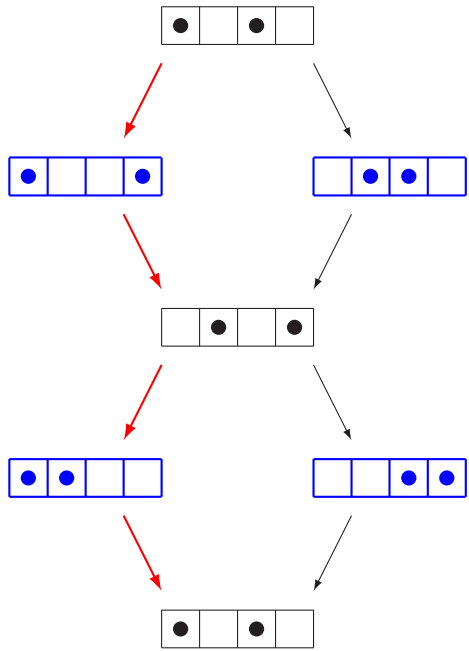
\begin{figure}
	\begin{center}
		\begin{picture}(60,92.5)
			\put(20,80){
				\multiput(0,0)(0,5){2}{\line(1,0){20}}
				\multiput(0,0)(5,0){5}{\line(0,1){5}}
				\put(2.5,2.5){\circle*{2}}
				\put(12.5,2.5){\circle*{2}}
			}
			\put(0,60){\thicklines\color{blue}
				\multiput(0,0)(0,5){2}{\line(1,0){20}}
				\multiput(0,0)(5,0){5}{\line(0,1){5}}
				\put(2.5,2.5){\circle*{2}}
				\put(17.5,2.5){\circle*{2}}
			}
			\put(40,60){\thicklines\color{blue}
				\multiput(0,0)(0,5){2}{\line(1,0){20}}
				\multiput(0,0)(5,0){5}{\line(0,1){5}}
				\put(7.5,2.5){\circle*{2}}
				\put(12.5,2.5){\circle*{2}}
			}
			\put(20,40){
				\multiput(0,0)(0,5){2}{\line(1,0){20}}
				\multiput(0,0)(5,0){5}{\line(0,1){5}}
				\put(7.5,2.5){\circle*{2}}
				\put(17.5,2.5){\circle*{2}}
			}
			\put(0,20){\thicklines\color{blue}
				\multiput(0,0)(0,5){2}{\line(1,0){20}}
				\multiput(0,0)(5,0){5}{\line(0,1){5}}
				\put(2.5,2.5){\circle*{2}}
				\put(7.5,2.5){\circle*{2}}
			}
			\put(40,20){\thicklines\color{blue}
				\multiput(0,0)(0,5){2}{\line(1,0){20}}
				\multiput(0,0)(5,0){5}{\line(0,1){5}}
				\put(12.5,2.5){\circle*{2}}
				\put(17.5,2.5){\circle*{2}}
			}
			\put(20,0){
				\multiput(0,0)(0,5){2}{\line(1,0){20}}
				\multiput(0,0)(5,0){5}{\line(0,1){5}}
				\put(2.5,2.5){\circle*{2}}
				\put(12.5,2.5){\circle*{2}}
			}
			\put(20,77.5){\thicklines\color{red}\vector(-1,-2){5}}
			\put(40,77.5){\vector(1,-2){5}}
			\put(15,57.5){\thicklines\color{red}\vector(1,-2){5}}
			\put(45,57.5){\vector(-1,-2){5}}
			\put(20,37.5){\thicklines\color{red}\vector(-1,-2){5}}
			\put(40,37.5){\vector(1,-2){5}}
			\put(15,17.5){\thicklines\color{red}\vector(1,-2){5}}
			\put(45,17.5){\vector(-1,-2){5}}
		\end{picture}
	\end{center}
	\caption{Graph of the dynamics for the unidirectional simple exclusion process with $N=2$ particles on $L=4$ sites. There is no Hamiltonian cycle on this graph. The maximal cycle has length $4$ (one is represented with red, thicker arrows), which is also the number of aperiodic states (blue, thicker boxes), in accordance with conjecture (\ref{d+ generic}).}
	\label{fig graph TASEP L=4 N=2}
\end{figure}

\begin{table}
	\begin{center}
		\begin{tabular}{c|ccccccccccccc}
		$L\,\backslash\,N$ & 1 & 2 & 3 & 4 & 5 & 6 & 7 & 8 & 9 & 10 & 11 & 12 & 13\\\hline
		2 & 1 & $\cdot$ & $\cdot$ & $\cdot$ & $\cdot$ & $\cdot$ & $\cdot$ & $\cdot$ & $\cdot$ & $\cdot$ & $\cdot$ & $\cdot$ & $\cdot$\\
		3 & 1 & 1 & $\cdot$ & $\cdot$ & $\cdot$ & $\cdot$ & $\cdot$ & $\cdot$ & $\cdot$ & $\cdot$ & $\cdot$ & $\cdot$ & $\cdot$\\
		4 & 1 & 1 & 1 & $\cdot$ & $\cdot$ & $\cdot$ & $\cdot$ & $\cdot$ & $\cdot$ & $\cdot$ & $\cdot$ & $\cdot$ & $\cdot$\\
		5 & 1 & 2 & 2 & 1 & $\cdot$ & $\cdot$ & $\cdot$ & $\cdot$ & $\cdot$ & $\cdot$ & $\cdot$ & $\cdot$ & $\cdot$\\
		6 & 1 & 2 & 3 & 2 & 1 & $\cdot$ & $\cdot$ & $\cdot$ & $\cdot$ & $\cdot$ & $\cdot$ & $\cdot$ & $\cdot$\\
		7 & 1 & 3 & 5 & 5 & 3 & 1 & $\cdot$ & $\cdot$ & $\cdot$ & $\cdot$ & $\cdot$ & $\cdot$ & $\cdot$\\
		8 & 1 & 3 & 7 & 8 & 7 & 3 & 1 & $\cdot$ & $\cdot$ & $\cdot$ & $\cdot$ & $\cdot$ & $\cdot$\\
		9 & 1 & 4 & 9 & 14 & 14 & 9 & 4 & 1 & $\cdot$ & $\cdot$ & $\cdot$ & $\cdot$ & $\cdot$\\
		10 & 1 & 4 & 12 & 20 & 25 & 20 & 12 & 4 & 1 & $\cdot$ & $\cdot$ & $\cdot$ & $\cdot$\\
		11 & 1 & 5 & 15 & 30 & 42 & 42 & 30 & 15 & 5 & 1 & $\cdot$ & $\cdot$ & $\cdot$\\
		12 & 1 & 5 & 18 & 40 & \color{red}66 & 75 & \color{red}66 & 40 & 18 & 5 & 1 & $\cdot$ & $\cdot$\\
		13 & 1 & 6 & 22 & 55 & 99 & 132 & 132 & 99 & 55 & 22 & 6 & 1 & $\cdot$\\
		14 & 1 & 6 & 26 & 70 & 143 & 212 & 245 & 212 & 143 & 70 & 26 & 6 & 1 \\
		\end{tabular}
	\end{center}
	\caption{Value of the degree $d_{+}$ in the variable $g$ of the characteristic polynomial $P(\lambda,g)$ for a simple exclusion process with $N$ particles on $L$ sites and generic transition rates. The two values in red, for $L=12$, $N=5,7$, do not agree with the corresponding value $d_{+}=65$ for TASEP, as explained in section~\ref{section d+ TASEP}, but do match with the expression (\ref{d+ generic}) conjectured for generic rates.}
	\label{table d+}
\end{table}

The value of $d_{+}$ can be computed easily for small systems. An expansion of (\ref{P[M,Mtot]}) in powers of $g$ with random choices $w_{C\to C'}\in\mathbb{Q}$ of transition rates up to $L=10$, supplemented with numerical computation of the eigenvalues of $M(g)$ for large numerical values of $g$ up to $L=14$, leads to the values in table~\ref{table d+}. We observe that these numbers match perfectly with the sequence A051168 from the on-line encyclopedia on integer sequences \cite{OEIS}, which suggests the exact expression
\begin{equation}
\label{d+ generic}
d_{+}=\frac{1}{L}\sum_{d|L\wedge N}\!{{L/d}\choose{N/d}}\,\mu(d)\;.
\end{equation}
The sum is over all divisors $d$ of both $L$ and $N$ (or equivalently divisors of the greatest common divisor $L\wedge N$ of $L$ and $N$), and $\mu$ is the M\"obius function, equal to $\mu(d)=0$ if $d$ is not square free (i.e. $d$ is divisible by the square of an integer strictly larger than one) and to $\mu(d)=(-1)^{n}$ if $d$ has $n$ distinct prime factors. The expression (\ref{d+ generic}) reduces to $\frac{1}{L}{{L}\choose{N}}$ if $L$ and $N$ are co-prime, which is then compatible with the existence of a Hamiltonian cycle on the graph of the dynamics in that case.

The numbers (\ref{d+ generic}) also have a combinatorial interpretation, with $L\,d_{+}$ being the number of aperiodic states $C\in\Omega$ (i.e. sets $C$ such that shifting the positions modulo $L$ of all the particles in $C$ by some amount $\ell$ does not give $C$ again except if $\ell$ is proportional to $L$), see e.g. \cite{S1962.1}. This combinatorial interpretation for $d_{+}$ is rather puzzling since maximal cycles on the graph of the dynamics do not seem to have anything to do with non-periodic configurations, see figure~\ref{fig graph TASEP L=4 N=2}. Bethe ansatz results for TASEP in section~\ref{section d+ TASEP} suggest that those non-periodic states are not physical states of the process but should rather be interpreted as labels for the eigenstates.

We emphasize that the conjecture (\ref{d+ generic}) is only expected to hold for generic transition rates: indeed, when all the rates are equal, cancellations happen and Bethe ansatz gives a different expression for $d_{+}$, see table~\ref{table d+} and section~\ref{section d+ TASEP}.
\end{subsubsection}

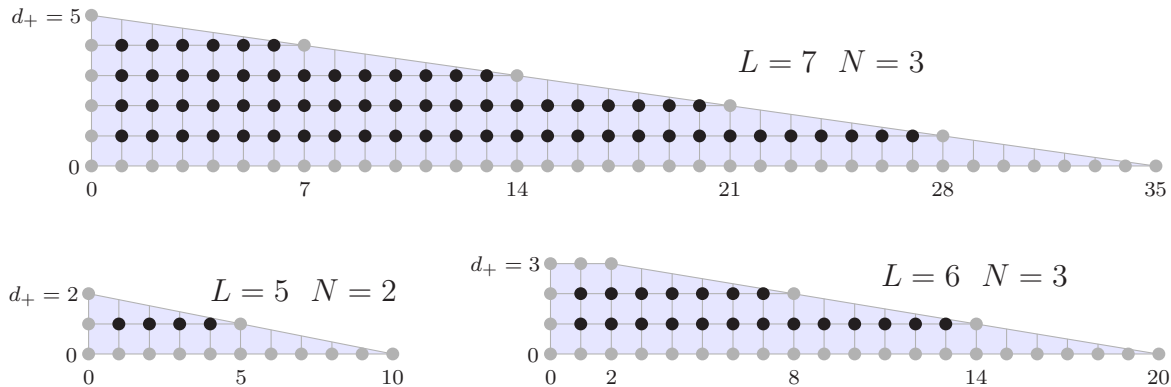
\begin{figure}
	\begin{center}
	\begin{tabular}{c}
		\begin{tabular}{l}
		\begin{picture}(140,20)(-5,0)
			{\color[rgb]{0.9,0.9,1}\polygon*(0,0)(140,0)(0,20)}%
			{\color[rgb]{0.7,0.7,0.7}
			\put(0,0){\line(1,0){140}}\put(0,4){\line(1,0){112}}\put(0,8){\line(1,0){84}}\put(0,12){\line(1,0){56}}\put(0,16){\line(1,0){28}}
			\put(0,0){\line(0,1){20}}\put(4,0){\line(0,1){19.4286}}\put(8,0){\line(0,1){18.8571}}\put(12,0){\line(0,1){18.2857}}\put(16,0){\line(0,1){17.7143}}\put(20,0){\line(0,1){17.1429}}\put(24,0){\line(0,1){16.5714}}\put(28,0){\line(0,1){16.}}\put(32,0){\line(0,1){15.4286}}\put(36,0){\line(0,1){14.8571}}\put(40,0){\line(0,1){14.2857}}\put(44,0){\line(0,1){13.7143}}\put(48,0){\line(0,1){13.1429}}\put(52,0){\line(0,1){12.5714}}\put(56,0){\line(0,1){12.}}\put(60,0){\line(0,1){11.4286}}\put(64,0){\line(0,1){10.8571}}\put(68,0){\line(0,1){10.2857}}\put(72,0){\line(0,1){9.71429}}\put(76,0){\line(0,1){9.14286}}\put(80,0){\line(0,1){8.57143}}\put(84,0){\line(0,1){8.}}\put(88,0){\line(0,1){7.42857}}\put(92,0){\line(0,1){6.85714}}\put(96,0){\line(0,1){6.28571}}\put(100,0){\line(0,1){5.71429}}\put(104,0){\line(0,1){5.14286}}\put(108,0){\line(0,1){4.57143}}\put(112,0){\line(0,1){4.}}\put(116,0){\line(0,1){3.42857}}\put(120,0){\line(0,1){2.85714}}\put(124,0){\line(0,1){2.28571}}\put(128,0){\line(0,1){1.71429}}\put(132,0){\line(0,1){1.14286}}\put(136,0){\line(0,1){0.571429}}
			\put(0,20){\line(7,-1){140}}
			}%
			\multiput(0,0)(4,0){36}{\color[rgb]{0.7,0.7,0.7}\circle*{1.7}}
			\multiput(4,4)(4,0){27}{\circle*{1.7}}
			\multiput(4,8)(4,0){20}{\circle*{1.7}}
			\multiput(4,12)(4,0){13}{\circle*{1.7}}
			\multiput(4,16)(4,0){6}{\circle*{1.7}}
			\multiput(0,4)(0,4){4}{\color[rgb]{0.7,0.7,0.7}\circle*{1.7}}
			\multiput(0,20)(28,-4){5}{\color[rgb]{0.7,0.7,0.7}\circle*{1.7}}
			\put(-3,-1){\scriptsize$0$}
			\put(-10.5,19){\scriptsize$d_{+}=5$}
			\put(-0.7,-4){\scriptsize$0$}
			\put(27.3,-4){\scriptsize$7$}
			\put(54.5,-4){\scriptsize$14$}
			\put(82.5,-4){\scriptsize$21$}
			\put(110.5,-4){\scriptsize$28$}
			\put(138.5,-4){\scriptsize$35$}
			\put(85,12.5){$L=7$ \; $N=3$}
		\end{picture}
		\end{tabular}\\\\\\
		\begin{tabular}{l}
		\begin{picture}(40,8)(-5,0)
			{\color[rgb]{0.9,0.9,1}\polygon*(0,0)(40,0)(0,8)}%
			{\color[rgb]{0.7,0.7,0.7}
			\put(0,0){\line(1,0){40}}\put(0,4){\line(1,0){20}}
			\put(0,0){\line(0,1){8}}\put(4,0){\line(0,1){7.2}}\put(8,0){\line(0,1){6.4}}\put(12,0){\line(0,1){5.6}}\put(16,0){\line(0,1){4.8}}\put(20,0){\line(0,1){4}}\put(24,0){\line(0,1){3.2}}\put(28,0){\line(0,1){2.4}}\put(32,0){\line(0,1){1.6}}\put(36,0){\line(0,1){0.8}}
			\put(0,8){\line(5,-1){40}}
			}%
			\multiput(0,0)(4,0){11}{\color[rgb]{0.7,0.7,0.7}\circle*{1.7}}
			\multiput(4,4)(4,0){4}{\circle*{1.7}}
			\put(0,4){\color[rgb]{0.7,0.7,0.7}\circle*{1.7}}
			\multiput(0,8)(20,-4){2}{\color[rgb]{0.7,0.7,0.7}\circle*{1.7}}
			\put(-3,-1){\scriptsize$0$}
			\put(-10.5,7){\scriptsize$d_{+}=2$}
			\put(-0.7,-4){\scriptsize$0$}
			\put(19.3,-4){\scriptsize$5$}
			\put(38.5,-4){\scriptsize$10$}
			\put(16,7){$L=5$ \; $N=2$}
		\end{picture}
		\hspace{18mm}
		\begin{picture}(80,12)(-5,0)
			{\color[rgb]{0.9,0.9,1}\polygon*(0,0)(80,0)(8,12)(0,12)}%
			{\color[rgb]{0.7,0.7,0.7}
			\put(0,0){\line(1,0){80}}\put(0,4){\line(1,0){56}}\put(0,8){\line(1,0){32}}\put(0,12){\line(1,0){8}}
			\put(0,0){\line(0,1){12}}\put(4,0){\line(0,1){12}}\put(8,0){\line(0,1){12}}\put(12,0){\line(0,1){11.3333}}\put(16,0){\line(0,1){10.6667}}\put(20,0){\line(0,1){10}}\put(24,0){\line(0,1){9.33333}}\put(28,0){\line(0,1){8.66667}}\put(32,0){\line(0,1){8}}\put(36,0){\line(0,1){7.33333}}\put(40,0){\line(0,1){6.66667}}\put(44,0){\line(0,1){6}}\put(48,0){\line(0,1){5.33333}}\put(52,0){\line(0,1){4.66667}}\put(56,0){\line(0,1){4}}\put(60,0){\line(0,1){3.33333}}\put(64,0){\line(0,1){2.66667}}\put(68,0){\line(0,1){2.}}\put(72,0){\line(0,1){1.33333}}\put(76,0){\line(0,1){0.666667}}
			\put(8,12){\line(6,-1){72}}
			}%
			\multiput(0,0)(4,0){21}{\color[rgb]{0.7,0.7,0.7}\circle*{1.7}}
			\multiput(4,4)(4,0){13}{\circle*{1.7}}
			\multiput(4,8)(4,0){7}{\circle*{1.7}}
			\put(4,12){\color[rgb]{0.7,0.7,0.7}\circle*{1.7}}
			\multiput(0,4)(0,4){3}{\color[rgb]{0.7,0.7,0.7}\circle*{1.7}}
			\multiput(8,12)(24,-4){3}{\color[rgb]{0.7,0.7,0.7}\circle*{1.7}}
			\put(-3,-1){\scriptsize$0$}
			\put(-10.5,11){\scriptsize$d_{+}=3$}
			\put(-0.7,-4){\scriptsize$0$}
			\put(7.3,-4){\scriptsize$2$}
			\put(31.3,-4){\scriptsize$8$}
			\put(54.5,-4){\scriptsize$14$}
			\put(78.5,-4){\scriptsize$20$}
			\put(43.5,9){$L=6$ \; $N=3$}
		\end{picture}
		\end{tabular}
	\end{tabular}
	\end{center}
	\caption{Newton polygon for the characteristic polynomial associated to the current of particles for unidirectional simple exclusion process with $N$ particles on $L$ sites, represented for the cases $L=7$, $N=3$ (top), $L=5$, $N=2$ (bottom left) and $L=6$, $N=3$ (bottom right). Powers of $\lambda$ are represented horizontally and powers of $g$ vertically. The gray dots are the points with integer coordinates on the boundary of the polygon, while the black dots are the ones in the interior of the polygon. The number of black dots, equal to $66$ for the case on top, $4$ for the one at bottom left and $20$ for the one at bottom right, is the genus $\mathrm{g}$ of the Riemann surface $\R$ for generic transition rates, which is given by (\ref{g generic unidirectional}).}
	\label{fig Newton polygon TASEP}
\end{figure}

\begin{subsubsection}{Genus}\hfill\\
Since the algebraic curve $\mathcal{A}$ is non-singular for generic transition rates, its genus $\mathrm{g}$ is equal to the number of points with integer coordinates in the interior of the Newton polygon. From (\ref{P[Pk,TASEP]}), the genus is then equal to $\mathrm{g}=\sum_{k=1}^{d_{+}-1}(d_{k}-1)$ with $d_{k}={{L}\choose{N}}-kL$ the degree of the polynomial $P_{k}$, see figure~\ref{fig Newton polygon TASEP}. Thus, one finds
\begin{equation}
\label{g generic unidirectional}
\mathrm{g}=(d_{+}-1)\Bigg({{L}\choose{N}}-\frac{L\,d_{+}}{2}-1\Bigg)\;.
\end{equation}
When $L$ and $N$ are co-prime, the conjecture (\ref{d+ generic}) gives in particular $d_{+}={{L}\choose{N}}/L$, and then $\mathrm{g}=({{L}\choose{N}}-2)({{L}\choose{N}}-L)/2L$, which grows at large $L$, $N$ with fixed density of particles $\rho=N/L$ as $\mathrm{g}\simeq(4\pi\rho(1-\rho)L^{2})^{-1}\,\rme^{-2L(\rho\log\rho+(1-\rho)\log(1-\rho))}$.

We show in the next section that for the special case of TASEP, where all the transition rates are equal, the genus is significantly smaller, because of the presence of a huge number of singular points on $\mathcal{A}$.
\end{subsubsection}

\end{subsection}

\begin{subsection}{Process with all transition rates equal: TASEP}
\label{section TASEP}
We consider in this section the special case of TASEP, where all the transition rates are equal to one, and whose algebraic curve is singular. We discuss the alternative description of the corresponding Riemann surface in terms of Bethe ansatz, from which exact results for current fluctuations with simple initial conditions have been obtained in \cite{P2020.2}.

\begin{subsubsection}{Riemann surface from Bethe ansatz}\hfill\\
TASEP is an integrable model, whose dynamics can be understood in terms of quasi-particles evolving by elastic scattering, and whose momenta $q_{j}$ are preserved in one dimension up to exchanges during interactions. Bethe ansatz then consists in looking for eigenstates as appropriate linear combinations of plane waves with momenta $q_{j}$. We refer to \cite{D1998.1,GM2006.1} for introductions to Bethe ansatz in the context of TASEP with periodic boundary conditions.

TASEP with $N$ particles is described in terms of $N$ quasi-particles, and periodic boundary condition leads to Bethe equations quantizing the $N$ momenta $q_{j}$. In terms of the more convenient variables $y_{j}=1-g^{-1/L}\,\rme^{\rmi q_{j}}$, called the \textit{Bethe roots} in the following, the quantization conditions read $R(y_{j},B)=0$,
with $R$ the polynomial
\begin{equation}
\label{R(y,B)}
R(y,B)=B\,(1-y)^{L}+(-y)^{N}\;,
\end{equation}
and where the parameter
\begin{equation}
\label{B[g,y]}
B=g\prod_{j}y_{j}
\end{equation}
will make in the following an especially nice parametrization of the Riemann surface $\R$.

The equation $R(y,B)=0$ does define an algebraic curve, whose associated Riemann surface is the Riemann sphere. This is \textit{not} the Riemann surface $\R$ we are interested in here, and which is defined below as the natural domain of definition for rational symmetric functions of $N$ distinct solutions $y_{j}$ of $R(y_{j},B)=0$.

Each appropriate solution of the Bethe equations above, consisting in a set of $N$ generically distinct Bethe roots, corresponds to an eigenstate of $M(g)$, and thus also to a point $p\in\R$. Eigenvalues and eigenvectors of $M(g)$ are given by explicit rational symmetric functions of the Bethe roots, in particular
\begin{equation}
\label{lambda[y] TASEP}
\lambda=\sum_{j}\frac{y_{j}}{1-y_{j}}
\end{equation}
for the eigenvalue.

The equation $R(y,B)=0$ has $L$ solutions for $y$, which we would like to label as $y_{j}(B)$, $j=1,\ldots,L$. The Bethe root functions $y_{j}(B)$ are not analytic in $B$, but have branch points. Solving $R(y,B)=R^{(1,0)}(y,B)=0$, one finds three branch points $B=0$, $B=\infty$ and $B=B_{*}$ with $B_{*}=-N^{N}(L-N)^{L-N}/L^{L}<0$. The Bethe root functions $y_{j}$ may thus be defined so as to be analytic in the domain $\C\setminus\mathbb{R}^{-}$, with branch cuts $(-\infty,B_{*})$ and $(B_{*},0)$ located on the negative real axis. Analytic continuation across these cuts leads to permutations of the functions $y_{j}$.

We use in the following the labelling of the Bethe root functions introduced in \cite{P2020.2}, such that crossing the cut $(-\infty,B_{*})$ from above sends $y_{L}$ to $y_{1}$ and $y_{j}$ to $y_{j+1}$, $j=1,\ldots,L-1$, while crossing the cut $(B_{*},0)$ from above sends $y_{N}$ to $y_{1}$, $y_{L}$ to $y_{N+1}$, and $y_{j}$ to $y_{j+1}$ for $j\neq N,L$. Analytic continuation thus induces cyclic permutations of the Bethe root functions, either in a single cycle of length $L$ or in two disjoint cycles of length $N$ and $L-N$ depending on where the cut $\mathbb{R}^{-}$ is crossed with respect to $B_{*}$. Analytic continuations of the Bethe root functions can then be formalized by introducing two permutations $a_{0}$ and $a_{\infty}$ of the set of integers $j\in\lb1,L\rb$ as
\begin{equation}
\fl\hspace{15mm}
\label{aj}
\left\{\!\!
\begin{array}{l}
a_{\infty}\,L=1\\
a_{\infty}\,j=j+1,\;1\leq j<L
\end{array}
\right.
\qquad\text{and}\qquad
\left\{\!\!
\begin{array}{l}
a_{0}\,N=1\\
a_{0}\,L=N+1\\
a_{0}\,j=j+1,\;j\neq N,L
\end{array}
\right.\;,
\end{equation}
such that $y_{j}$ becomes $y_{a_{\infty}j}$ (respectively $y_{a_{0}j}$) when the cut $(-\infty,B_{*})$ (resp. $(B_{*},0)$) is crossed from above. The permutations $a_{0}$ and $a_{\infty}$ do not commute for $1\leq N\leq L-1$.

From (\ref{lambda[y] TASEP}) and (\ref{B[g,y]}), the eigenvalue $\lambda$ and the parameter $g$ are both rational symmetric functions of $N$ Bethe roots with coefficients rational in the variable $B$. Eigenstates may thus be parametrized by complex values of $B$ and sets $J\subset\lb1,L\rb$ with $|J|=N$ elements, in such a way that the $N$ Bethe roots characterizing the eigenstate are the $y_{j}(B)$, $j\in J$. This means that the points $p$ of the Riemann surface $\R$ may be uniquely labelled as $p=[B,J]$, except at branch points $B\in\{0,B_{*},\infty\}$ where several sets $J$ correspond to the same point on $\R$.

The Riemann surface $\R$ then consists of $|\Omega|={{L}\choose{N}}$ \textit{sheets} $\C_{J}$, copies of the complex plane for $B$ slit along $\mathbb{R}^{-}$, glued together along the cuts $(-\infty,B_{*})$ and $(B_{*},0)$ according to the action of (\ref{aj}) on sets $J$, and made compact by adding the points with $B=\infty$. This Riemann surface, built from Bethe ansatz, is expected to be the same as the one corresponding to the algebraic curve $\mathcal{A}$ from section~\ref{section counting processes}.

We emphasize that by the construction above, any rational symmetric function $s([B,\{j_{1},\ldots,j_{N}\}])=s(B,y_{j_{1}}(B),\ldots,y_{j_{N}}(B))$ of $N$ distinct Bethe root functions with coefficients rational in $B$ is meromorphic on $\R$. In particular, at the point $p=[B,J]\in\R$, the functions $\lambda(p)=\sum_{j\in J}\frac{y_{j}(B)}{1-y_{j}(B)}$, $g(p)=B/\prod_{j\in J}y_{j}(B)$ and $B(p)=B$ are indeed meromorphic on $\R$. We recall that meromorphic functions on the Riemann surface $\R$ associated to the algebraic curve $\mathcal{A}$ defined by $P(\lambda,g)=0$ can always be written as rational functions of $\lambda$ and $g$, if $\mathcal{A}$ is non-singular. From the equations above, it does not seem possible to write $B$ in such a way in general, which hints at $\mathcal{A}$ being singular. We confirm this in section~\ref{section genus TASEP} by looking at the genus of $\R$.
\end{subsubsection}

\begin{figure}
	\begin{center}
	\begin{tabular}{l}
		\begin{picture}(140,25)
			\multiput(10,10)(60,0){3}{\color{red}\circle*{3}}
			\put(10,10){\thicklines\color{red}\line(1,0){120}}
			\put(8,5.5){\color{red}$\infty$}
			\put(68,5){\color{red}$B_{*}$}
			\put(129,5){\color{red}$0$}
			\put(10,10){\circle{20}}\put(10,10){\color{white}\polygon*(0,-5)(0,-11)(10,-11)}\put(11,0){\thicklines\vector(1,0){0.1}}\put(-5,0){$a_{\infty}^{-1}$}
			\put(70,10){\circle{15}}\put(70,10){\color{white}\polygon*(0,-6)(0,-10)(10,-10)(10,-6)}\put(71,2.5){\thicklines\vector(1,0){0.1}}\put(66,0){\small$a_{\infty}a_{0}^{-1}$}
			\put(70,10){\circle{30}}\put(70,15){\color{white}\polygon*(0,5)(0,11)(-10,11)}\put(69,25){\thicklines\vector(-1,0){0.1}}\put(47,21){\small$a_{0}^{-1}a_{\infty}$}
			\put(130,10){\circle{20}}\put(130,10){\color{white}\polygon*(0,-5)(0,-11)(10,-11)}\put(131,0){\thicklines\vector(1,0){0.1}}\put(138.5,18){$a_{0}$}
			\put(35,20){\color{blue}\vector(0,-1){20}}\put(29,18){\color{blue}$a_{\infty}$}
			\put(40,0){\color{blue}\vector(0,1){20}}\put(41,1){\color{blue}$a_{\infty}^{-1}$}
			\put(100,20){\color{blue}\vector(0,-1){20}}\put(95.5,18){\color{blue}$a_{0}$}
			\put(105,0){\color{blue}\vector(0,1){20}}\put(106,1){\color{blue}$a_{0}^{-1}$}
		\end{picture}\\\\[2mm]
		\begin{picture}(140,20)
			\multiput(10,10)(60,0){3}{\color{red}\circle*{3}}
			\put(70,10){\color{white}\polygon*(-5,0)(5,0)(0,-5)}
			\put(10,10){\thicklines\color{red}\line(1,0){120}}
			\put(8,5.5){\color{red}$\infty$}
			\put(68,5){\color{red}$B_{*}$}
			\put(129,5){\color{red}$0$}
			\put(10,10){\circle{20}}\put(10,10){\color{white}\polygon*(0,-5)(0,-11)(10,-11)}\put(11,0){\thicklines\vector(1,0){0.1}}\put(-5,0){$a_{\infty}^{-1}$}
			\put(70,10){\circle{20}}\put(70,10){\color{white}\polygon*(0,5)(0,11)(-10,11)}\put(69,20){\thicklines\vector(-1,0){0.1}}\put(57,18){$a_{\infty}$}
			\put(130,10){\circle{20}}\put(130,10){\color{white}\polygon*(0,-5)(0,-11)(10,-11)}\put(131,0){\thicklines\vector(1,0){0.1}}\put(138.5,18){$a_{0}$}
			\put(37.5,20){\color{blue}\vector(0,-1){20}}\put(31.5,18){\color{blue}$a_{\infty}$}
			\put(92,3){\color{blue}\vector(0,1){17}}\put(80.4,0){\color{blue}$a_{0}^{-1}J=a_{\infty}^{-1}J$}
			\put(110,20){\color{blue}\vector(0,-1){20}}\put(111,18){\color{blue}$a_{0}$}
		\end{picture}\\\\
		\begin{picture}(140,20)
			\multiput(10,10)(60,0){3}{\color{red}\circle*{3}}
			\put(70,10){\color{white}\polygon*(-5,0)(5,0)(0,5)}
			\put(10,10){\thicklines\color{red}\line(1,0){120}}
			\put(8,5.5){\color{red}$\infty$}
			\put(68,5){\color{red}$B_{*}$}
			\put(129,5){\color{red}$0$}
			\put(10,10){\circle{20}}\put(10,10){\color{white}\polygon*(0,-5)(0,-11)(10,-11)}\put(11,0){\thicklines\vector(1,0){0.1}}\put(-5,0){$a_{\infty}^{-1}$}
			\put(70,10){\circle{20}}\put(70,10){\color{white}\polygon*(0,-5)(0,-11)(10,-11)}\put(71,0){\thicklines\vector(1,0){0.1}}\put(58,18){$a_{0}^{-1}$}
			\put(130,10){\circle{20}}\put(130,10){\color{white}\polygon*(0,-5)(0,-11)(10,-11)}\put(131,0){\thicklines\vector(1,0){0.1}}\put(138.5,18){$a_{0}$}
			\put(37.5,0){\color{blue}\vector(0,1){20}}\put(38.5,1){\color{blue}$a_{\infty}^{-1}$}
			\put(92,17){\color{blue}\vector(0,-1){17}}\put(82.7,18){\color{blue}$a_{0}J=a_{\infty}J$}
			\put(110,0){\color{blue}\vector(0,1){20}}\put(111,1){\color{blue}$a_{0}^{-1}$}
		\end{picture}\\\\
		\begin{picture}(140,20)
			\multiput(10,10)(120,0){2}{\color{red}\circle*{3}}
			\put(10,10){\thicklines\color{red}\line(1,0){120}}
			\put(8,5.5){\color{red}$\infty$}
			\put(129,5){\color{red}$0$}
			\put(10,10){\circle{20}}\put(10,10){\color{white}\polygon*(0,-5)(0,-11)(10,-11)}\put(11,0){\thicklines\vector(1,0){0.1}}\put(-5,0){$a_{\infty}^{-1}$}
			\put(130,10){\circle{20}}\put(130,10){\color{white}\polygon*(0,-5)(0,-11)(10,-11)}\put(131,0){\thicklines\vector(1,0){0.1}}\put(138.5,18){$a_{0}$}
			\put(60,17){\color{blue}\vector(0,-1){17}}\put(50.7,18){\color{blue}$a_{0}J=a_{\infty}J$}
			\put(80,3){\color{blue}\vector(0,1){17}}\put(68.4,0){\color{blue}$a_{0}^{-1}J=a_{\infty}^{-1}J$}
		\end{picture}\\\\
		\begin{picture}(140,20)
			\multiput(10,10)(60,0){2}{\color{red}\circle*{3}}
			\put(10,10){\thicklines\color{red}\line(1,0){60}}
			\put(8,5.5){\color{red}$\infty$}
			\put(68,5){\color{red}$B_{*}$}
			\put(10,10){\circle{20}}\put(10,10){\color{white}\polygon*(0,-5)(0,-11)(10,-11)}\put(11,0){\thicklines\vector(1,0){0.1}}\put(-5,0){$a_{\infty}^{-1}$}
			\put(70,10){\circle{20}}\put(70,10){\color{white}\polygon*(0,5)(0,11)(-10,11)}\put(69,20){\thicklines\vector(-1,0){0.1}}\put(79,2){$a_{\infty}$}
			\put(35,20){\color{blue}\vector(0,-1){20}}\put(29,18){\color{blue}$a_{\infty}$}
			\put(40,0){\color{blue}\vector(0,1){20}}\put(41,1){\color{blue}$a_{\infty}^{-1}$}
			\put(100,20){\color{blue}\vector(0,-1){20}}\put(100,0){\color{blue}\vector(0,1){20}}\put(101,9){\color{blue}$a_{0}J=a_{0}^{-1}J=J$}
		\end{picture}
	\end{tabular}
	\end{center}
	\caption{Monodromy operators acting on sheet labels $J\subset\lb1,L\rb$ for small loops around the branch points $\infty$, $B_{*}$, $0$ (red dots, or half-dots if the branch point exists only on one side of the cut) of the parameter $B$. The red, horizontal line represents the cuts along which sheets $\C_{J}$ are glued together within $\R$, and the blue, vertical arrow crossing the cut on the left (respectively right) indicates analytic continuation to the sheet $\C_{a_{\infty}J}$ (resp. $\C_{a_{0}J}$). The four first situations represented correspond from top to bottom to analytic continuations starting from a sheet $\C_{J}$ with $(|J\cap\{N,L\}|,|J\cap\{1,N+1\}|)$ respectively equal to $(1,1)$, $(1,0\;\text{or}\;2)$, $(0\;\text{or}\;2,1)$, $(0\;\text{or}\;2,0\;\text{or}\;2)$. The special case at the bottom corresponds to analytic continuation from the sheet $J=\lb1,N\rb$.}
	\label{fig monodromy B TASEP}
\end{figure}
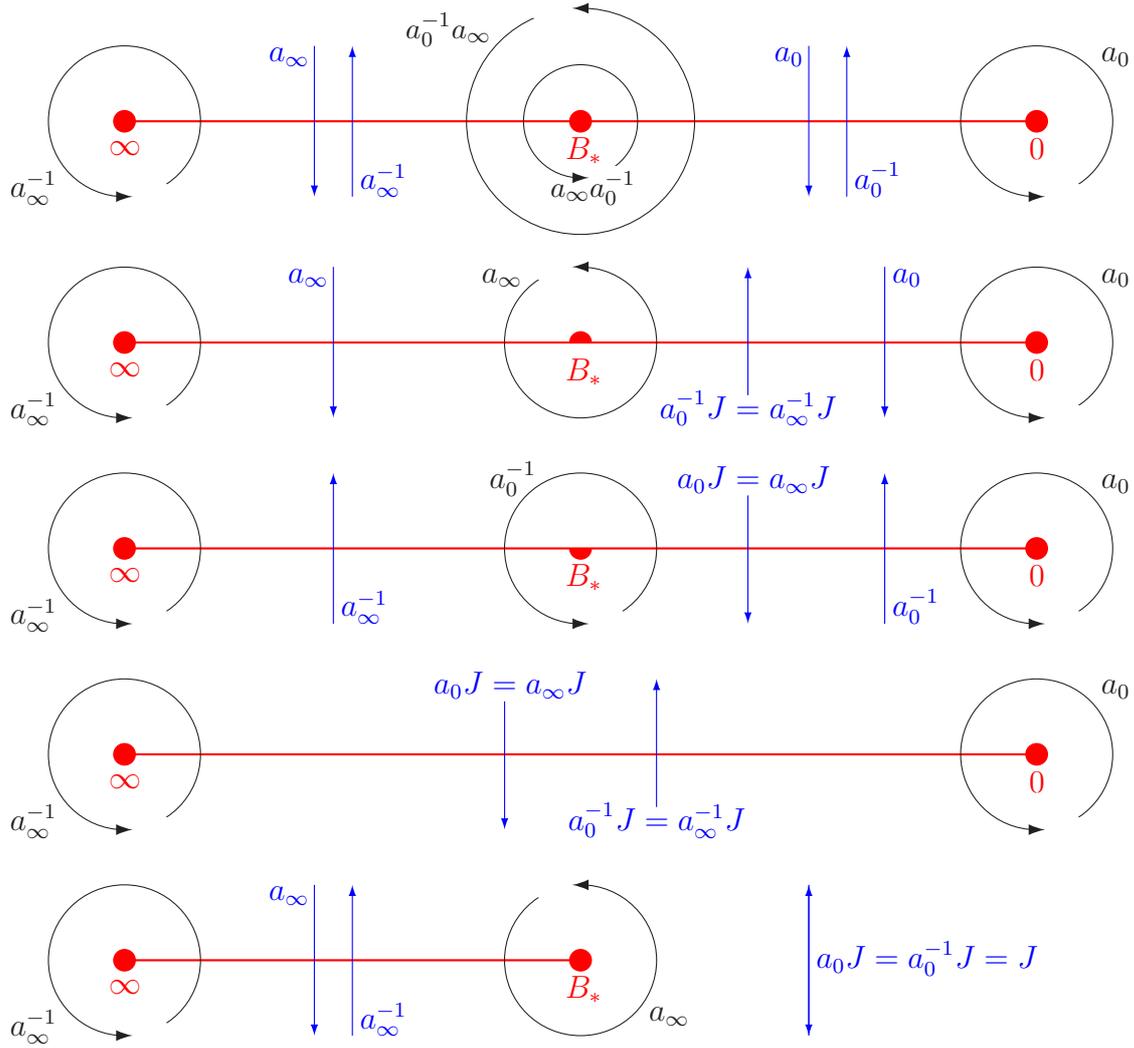

\begin{subsubsection}{Connected components, ramification}\hfill\\
\label{section ramification B TASEP}
Connectivity of the sheets $\C_{J}$ of $\R$ with respect to the parametrization by $B$ is fully encoded in the action (\ref{aj}) of the group $G$ generated by $a_{\infty}$ and $a_{0}$ on sets of $N$ distinct integers between $1$ and $L$. The connected components of $\R$ are in particular described by orbits of this group action, and their number $K$ is equal to $K=1$ if and only if $L$ and $N$ are co-prime. $\R$ has always a single connected component for $N=1$, while for $N=2$, one has $K=1$ (respectively $K=2$) for $L$ odd (resp. even). Additionally, particle-hole symmetry, which consists in the Bethe ansatz formalism in replacing the sheet label $J$ by its complement $\lb1,L\rb\setminus J$, implies that $K$ is invariant under $N\to L-N$.

Analytic continuation in the variable $B$ along a small circle with positive orientation enclosing a branch point $B_{0}\in\{0,B_{*},\infty\}$ sends a sheet $\C_{J}$ to a sheet $\C_{aJ}$, with monodromy operator $a\in G$ associated to the branch point. The ramification index of the corresponding ramification point $p_{0}=[B_{0},J]$ is then the smallest integer $m\geq1$ such that $a^{m}J=J$, and $[B_{0},a^{k}J]=p_{0}$ for any $k\in\mathbb{Z}$. The monodromy operators of the branch points $0$ and $\infty$ are respectively $a_{0}$ and $a_{\infty}^{-1}$, see figure~\ref{fig monodromy B TASEP}. For the branch point $B_{*}$, which lies on the middle of the line where sheets are cut, one has to distinguish between analytic continuations starting from either side of the cut, corresponding to distinct ramification points $[B_{*}+\rmi0^{+},J]$ and $[B_{*}-\rmi0^{+},J]$: this leads to two monodromy operators, $a_{0}^{-1}a_{\infty}$, which is the transposition between $N$ and $L$, and $a_{\infty}a_{0}^{-1}$, which is the transposition between $1$ and $N+1$.

Ramification indices for the variable $B$ depend on the ramification point and hence both on the branch point $B_{0}\in\{0,B_{*},\infty\}$ and on the sheet $J$. From the monodromy operators above, the point $[0,\lb1,N\rb]$ representing the stationary state of the model, see below, is never a ramification point since $a_{0}\lb1,N\rb=\lb1,N\rb$, while ramification indices $m_{B}^{0}$ for other points $[0,J]$ depend on $J$. All the points $[\infty,J]$ are ramification points, whose ramification index $m_{B}^{\infty}$ depends on $J$ in general. Finally, the points $[B_{+}+\rmi0^{+},J]$ (respectively $[B_{+}-\rmi0^{+},J]$) have ramification index $2$ if $J\cap\{N,L\}$ (resp. $J\cap\{1,N+1\}$) has a single element (and one has the identification $[B_{*}+\rmi0^{+},J]=[B_{*}-\rmi0^{+},a_{\infty}J]=[B_{*}+\rmi0^{+},a_{0}^{-1}a_{\infty}J]=[B_{*}-\rmi0^{+},a_{\infty}a_{0}^{-1}a_{\infty}J]$), and are not ramification points otherwise.

\begin{table}
	\begin{tabular}{|cccc|}\hline
	Point $p\in\R$ & Bethe roots & \begin{tabular}{c}Ramification\\index for $B$\end{tabular} & \begin{tabular}{c}Ramification\\index for $g$\end{tabular}\\\hline
	\begin{tabular}{c}$g=1$, $B=0$\\$J=\lb1,N\rb$\end{tabular} & all $y_{j}=0$ & $1$ & $1$\\[7mm]
	$g=B=0$ & all $y_{j}\in\{0,\infty\}$ & $m_{B}^{0}$ & $m_{g}^{0}$\\[7mm]
	$g=B=\infty$ & all $y_{j}=1$ & $m^{\infty}$ & $m^{\infty}$\\[7mm]
	\begin{tabular}{c}$B=B_{*}+\rmi0^{+}$\\$|J\cap\{N,L\}|=1$\end{tabular} & \begin{tabular}{c}one Bethe root\\equal to $-\frac{N}{L-N}$\end{tabular} & $2$ & $1$\\[5mm]
	\begin{tabular}{c}$B=B_{*}-\rmi0^{+}$\\$|J\cap\{1,N+1\}|=1$\end{tabular} & \begin{tabular}{c}one Bethe root\\equal to $-\frac{N}{L-N}$\end{tabular} & $2$ & $1$\\[5mm]
	\begin{tabular}{c}$B=B_{*}+\rmi0^{+}$\\$|J\cap\{N,L\}|=2$\end{tabular} & \begin{tabular}{c}two Bethe roots\\equal to $-\frac{N}{L-N}$\end{tabular} & $1$ & $1$\\[5mm]
	\begin{tabular}{c}$B=B_{*}-\rmi0^{+}$\\$|J\cap\{1,N+1\}|=2$\end{tabular} & \begin{tabular}{c}two Bethe roots\\equal to $-\frac{N}{L-N}$\end{tabular} & $1$ & $1$\\[5mm]
	$\kappa=0$, $g\neq1$ & unremarkable & $1$ & $2$\\[2mm]\hline
	\end{tabular}
	\caption{Bethe roots and ramification indices of some special points on the Riemann surface $\R$ for the current of TASEP.}
	\label{table points TASEP}
\end{table}

At ramification points for $B$, several Bethe roots necessarily coincide, see table~\ref{table points TASEP} for a summary and \cite{P2020.2} for detailed derivations. When $B\to\infty$, which is equivalent to $g\to\infty$, all the $y_{j}$ converge to $1$ as
\begin{equation}
\label{yj Binfinity}
1-y_{j}(B)\sim B^{-1/L}\;,
\end{equation}
and one has $g\simeq B$. When $B\to0$, our labelling of the Bethe root functions implies that all the $y_{j}(B)$ with $j\in\lb1,N\rb$ converge to $0$ while all the $y_{j}$ with $j\in\lb N+1,L\rb$ go to $\infty$, as
\begin{equation}
\label{yj B0}
y_{j}\sim\Bigg\{
\begin{array}{llc}
	B^{1/N} && 1\leq j\leq N\\
	B^{-1/(L-N)} && N+1\leq j\leq L\\
\end{array}\;.
\end{equation}
In particular, the stationary point $o=[0,\lb1,N\rb]\in\R$ with $g=1$ corresponds to all the Bethe roots equal to $0$, while $B=0$ with $J\neq\lb1,N\rb$ is equivalent to $g=0$. Finally, Bethe root functions $y_{j}(B)\neq-\frac{N}{L-N}$ everywhere, except at $B=B_{*}+\rmi0^{+}$ (respectively $B=B_{*}-\rmi0^{+}$), where the Bethe root functions $y_{N}$ and $y_{L}$ (resp. $y_{1}$ and $y_{N+1}$) are equal to $-\frac{N}{L-N}$.
\end{subsubsection}

\begin{subsubsection}{Genus}\hfill\\
\label{section genus TASEP}
By construction, the meromorphic function $B$ on $\R$ has degree $|\Omega|$, and the genus of $\R$ can be computed from the Riemann-Hurwitz formula (\ref{Riemann-Hurwitz}). Particle-hole symmetry implies that $\mathrm{g}$ is invariant under $N\to L-N$. For $N=1$, we observe that there is a single ramification point for each branch point, with ramification index respectively $L$, $L-1$, $2$ for $B=\infty,0,B_{*}$. The Riemann-Hurwitz formula then gives $\mathrm{g}=-L+1+\frac{L-1}{2}+\frac{L-2}{2}+\frac{1}{2}=0$, and $\R$ is then the Riemann sphere. There should then exist a global parametrization $z$ in terms of which any meromorphic function on $\R$ can be expressed as a rational function of $z$. Since meromorphic functions on $\R$ are defined here as rational functions of $B$ and the Bethe root $y$, the relation $B=y/(1-y)^{L}$ implies that one can choose $z=y$. For $N=2$, one can show that the Riemann-Hurwitz formula still implies $\mathrm{g}=0$, and $\R$ is either one Riemann sphere for $L$ odd or two Riemann spheres for $L$ even. One can then choose $z=\sqrt{\frac{1-y_{1}}{1-y_{2}}}$ for a parameter in terms of which any meromorphic function on $\R$ is rational.

For general $L$, $N$, the ramification data for the variable $B$ is rather involved, and we focus on the case where $L$ and $N$ are co-prime, corresponding to $\R$ having a single connected component. Then, the orbits of $a_{\infty}$ have length $L$, which implies that the ${{L}\choose{N}}/L$ distinct ramification points with $B=\infty$ all have ramification index $m_{B}^{\infty}=L$, and contribute $\frac{L-1}{2L}{{L}\choose{N}}$ to the genus. Furthermore, the ramification points with $B=B_{*}$, corresponding e.g. to the number of ways to choose a sheet label $J$ containing $N$ but not $L$, contribute ${{L-2}\choose{N-1}}$ to the genus (the set of ramification points of the form $[B_{*}+\rmi0^{+},J]$, $|J\cap\{N,L\}|=1$ and those of the form $[B_{*}-\rmi0^{+},J]$, $|J\cap\{1,N+1\}|=1$ are the same, and should not be counted twice).

Finally, the ramification points with $B=0$ are more complicated, and we further restrict to the case where both $N$ and $L-N$ are prime numbers for simplicity. The ramification points with $B=0$ contribute $\tfrac{1}{2}\sum_{p\in B^{-1}(0)}(e_{p}-1)=\tfrac{1}{2}(|\Omega|-|B^{-1}(0)|)$. The monodromy operator $a_{0}$ acts independently on $\lb1,N\rb$ and $\lb N+1,L\rb$ and preserves $k=|J\cap\lb1,N\rb|$. The cardinal of $B^{-1}(0)$, which is also the number of orbits of $a_{0}$, is then equal to $\frac{1}{N}{{N}\choose{k}}\times\frac{1}{L-N}{{L-N}\choose{N-k}}$ if $0<k<N<L-N$ and $N$ and $L-N$ are prime. Adding the cases $k=0$ and $k=N$, summing over $k$ and treating separately the case $N>L-N$ then leads to $|B^{-1}(0)|=1+\frac{1}{N(L-N)}({{L}\choose{N}}-1)+(\frac{1}{\max(N,L-N)}-\frac{1}{N(L-N)}){{\max(N,L-N)}\choose{\min(N,L-N)}}$.

Gathering the contributions of all the branch points, the Riemann-Hurwitz formula (\ref{Riemann-Hurwitz}) finally gives the genus for TASEP with $N$, $L-N$ prime numbers and $L$ and $N$ co-prime as
\begin{eqnarray}
\label{g TASEP}
&& \mathrm{g}=\frac{1}{2}+\frac{1}{2N(L-N)}+\frac{1}{2L}\Big(\frac{N(L-N)}{L-1}-1-\frac{1}{N}-\frac{1}{L-N}\Big){{L}\choose{N}}\\
&& \hspace{10mm}-\frac{1}{2}\Big(\frac{1}{\max(N,L-N)}-\frac{1}{N(L-N)}\Big){{\max(N,L-N)}\choose{\min(N,L-N)}}\;.\nonumber
\end{eqnarray}
This expression for the genus grows at large $L$, $N$ with fixed density of particles $\rho=N/L$ as $\mathrm{g}\simeq(\frac{\rho(1-\rho)}{8\pi L})^{1/2}\,\rme^{-L(\rho\log\rho+(1-\rho)\log(1-\rho))}$, which is much smaller than the genus with generic transition rates (\ref{g generic unidirectional}), indicating the presence of a huge number of singular points on the algebraic curve $\mathcal{A}$ for TASEP in the thermodynamic limit.

For example, for $L=5$, $N=2$, the genus for TASEP is equal to zero while the interior of the Newton polygon, represented in figure~\ref{fig Newton polygon TASEP}, has $4$ points, corresponding to a genus equal to $4$ for generic transition rates. Solving $P(\lambda,g)=P^{(1,0)}(\lambda,g)=P^{(0,1)}(\lambda,g)=0$ gives $3$ singular points $(\lambda,g)\in\mathcal{A}$: $(-\frac{3}{2}\pm\frac{1}{2\sqrt{5}},\pm\frac{1}{25\sqrt{5}})$, which are nodal points and both reduce the genus by one compared to the case with generic transition rates, and $(-1,0)$, which is non-nodal and is responsible for a further decrease of the genus by two.
\end{subsubsection}

\begin{subsubsection}{Degree of the characteristic polynomial}\hfill\\
\label{section d+ TASEP}
The characteristic polynomial $P(\lambda,g)$ of the matrix $M(g)$ does not appear in the Bethe ansatz construction of the Riemann surface $\R$. Its degree $d_{+}$ in the variable $g$ can however be computed directly from the behaviour of the eigenvalues of $M(g)$ at large $g$, which is useful to compare with the conjectured expression (\ref{d+ generic}) for generic transition rates.

With our choice of labelling of the Bethe root functions \cite{P2020.2}, one has the expansion $y_{j}(B)\simeq1-\omega_{j}^{-1}B^{-1/L}+\frac{N}{L}\,\omega_{j}^{-2}B^{-2/L}$ for $|B|\to\infty$ with $-\pi<\arg B<\pi$, where $\omega_{j}=\rme^{\frac{2\rmi\pi}{L}(j-\frac{N+1}{2})}$ and non-integer powers are defined with branch cut $\mathbb{R}^{-}$. After some calculations, one finds for the eigenvalue
\begin{equation}
\label{lambda ginf TASEP}
\lambda([B,J])\simeq g^{1/L}\sum_{j\in J}\omega_{j}-\frac{N(L-N)+(\sum_{j\in J}\omega_{j})(\sum_{j\in J}\omega_{j}^{-1})}{L}\;.
\end{equation}
We observe that at large $|g|$, the eigenvalue on the sheet $\C_{J}$ either grows as $g^{1/L}$ if $\sum_{j\in J}\rme^{2\rmi\pi j/L}\neq0$ or converges to a non-zero constant if $\sum_{j\in J}\rme^{2\rmi\pi j/L}=0$. Since the characteristic polynomial is equal to $P(\lambda,g)=\prod_{r=1}^{|\Omega|}(\lambda-\lambda_{r}(g))$ with $\lambda_{r}(g)$ the eigenvalues of $M(g)$, the degree $d_{+}$ appearing in (\ref{P[Pk,TASEP]}) is equal to $1/L$ times the number of sheet labels $J$ such that $\sum_{j\in J}\rme^{2\rmi\pi j/L}\neq0$.

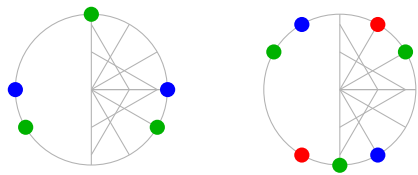
\begin{figure}
	\begin{center}
	\begin{tabular}{lll}
		\begin{tabular}{l}
		\begin{picture}(20,20)(0,2)
		\put(10,10){\color[rgb]{0.7,0.7,0.7}\circle{20}}
		\put(10,10){\rotatebox{0}{\color[rgb]{0.7,0.7,0.7}\line(1,0){10}}}
		\put(10,10){\rotatebox{30}{\color[rgb]{0.7,0.7,0.7}\line(1,0){10}}}
		\put(10,10){\rotatebox{60}{\color[rgb]{0.7,0.7,0.7}\line(1,0){10}}}
		\put(10,10){\rotatebox{90}{\color[rgb]{0.7,0.7,0.7}\line(1,0){10}}}
		\put(10,10){\rotatebox{120}{\color[rgb]{0.7,0.7,0.7}\line(1,0){10}}}
		\put(10,10){\rotatebox{150}{\color[rgb]{0.7,0.7,0.7}\line(1,0){10}}}
		\put(10,10){\rotatebox{180}{\color[rgb]{0.7,0.7,0.7}\line(1,0){10}}}
		\put(10,10){\rotatebox{210}{\color[rgb]{0.7,0.7,0.7}\line(1,0){10}}}
		\put(10,10){\rotatebox{240}{\color[rgb]{0.7,0.7,0.7}\line(1,0){10}}}
		\put(10,10){\rotatebox{270}{\color[rgb]{0.7,0.7,0.7}\line(1,0){10}}}
		\put(10,10){\rotatebox{300}{\color[rgb]{0.7,0.7,0.7}\line(1,0){10}}}
		\put(10,10){\rotatebox{330}{\color[rgb]{0.7,0.7,0.7}\line(1,0){10}}}
		\put(0,10){\color{blue}\circle*{2}}
		\put(20,10){\color{blue}\circle*{2}}
		\put(10,20){\color[rgb]{0,0.7,0}\circle*{2}}
		\put(18.66,5){\color[rgb]{0,0.7,0}\circle*{2}}
		\put(1.34,5){\color[rgb]{0,0.7,0}\circle*{2}}
		\end{picture}
		\end{tabular}
		&&
		\begin{tabular}{l}
		\begin{picture}(20,20)(0,2)
		\put(10,10){\color[rgb]{0.7,0.7,0.7}\circle{20}}
		\put(10,10){\rotatebox{0}{\color[rgb]{0.7,0.7,0.7}\line(1,0){10}}}
		\put(10,10){\rotatebox{30}{\color[rgb]{0.7,0.7,0.7}\line(1,0){10}}}
		\put(10,10){\rotatebox{60}{\color[rgb]{0.7,0.7,0.7}\line(1,0){10}}}
		\put(10,10){\rotatebox{90}{\color[rgb]{0.7,0.7,0.7}\line(1,0){10}}}
		\put(10,10){\rotatebox{120}{\color[rgb]{0.7,0.7,0.7}\line(1,0){10}}}
		\put(10,10){\rotatebox{150}{\color[rgb]{0.7,0.7,0.7}\line(1,0){10}}}
		\put(10,10){\rotatebox{180}{\color[rgb]{0.7,0.7,0.7}\line(1,0){10}}}
		\put(10,10){\rotatebox{210}{\color[rgb]{0.7,0.7,0.7}\line(1,0){10}}}
		\put(10,10){\rotatebox{240}{\color[rgb]{0.7,0.7,0.7}\line(1,0){10}}}
		\put(10,10){\rotatebox{270}{\color[rgb]{0.7,0.7,0.7}\line(1,0){10}}}
		\put(10,10){\rotatebox{300}{\color[rgb]{0.7,0.7,0.7}\line(1,0){10}}}
		\put(10,10){\rotatebox{330}{\color[rgb]{0.7,0.7,0.7}\line(1,0){10}}}
		\put(5,18.66){\color{blue}\circle*{2}}
		\put(15,1.34){\color{blue}\circle*{2}}
		\put(5,1.34){\color{red}\circle*{2}}
		\put(15,18.66){\color{red}\circle*{2}}
		\put(10,0){\color[rgb]{0,0.7,0}\circle*{2}}
		\put(18.66,15){\color[rgb]{0,0.7,0}\circle*{2}}
		\put(1.34,15){\color[rgb]{0,0.7,0}\circle*{2}}
		\end{picture}
		\end{tabular}
	\end{tabular}
	\end{center}
	\caption{Aperiodic sets $J$ such that $\sum_{j\in J}\rme^{2\rmi\pi j/L}=0$ for $L=12$, $N=5$ (left) and $N=7$ (right). The values $\rme^{2\rmi\pi j/L}$, $j\in J$ are represented by dots, such that the sum of the $\rme^{2\rmi\pi j/L}$ cancels for the dots of same colour. There are $12$ such sets $J$ for each value of $N$, corresponding to shifts of $J$ modulo $L$. These sets are responsible for a discrepancy between TASEP and the model with generic transition rates about the degree $d_{+}$ of the characteristic polynomial $P(\lambda,g)$ in the variable $g$.}
	\label{fig non-periodic J sum 0}
\end{figure}

Writing $d_{+}^{\text{TASEP}}$ (respectively $d_{+}^{\text{generic}}$) for the value taken by $d_{+}$ for TASEP (resp. for generic transition rates), one has necessarily $d_{+}^{\text{TASEP}}\leq d_{+}^{\text{generic}}$ since cancellations in the characteristic polynomial can only decrease $d_{+}$. This is consistent with the conjecture that $d_{+}^{\text{generic}}$ is equal to the number (\ref{d+ generic}) of aperiodic sets $J$ modulo $L$ since periodic sets $J$ always have $\sum_{j\in J}\rme^{2\rmi\pi j/L}=0$. The smallest system with $d_{+}^{\text{TASEP}}\neq d_{+}^{\text{generic}}$ has $L=12$ and either $N=5$ or $N=7$. The mismatch comes from the existence of an aperiodic $J$ with $\sum_{j\in J}\rme^{2\rmi\pi j/L}=0$, see figure~\ref{fig non-periodic J sum 0}. Since the next system size where such a set $J$ appears is for $L=18$, we were only able to confirm numerically this discrepancy for the cases with $L=12$.
\end{subsubsection}

\begin{subsubsection}{Ramification structure in the variable $g$}\hfill\\
\label{section ramification g TASEP}
Bethe ansatz for TASEP gives a natural parametrization of $\R$ in terms of the function $B$, whose ramification data follows from the action of the operators (\ref{aj}). Ramification in the variable $g$ is however more natural from the point of view of the generator $M(g)$ of the counting process.

The relation between the ramification data for the variables $B$ and $g$ is determined by the function $\rmd g/\rmd B$. From (\ref{B[g,y]}) and (\ref{R(y,B)}), one has
\begin{equation}
\label{dg[dB]}
\frac{\rmd g}{g}=\kappa\,\frac{\rmd B}{B}
\end{equation}
with
\begin{equation}
\label{kappa(B,J)}
\kappa([B,J])=\frac{L}{N}\sum_{j\in J}\frac{y_{j}(B)}{N+(L-N)y_{j}(B)}\;.
\end{equation}

Let $p_{0}$ be a point on $\R$, and $m_{B}$ (respectively $m_{g}$) its ramification index for the variable $B$ (resp. $g$), taken equal to $1$ if $p_{0}$ is not a ramification point. If $B(p_{0})\neq\infty$, which is equivalent to $g(p_{0})\neq\infty$, see section~\ref{section ramification B TASEP}, there exists a local parameter $z$ in a neighbourhood of $p_{0}$ such that $B(p)-B(p_{0})\sim z^{m_{B}}$ and $g(p)-g(p_{0})\sim z^{m_{g}}$. If both $B(p_{0})$ and $g(p_{0})$ are additionally non-zero, one has $\kappa(p)\sim\rmd g/\rmd B\sim z^{m_{g}-m_{B}}$ for $p$ close to $p_{0}$.

The points on $\R$ with finite non-zero $B$ and $g$ at which ramification indices for the variables $B$ and $g$ differ are thus the poles and zeroes of $\kappa$. Since Bethe root functions may only be equal to $-\frac{N}{L-N}$ at $B=B_{*}$, the function $\kappa$ can only have poles there. More precisely, a detailed study shows that $\kappa$ has only simple poles, located at the ramification points for $B$ with $B=B_{*}$. These points are thus not ramified for $g$. Conversely, the zeroes of $\kappa$ with non-zero $B$, which are simple zeroes, are ramified twice for $g$.

The function $\kappa$ has an additional simple zero at the stationary point $o=[0,\lb1,N\rb]$, where all the Bethe roots vanish, and which is neither ramified for $B$ (since $a_{0}\lb1,N\rb=\lb1,N\rb$) nor for $g$ (the corresponding eigenvalue of $M$ is not degenerate by the Perron-Frobenius theorem). Indeed, $B(p)\sim z$ and $g(p)-1\sim z$ implies $\kappa(p)\sim z$ around that point. Finally, $\kappa$ has no poles or zeroes at points with $B=g=0$ or at points with $B=g=\infty$. Since $\kappa$ must have as many zeroes as poles, the number of zeroes $p\neq o$ of $\kappa$, equal to the number of ramification points for $g$ and hence of Jordan blocks of $M(g)$ with $g\not\in\{0,\infty\}$, is then equal to ${{L-2}\choose{N-1}}-1$. This was checked directly by solving numerically $P(\lambda,g)=P^{(1,0)}(\lambda,g)=0$ up to $L=7$, where $P$ is the characteristic polynomial of $M(g)$.

The fact that $\kappa$ has no poles or zeroes with $B=g=0$ and $B=g=\infty$ does not say anything about ramification there since $\rmd g/g\sim z\sim\rmd B/B$ at those points independently of ramification indices. The points with $B=g=\infty$ have in fact the same ramification indices $m_{B}^{\infty}=m_{g}^{\infty}$ (simply written $m^{\infty}$ in the following) for $B$ and $g$ since $g\simeq B$ when $|B|\to\infty$. On the other hand, ramification indices $m_{g}^{0}$ for $g$ at the points with $B=g=0$ can be deduced from the behaviour (\ref{yj B0}) of the Bethe root functions when $B\to0$, which implies $g\sim B^{\frac{L|J\cap\lb N+1,L\rb|}{N(L-N)}}$ on the sheet $\C_{J}$, and thus $m_{g}^{0}/m_{B}^{0}=\frac{L|J\cap\lb N+1,L\rb|}{N(L-N)}$.

For example, in the case with $L=5$, $N=2$ and $|\Omega|=10$ states, there are two distinct points on $\R$ with $g=\infty$, both ramified $m^{\infty}=5$ times for $B$ and for $g$, corresponding respectively to sets $J=\{1,2\}$ and $J=\{1,3\}$ modulo $L$, with Bethe roots equal to $1$ and eigenvalue $\lambda=\infty$. There are also two points on $\R$ with $g=0$: a single point corresponding to all $6$ sets $J$ such that $|J\cap\{3,4,5\}|=1$, ramified $m_{B}^{0}=6$ times for $B$ and $m_{g}^{0}=5$ times for $g$, with one Bethe root equal to $0$, the other one to $\infty$ and eigenvalue $\lambda=-1$, and another point corresponding to all $3$ sets $J$ such that $J\subset\{3,4,5\}$, ramified $m_{B}^{0}=3$ times for $B$ and $m_{g}^{0}=5$ times for $g$, with both Bethe roots infinite and eigenvalue $\lambda=-2$ (for the remaining set $J=\{1,2\}$, the point at $B=0$ is the stationary point $o$ and corresponds to $g=1$ and not $g=0$). Finally, the function $\kappa$ has two zeroes $p_{0}\neq o$, with Bethe roots $\{y_{j},j\in J\}=\{3(1\pm\rmi)/2+\sqrt{1\pm11\,\rmi/2},\;3(1\pm\rmi)/2-\sqrt{1\pm11\,\rmi/2}\}$, corresponding to $B(p_{0})=-\frac{79\pm3\rmi}{3125}$, $\lambda(p_{0})=\frac{-7\pm\rmi}{5}$ and $g(p_{0})=\frac{41\mp38\rmi}{3125}$, which are ramified twice for $g$ but are regular points for $B$. All this is consistent with a direct calculation at the level of the algebraic curve $P(\lambda,g)$, solving $P(\lambda,g)=P^{(1,0)}(\lambda,g)=0$ with $P^{(0,1)}(\lambda,g)\neq0$, and the matrix $M(g)$ indeed has Jordan blocks at the branch points for $g$.
\end{subsubsection}

\begin{table}
	\begin{tabular}{|lc|cccc|}\hline
		Point $p\in\R$ & local parameter $z$ & $\frac{\rmd B}{B}$ & $\kappa\,\frac{\rmd B}{B}$ & $\kappa^{2}\,\frac{\rmd B}{B}$ & $\frac{\rmd g}{g-1}$\\\hline
		$g=1$, $B=0$ & $B\sim g-1$ & $(-1)_{1}$ & $\cdot$ & $1$ & $(-1)_{1}$\\[2mm]
		$g=1$, $B\neq0$ & $B\sim g-1$ & $\cdot$ & $\cdot$ & $\cdot$ & $(-1)_{1}$\\[2mm]
		$B=g=0$ & $B^{1/m_{B}^{0}}\sim g^{1/m_{g}^{0}}$ & $(-1)_{m_{B}^{0}}$ & $(-1)_{m_{g}^{0}}$ & $(-1)_{\frac{(m_{g}^{0})^{2}}{m_{B}^{0}}}$ & $m_{g}^{0}-1$\\[2mm]
		$B=g=\infty$ & $B^{-1/m^{\infty}}\simeq g^{-1/m^{\infty}}$ & $\!(-1)_{-m^{\infty}}$ & $\!(-1)_{-m^{\infty}}$ & $(-1)_{-m^{\infty}}$ & $\!(-1)_{-m^{\infty}}$\\[2mm]
		$\kappa=\infty$, $B=B_{*}$ & $\sqrt{B-B_{*}}\sim g-g_{0}$ & $1$ & $\cdot$ & $(-1)_{-\frac{L}{N(L-N)}}$ & $\cdot$\\[2mm]
		$\kappa=0$, $g\neq1$ & $B-B_{0}\sim\sqrt{g-g_{0}}$ & $\cdot$ & $1$ & $2$ & $1$\\[2mm]\hline
	\end{tabular}\vspace{2mm}\\
	\begin{tabular}{|l|ccc|}\hline
		Point $p\in\R$ & $\kappa$ & $\mathcal{N}_{\text{stat}}$ & $\mathcal{N}_{\text{dw}_{i}}$\\\hline
		$g=1$, $B=0$ & $1$ & $\cdot$ & $\cdot$\\[2mm]
		$g=1$, $B\neq0$ & $\cdot$ & $2$ & $1$\\[2mm]
		$B=g=0$ & $\cdot$ & $(k-1)\,m_{g}^{0}-m_{B}^{0}$ & $(k-1-\frac{iN}{L})m_{g}^{0}$\\[2mm]
		$B=g=\infty$ & $\cdot$ & $-\frac{N(L-N)}{L}\,m^{\infty}$ & $-\frac{N(L-N-i)}{L}\,m^{\infty}$\\[2mm]
		$\kappa=\infty$, $B=B_{*}$ & $-1$ & $\cdot$ & $\cdot$\\[2mm]
		$\kappa=0$, $g\neq1$ & $1$ & $-1$ & $-1$\\[2mm]\hline
	\end{tabular}
	\caption{Zeroes and poles of some meromorphic differentials (top) and functions (bottom) on the Riemann surface $\R$ for the current of TASEP. The positive integers are the orders of the zeroes, the negative integers minus the orders of the poles, and a dot indicates that the point is neither a pole nor a zero. Residues of the poles are shown in index for the differentials. The ramification indices $m^{\infty}$, $m_{B}^{0}$, $m_{g}^{0}$ and the integer $k=|J\cap\lb N+1,L\rb|$ depend on the sheet $\C_{J}$ to which the point $p$ belongs. One can check that the differentials have total residue $0$, and that the functions have the same number of poles and zeroes.}
	\label{table poles diff TASEP}
\end{table}

\begin{subsubsection}{Explicit differentials for simple initial conditions}\hfill\\
We finally consider the probability of the current $Q_{t}$ for TASEP. The general expression (\ref{Prob[int R]}) applies. Since the variable $B$ appears to parametrize $\R$ in a simpler way than $g$, we use (\ref{dg[dB]}) in order to replace the differential $\rmd g$ by $\rmd B$, which introduces the function $\kappa$ defined in (\ref{kappa(B,J)}). Then, when $L$ and $N$ are co-prime, $\R$ has a single connected component, and one can write from (\ref{Prob[int R]})
\begin{equation}
\label{Prob TASEP}
\P(Q_{t}=Q)=\oint_{\Gamma}\frac{\rmd B}{2\rmi\pi B}\,\rme^{\int_{o}^{p}(\rmd\log(\kappa\mathcal{N})+t\,\rmd\lambda-Q\,\rmd g/g)}\;.
\end{equation}
Explicit expressions were obtained from Bethe ansatz in \cite{P2020.2} for the differential $\rmd\log(\kappa\mathcal{N})$ with special initial conditions. For stationary initial condition, one has
\begin{equation}
\label{dlogkappaN stat}
\rmd\log(\kappa\,\mathcal{N})_{\text{stat}}=\frac{N(L-N)}{L}\,\kappa^{2}\,\frac{\rmd B}{B}+\frac{2\,\rmd g}{g-1}-\frac{\rmd g}{g}-\frac{\rmd B}{B}\;,
\end{equation}
while domain wall initial condition with particles located at positions $L-N-i+j$, $j=1,\ldots,N$ with $0\leq i\leq L-N$ leads to
\begin{equation}
\label{dlogkappaN dw}
\rmd\log(\kappa\,\mathcal{N})_{\text{dw}_{i}}=\frac{N(L-N)}{L}\,\kappa^{2}\,\frac{\rmd B}{B}+\frac{\rmd g}{g-1}-\Big(1+\frac{iN}{L}\Big)\frac{\rmd g}{g}\;.
\end{equation}
In both cases, we observe that the poles of $\rmd\log(\kappa\mathcal{N})$ (and thus also the poles and zeroes of $\kappa\mathcal{N}$) are located only at points with $g\in\{0,1,\infty\}$ and at the ramification points for $B$ with $B=B_{*}$. The poles and zeroes of the function $\mathcal{N}$ are then located only at points with $g\in\{0,1,\infty\}$ and at the ramification points for $g$, see table~\ref{table poles diff TASEP}. We emphasize that for general initial condition, the poles and zeroes of $\mathcal{N}$ are not expected to lie at such simple locations: this is a very special feature of the initial conditions above.

For joint statistics of $Q_{t}$ at multiple times, the additional scalar products $\langle\psi(p_{\ell+1})|\psi(p_{\ell})\rangle$ appearing in the integrand for the probability (\ref{Prob[int R] multiple-time}) give from \cite{P2020.2} explicit factors depending again on $\kappa$ only, $\exp(-\frac{N(L-N)}{L}\int_{0}^{1}\frac{\rmd u}{u}\,\kappa([uB_{\ell},\cdot])\kappa([uB_{\ell+1},\cdot]))$, where the path of integration between $0$ and $1$ is such that $([uB_{\ell},\cdot],[uB_{\ell+1},\cdot])$ lifts to a path from $(o,o)$ to $(p_{\ell},p_{\ell+1})$ on the fibre product $\R*\R$ generated by analytic continuations, which is a connected space here.

It was shown in \cite{P2020.2} that the expression (\ref{Prob TASEP}) with either (\ref{dlogkappaN stat}) or (\ref{dlogkappaN dw}) is particularly suitable for asymptotic analysis to the KPZ fixed point with periodic boundaries, allowing to recover earlier results \cite{P2016.1,BL2018.1} in a much cleaner way. At large $L$, $N$ with fixed density of particles $\rho=N/L$, one finds on the sheet containing the point $o$ the asymptotics $\frac{N(L-N)}{L}\,\kappa([B,\lb1,N\rb])\simeq-\frac{\Li_{1/2}(B/B_{*})}{\sqrt{2\pi}}$ for $B\sim B_{*}$, where $\Li_{1/2}(z)=\sum_{n=1}^{\infty}\frac{z^{n}}{\sqrt{n}}$ is the polylogarithm of index $1/2$ characterizing stationary large deviations of the current \cite{DL1998.1}. Under analytic continuations, the domain of $\Li_{1/2}(z)$ can be extended to a non-compact Riemann surface $\R_{\text{KPZ}}$, whose ramification data in the variable $z$ is analogue to that of $\R$ in the variable $B$ (up to some additional complications coming from the fact that $\R$ splits into several connected components in the KPZ scaling limit). More precisely, $\Li_{1/2}(z)$ is analytic on $\C\setminus(1,\infty)$, with branch point $z=1$, and analytic continuation across $(1,\infty)$ gives the additional branch point $z=0$, leading to two cuts $(0,1)$ and $(1,\infty)$ for the other branches of $\Li_{1/2}$. This is consistent with the branch cut structure of the sheets $\C_{J}$ of $\R$, see figure~\ref{fig monodromy B TASEP}, and the KPZ scaling limit thus preserves the local connectivity of $\R$.
\end{subsubsection}

\end{subsection}

\end{section}

\begin{section}{Current for bidirectional simple exclusion process in one dimension}
\label{section single file bidirectional}
In this section we consider the integer counting process equal to the local time-integrated current of particles for a simple exclusion process on a one-dimensional periodic lattice, where particles of a single species move in a single file by local hops in both directions between neighbouring sites $i$ and $i+1$. In a first part, we focus on the model with generic transition rates $w_{C\to C'}$ between allowed states. In a second part, we consider the asymmetric simple exclusion process (ASEP), where all the transition rates from $i$ to $i+1$ (respectively from $i+1$ to $i$) are equal to $1$ (resp. $q<1$).

\begin{subsection}{Process with generic transition rates}
We study in this section the model with generic transition rates, whose algebraic curve is non-singular.

\begin{subsubsection}{Definition of the model}\hfill\\
We consider again an exclusion process with $N$ particles on a periodic one-dimensional lattice of $L$ sites, corresponding to the same set of states $\Omega$ as in section~\ref{section single file unidirectional} with unidirectional hopping. Particles can hop forward from any site $i$ to the site $i+1$. Compared to section~\ref{section single file unidirectional}, particles can also hop backward from any site $i$ to the site $i-1$. All the allowed transition rates $w_{C\to C'}$ are assumed to be generic, and in particular non-zero.

We are interested again in the local time-integrated current of particles $Q_{t}$ between site $L$ and site $1$ up to time $t$. The current $Q_{t}$ increases by one each time a particle hops from site $L$ to site $1$ and decreases by one each time a particle hops from site $1$ to site $L$. We call $M(g)$ the generator of this integer counting process. In terms of the Markov matrix $M$, one has $\langle C'|M(g)|C\rangle=g\langle C'|M|C\rangle$ (respectively $\langle C'|M(g)|C\rangle=g^{-1}\langle C'|M|C\rangle$) if one can go from $C$ to $C'$ by moving one particle from site $L$ to site $1$ (resp. from site $1$ to site $L$), and $\langle C'|M(g)|C\rangle=\langle C'|M|C\rangle$ otherwise.

As before, it is useful to consider also the total current $Q_{t}^{\text{tot}}$, which increases by one (respectively decreases by one) each time a particle moves forward (resp. backward) anywhere in the system. We observe that the generator $M_{\text{tot}}(g^{1/L})$ of $Q_{t}^{\text{tot}}/L$ is related to $M(g)$ by the same similarity transformation (\ref{M[Mtot]}) as in the unidirectional case, and the identity (\ref{P[M,Mtot]}) for the characteristic polynomial still holds.
\end{subsubsection}

\begin{subsubsection}{Degree of the characteristic polynomial}\hfill\\
By symmetry between forward an backward hopping, the characteristic polynomial of $M(g)$ has the form
\begin{equation}
P(\lambda,g)=\sum_{k=d_{-}}^{d_{+}}P_{k}(\lambda)g^{k}
\end{equation}
with $d_{-}=-d_{+}$ and $P_{k}$ of degree $d_{k}={{L}\choose{N}}-|k|L$ from (\ref{P[M,Mtot]}).

From numerics up to $L=14$, the degree $d_{+}$ appears to be the same as in section~\ref{section single file unidirectional generic} where only forward hopping was allowed, and the conjecture (\ref{d+ generic}) still stands with bidirectional hopping. This is not entirely obvious since a few backward transitions might allow for longer cycles on the graph of the dynamics.
\end{subsubsection}

\begin{figure}
	\begin{center}
		\hspace{10mm}
		\begin{picture}(40,16)
			{\color[rgb]{0.9,0.9,1}\polygon*(0,0)(40,8)(0,16)}%
			{\color[rgb]{0.7,0.7,0.7}
			\put(0,0){\line(5,1){40}}\put(0,16){\line(5,-1){40}}
			\put(0,4){\line(1,0){20}}\put(0,8){\line(1,0){40}}\put(0,12){\line(1,0){20}}
			\put(0,0){\line(0,1){16}}\put(4,0.8){\line(0,1){14.4}}\put(8,1.6){\line(0,1){12.8}}\put(12,2.4){\line(0,1){11.2}}\put(16,3.2){\line(0,1){9.6}}\put(20,4){\line(0,1){8}}\put(24,4.8){\line(0,1){6.4}}\put(28,5.6){\line(0,1){4.8}}\put(32,6.4){\line(0,1){3.2}}\put(36,7.2){\line(0,1){1.6}}
			}%
			\multiput(4,4)(4,0){4}{\circle*{1.7}}
			\multiput(4,8)(4,0){9}{\circle*{1.7}}
			\multiput(4,12)(4,0){4}{\circle*{1.7}}
			\multiput(0,4)(0,4){3}{\color[rgb]{0.7,0.7,0.7}\circle*{1.7}}
			\multiput(0,0)(20,4){3}{\color[rgb]{0.7,0.7,0.7}\circle*{1.7}}
			\multiput(0,16)(20,-4){2}{\color[rgb]{0.7,0.7,0.7}\circle*{1.7}}
			\put(-13,-1){\scriptsize$d_{-}=-2$}
			\put(-3,7){\scriptsize$0$}
			\put(-10.5,15){\scriptsize$d_{+}=2$}
			\put(-0.7,-4){\scriptsize$0$}
			\put(19.3,-4){\scriptsize$5$}
			\put(38.5,-4){\scriptsize$10$}
		\end{picture}
		\hspace{30mm}
		\begin{picture}(60,16)
			{\color[rgb]{0.9,0.9,1}\polygon*(0,0)(12,0)(60,8)(12,16)(0,16)}%
			{\color[rgb]{0.7,0.7,0.7}
			\put(12,0){\line(6,1){48}}\put(12,16){\line(6,-1){48}}
			\put(0,0){\line(1,0){12}}\put(0,4){\line(1,0){36}}\put(0,8){\line(1,0){60}}\put(0,12){\line(1,0){36}}			\put(0,16){\line(1,0){12}}
			\put(0,0){\line(0,1){16}}\put(4,0){\line(0,1){16}}\put(8,0){\line(0,1){16}}\put(12,0){\line(0,1){16}}\put(16,0.666667){\line(0,1){14.6667}}\put(20,1.33333){\line(0,1){13.3333}}\put(24,2){\line(0,1){12}}\put(28,2.66667){\line(0,1){10.6667}}\put(32,3.33333){\line(0,1){9.33333}}\put(36,4){\line(0,1){8}}\put(40,4.66667){\line(0,1){6.66667}}\put(44,5.33333){\line(0,1){5.33333}}\put(48,6){\line(0,1){4}}\put(52,6.66667){\line(0,1){2.66667}}\put(56,7.33333){\line(0,1){1.33333}}
			}%
			\multiput(4,4)(4,0){8}{\circle*{1.7}}
			\multiput(4,8)(4,0){14}{\circle*{1.7}}
			\multiput(4,12)(4,0){8}{\circle*{1.7}}
			\multiput(0,4)(0,4){3}{\color[rgb]{0.7,0.7,0.7}\circle*{1.7}}
			\multiput(0,0)(4,0){3}{\color[rgb]{0.7,0.7,0.7}\circle*{1.7}}
			\multiput(0,16)(4,0){3}{\color[rgb]{0.7,0.7,0.7}\circle*{1.7}}
			\multiput(12,0)(24,4){3}{\color[rgb]{0.7,0.7,0.7}\circle*{1.7}}
			\multiput(12,16)(24,-4){2}{\color[rgb]{0.7,0.7,0.7}\circle*{1.7}}
			\put(-13,-1){\scriptsize$d_{-}=-2$}
			\put(-3,7){\scriptsize$0$}
			\put(-10.5,15){\scriptsize$d_{+}=2$}
			\put(-0.7,-4){\scriptsize$0$}
			\put(11.3,-4){\scriptsize$3$}
			\put(35.3,-4){\scriptsize$9$}
			\put(59,-4){\scriptsize$15$}
		\end{picture}
	\end{center}
	\caption{Newton polygon for the characteristic polynomial associated to the current of particles for bidirectional simple exclusion process with $N$ particles on $L$ sites, for $L=5$, $N=2$ (left) and $L=6$, $N=2$ (right). Powers of $\lambda$ are represented horizontally and powers of $g$ vertically. Gray dots represent the points with integer coordinates on the boundary of the polygon, while black dots represent the ones in the interior of the polygon. The number of black dots, equal to $17$ on the left and to $30$ on the right, is the genus $\mathrm{g}$ of the Riemann surface $\R$ for generic transition rates, given by (\ref{g generic bidirectional}).}
	\label{fig Newton polygon ASEP}
\end{figure}
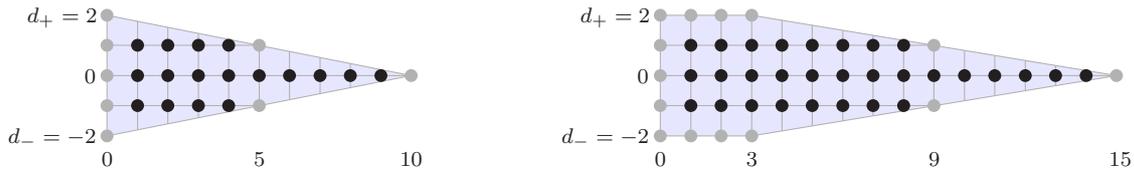

\begin{subsubsection}{Genus}\hfill\\
Since the polynomial $P_{k}$ has degree $d_{k}={{L}\choose{N}}-|k|L$, the Newton polygon for bidirectional hopping is essentially two copies of the Newton polygon for unidirectional hopping put together, compare figures~\ref{fig Newton polygon TASEP} and \ref{fig Newton polygon ASEP}. For generic transition rates, the genus of the Riemann surface $\R$ associated to $M(g)$ is then equal to $\mathrm{g}=\sum_{k=d_{-}+1}^{d_{+}-1}(d_{k}-1)$. This gives
\begin{equation}
\label{g generic bidirectional}
\mathrm{g}=(2d_{+}-1)\Bigg({{L}\choose{N}}-1\Bigg)-L\,d_{+}(d_{+}-1)\;.
\end{equation}
The conjecture (\ref{d+ generic}) for $d_{+}$ then gives a complete formula for the genus with generic transition rates. At large $L$, $N$ with fixed density of particles $\rho=N/L$, one has $\mathrm{g}\simeq(2\pi\rho(1-\rho)L^{2})^{-1}\,\rme^{-2L(\rho\log\rho+(1-\rho)\log(1-\rho))}$, which grows as expected as twice the genus for unidirectional hopping.
\end{subsubsection}

\end{subsection}

\begin{subsection}{Process with all transition rates equal: ASEP}
\label{section ASEP}
We consider in this section the special case of ASEP, where all the forward transition rates are equal to one and all the backward transition rates are equal to $q<1$. As for TASEP in section~\ref{section TASEP}, the corresponding algebraic curve is singular, but there is an alternative description of the corresponding Riemann surface in terms of Bethe ansatz.

\begin{subsubsection}{Gallavotti-Cohen symmetry}\hfill\\
The generator $M_{\text{tot}}(g)$ verifies $M_{\text{tot}}(q/g)=M_{\text{tot}}(g)^{\top}$ with $^{\top}$ indicating transposition. The characteristic polynomial has thus from (\ref{P[M,Mtot]}) the symmetry
\begin{equation}
P(\lambda,g)=P(\lambda,q^{L}/g)\;,
\end{equation}
and the spectrum of $M(g)$ and $M(q^{L}/g)$ are identical. This corresponds to a symmetry of the algebraic curve $\mathcal{A}$ defined by $P(\lambda,g)=0$: if $(\lambda,g)$ belongs to $\mathcal{A}$ then $(\lambda,q^{L}/g)$ also does. At the level of the corresponding Riemann surface, this symmetry indicates the existence of an analytic automorphism $\varphi:\R\to\R$ with analytic inverse, such that $\lambda(\varphi(p))=\lambda(p)$ and $g(\varphi(p))=q^{L}/g(p)$.

For $g>0$, the eigenvalue with largest real part $\lambda_{0}(g)$ verifies in particular $\lambda_{0}(g)=\lambda_{0}(q^{L}/g)$, which can be interpreted as a symmetry of stationary large deviations of the current. Indeed, in the long time limit, one has $\langle g^{Q_{t}}\rangle\sim\rme^{t\lambda_{0}(g)}$, which is equivalent to $\P(Q_{t}=jt)\sim\rme^{-tG(j)}$, where the large deviation function $G$ is the Legendre transform $G(j)=\max_{g>0}(j\log g-\lambda_{0}(g))$ of $\lambda_{0}$. The symmetry above for $\lambda_{0}$ is then equivalent to the Gallavotti-Cohen symmetry \cite{LS1999.1} $G(j)-G(-j)=j\log(q^{L})$.

From (\ref{M[Mtot]}), we observe that the symmetry $M_{\text{tot}}(q/g)=M_{\text{tot}}(g)^{\top}$ translates for $M(g)$ to $q^{S}M(q^{L}/g)\,q^{-S}=M(g)^{\top}$, with $S$ the diagonal matrix with $\langle C|S|C\rangle$ equal to the sum of the positions of the particles, counted from site $1$. Then, the generating function of $Q_{t}$ with initial condition $P_{0}$, given in terms of $M$ by (\ref{GF[M]}), verifies $\langle g^{Q_{t}}\rangle=\langle P_{0}|q^{S}\rme^{tM(q^{L}/g)}q^{-S}\sum_{C\in\Omega}|C\rangle$, with $\langle P_{0}|=\sum_{C\in\Omega}P_{0}(C)\langle C|$. We observe that for the initial condition $P_{0}(C)\propto q^{-\langle C|S|C\rangle}$, the generating function then simply verifies $\langle g^{Q_{t}}\rangle=\langle (q^{L}/g)^{Q_{t}}\rangle$. The probability of $Q_{t}$ can be extracted from (\ref{Prob[M]}), and one finds
\begin{equation}
\frac{\P(Q_{t}=-Q)}{\P(Q_{t}=Q)}=q^{LQ}\;,
\end{equation}
i.e. the Gallavotti-Cohen symmetry holds at any time $t$ for this special initial condition, that we refer to as GC in the following.

In the combinatorial identity $\sum_{N=0}^{L}t^{N}\sum_{0\leq a_{1}<\ldots<a_{N}<L}q^{\sum_{j=1}^{N}a_{j}}=\prod_{i=0}^{L-1}(1+q^{i}t)$, the coefficient of $t^{N}$ can be extracted using the q-binomial theorem $\prod_{i=0}^{L-1}(1+q^{i}t)=\sum_{N=0}^{L}{{L}\choose{N}}_{\!q}\,t^{N}q^{\frac{N(N-1)}{2}}$. This implies that the initial condition GC can be normalized as
\begin{equation}
\label{PGC}
P_{\text{GC}}(C)=\frac{q^{\sum_{j=1}^{N}(L+1-j-x_{j}(C))}}{{{L}\choose{N}}_{\!q}}\;,
\end{equation}
where the positions of the particles $x_{j}(C)$ are taken between $1$ and $L$ and correspond to the labels of the sites that are occupied for the state $C$. When $q\to0$, the initial condition (\ref{PGC}) converges to the domain wall initial condition $\mathrm{dw_{0}}$ from (\ref{dlogkappaN dw}), such that the state with particles at positions $L-N+1,\ldots,L$ has probability one, and $P_{\text{GC}}$ is then a natural candidate for a possible generalization of the exact result (\ref{dlogkappaN dw}) to ASEP.
\end{subsubsection}

\begin{subsubsection}{Bethe ansatz}\hfill\\
Like for TASEP, Bethe ansatz consists in looking for eigenstates of $M(g)$ as linear combinations of plane waves. Periodic boundary condition implies that the momenta $q_{j}$, $j=1,\ldots,N$ of the quasi-particles are quantized: the variables $y_{j}=\frac{1-g^{-1/L}\rme^{\rmi q_{j}}}{1-qg^{-1/L}\rme^{\rmi q_{j}}}$ must satisfy the Bethe equations \footnote{Unlike for TASEP, where there is a clear partition of $\R$ into sheets labelled by a set $J\subset\lb1,L\rb$ and Bethe roots are written as $y_{j}$, $j\in J$, for ASEP, we simply label the Bethe roots as $y_{j}$, $j=1,\ldots,N$ for lack of a better choice.}
\begin{equation}
\label{Bethe equations ASEP}
g\Big(\frac{1-y_{j}}{1-qy_{j}}\Big)^{L}=(-1)^{N-1}\prod_{k=1}^{N}\frac{y_{j}-qy_{k}}{y_{k}-qy_{j}}
\end{equation}
for any $j=1,\ldots,N$, see e.g. \cite{D1998.1,GM2006.1} for a derivation. The eigenvalue of the corresponding eigenstate is then given in terms of the Bethe roots $y_{j}$ by
\begin{equation}
\label{lambda[y] ASEP}
\lambda=(1-q)\sum_{j=1}^{N}\Big(\frac{1}{1-y_{j}}-\frac{1}{1-qy_{j}}\Big)\;,
\end{equation}
and eigenvectors may be expressed as symmetric functions of the $y_{j}$. Taking the TASEP limit $q\to0$, the Bethe equations become $g(1-y_{j})^{L}\prod_{k=1}^{N}y_{k}+(-y_{j})^{N}=0$, which is indeed $R(y_{j},B)=0$ with $R$ defined in (\ref{R(y,B)}) and $B$ in (\ref{B[g,y]}), and the eigenvalue matches with (\ref{lambda[y] TASEP}).

For a given value of $g\in\Ch$, a solution of the Bethe equations (\ref{Bethe equations ASEP}) is a set $Y=\{y_{1},\ldots,y_{N}\}$: the ordering of the Bethe roots does not matter. Additionally, physical solutions corresponding to eigenstates of $M(g)$ must have distinct $y_{j}$ at generic values of $g\in\C$.

For any $j=1,\ldots,N$, eliminating all the $y_{k}$, $k\neq j$ from the Bethe equations leads to a polynomial equation for $g$ and $y_{j}$ only, which is independent of $j$ by symmetry of the Bethe equations, and individual Bethe roots thus only have algebraic branch points for $g$. These branch points $g_{*}$ are of two types, depending on whether the Bethe roots of a solution $Y=\{y_{1},\ldots,y_{N}\}$ of the Bethe equations are simply permuted among themselves or not under analytic continuation along a small loop around $g_{*}$. In the former case, the space of solutions of the Bethe equations, which is not ramified at that point since $Y$ is left unchanged, can be identified locally as a neighbourhood of $g_{*}$ in the complex plane. In the latter case, the set $Y$ becomes another solution $Y'$ of the Bethe equations after analytic continuation, and the space of solutions of the Bethe equations then has a ramification point $Y_{*}$ at $g=g_{*}$, with ramification index equal to the number of distinct solutions of the Bethe equations obtained by analytic continuations around $g_{*}$.

Under analytic continuation in the variable $g$, the space of solutions $Y$ of the Bethe equations can then be identified as a Riemann surface $\R_{q}$ (possibly with several connected components if some sectors of solutions of the Bethe equations can not be reached from one another by analytic continuations), on which rational symmetric functions of the Bethe roots are meromorphic by construction. Under the assumption that Bethe ansatz is complete, i.e. any eigenstate of $M(g)$ for generic $g$ can be represented by a physical solution of the Bethe equations, the points of $\R_{q}$ which are non-ramified for $g$ are thus in one to one correspondence with non-degenerate eigenstates of $M(g)$, and $\R_{q}$ is then identical to the Riemann surface $\R$ built from the characteristic equation of $M(g)$, see section~\ref{section eigenvectors on R}.

The Bethe equations for ASEP do not have the same mean field structure as the ones for TASEP, where all the Bethe roots are only coupled through the parameter $B$, and it is important to understand the differences about the structure of the Riemann surface between both cases. A crucial feature for TASEP is that $\R$ has only three branch points ($0$, $\infty$ and $B_{*}$) for the variable $B$, which leads to the same two cuts on all the sheets and gives a simple description of the global structure of $\R$. The existence of a meromorphic function $B$ with only three branch points on a compact Riemann surface $\R$ is in fact not guaranteed: Belyi's theorem \cite{S2001.3} asserts that such a function can be found if and only if there exists a non-singular algebraic curve with coefficients in the set of rational numbers $\mathbb{Q}$ whose corresponding Riemann surface is $\R$. Since the Riemann surface $\R_{q}$ for ASEP depends continuously on the parameter $q$, the minimal number of branch points of any meromorphic function on $\R_{q}$ thus appears to be at least equal to four, but an explicit construction of a function with few branch points, from which the global structure of $\R_{q}$ could be better understood, is still lacking. Therefore, in the following, we rely heavily on numerics, using the Wronskian formulation of the Bethe equations presented in the next section.
\end{subsubsection}

\begin{subsubsection}{Functional equations}\hfill\\
The Bethe equations (\ref{Bethe equations ASEP}) can also be formulated as a functional equation for the polynomial $Q(z)=\prod_{j=1}^{N}(z-y_{j})$, whose zeroes are the Bethe roots. Indeed, the Bethe equations imply that $g(1-z)^{L}Q(qz)+q^{N}(1-qz)^{L}Q(z/q)$ vanishes when $z$ is a Bethe root, and must then be divisible by $Q(z)$. This gives Baxter's equation \cite{B1982.1}
\begin{equation}
\label{Baxter equation ASEP}
T(z)Q(z)=g\,(1-z)^{L}Q(qz)+q^{N}(1-qz)^{L}Q(z/q)\;,
\end{equation}
with $T$ a polynomial of degree $L$. The fact that both $Q$ and $T$ must be polynomials gives enough constraints so that (\ref{Baxter equation ASEP}) has only discrete solutions.

When $q\to1$, Baxter's equation reduces to a second order ordinary differential equation for $Q$, whose space of solutions is two dimensional, and another solution independent from $Q$ must exist. This is also the case for $q\neq1$, and one can build \cite{PS1999.1,P2010.1} another polynomial $P(z)=\prod_{j=1}^{L-N}(z-\tilde{y}_{j})$, whose zeroes are related to the Bethe roots for the system with particles and empty sites exchanged, and which verifies
\begin{equation}
T(z)P(z)=q^{N}(1-z)^{L}P(qz)+g\,(1-qz)^{L}P(z/q)\;,
\end{equation}
with $T$ the same as in (\ref{Baxter equation ASEP}). As in the case of differential equations $q\to1$, the Wronskian of $Q$ and $P$ then has a simple expression,
\begin{equation}
\label{Wronskian ASEP}
g\,Q(z)P(z/q)-q^{N}Q(z/q)P(z)=Q(0)P(0)(g-q^{N})(1-z)^{L}\;.
\end{equation}
Again, the requirement that $Q$ and $P$ must be polynomials of respective degrees $N$ and $L-N$ ensures that only a discrete number of solutions exist. Compared to Baxter's equation (\ref{Baxter equation ASEP}), which has many non-physical solutions that do not correspond to eigenstates of $M(g)$, the Wronskian equation (\ref{Wronskian ASEP}) appears to have exactly ${L}\choose{N}$ distinct solutions for generic values of $g$, corresponding to distinct eigenstates of $M(g)$. This makes the Wronskian equation particularly suitable for numerics, and all the numerics in the following are done using (\ref{Wronskian ASEP}).
\end{subsubsection}

\begin{subsubsection}{Special points on \texorpdfstring{$\R_{q}$}{Rq}}\hfill\\
We identify in this section points on $\R_{q}$ that are ramification points for the variable $g$, and points whose Bethe roots take special values, see table~\ref{table points ASEP}.

When $g\to\infty$, the physical solutions $Y=\{y_{1},\ldots,y_{N}\}$ of the Bethe equations (\ref{Bethe equations ASEP}) are such that all the $y_{j}$ converge to $1$ in a different direction. Around $g=\infty$, the solutions $Y$ are then labelled by a set of $N$ distinct integers between $1$ and $L$, just as for TASEP, or equivalently by as set of $N$ distinct roots $\omega_{j}$ of $(-1)^{N-1}$, and one finds the asymptotic expansion
\begin{equation}
\label{yj ginfinity ASEP}
\fl\hspace{2mm}
y_{j}\simeq1-(1-q)\,\omega_{j}^{-1}g^{-1/L}+\frac{1-q}{L}\Big((N-q(L-N))\,\omega_{j}^{-2}-(1+q)\,\omega_{j}^{-1}\sum_{k=1}^{N}\omega_{k}^{-1}\Big)\;.
\end{equation}
This gives for the eigenvalue (\ref{lambda[y] ASEP})
\begin{equation}
\label{lambda ginf ASEP}
\lambda\simeq g^{1/L}\sum_{j=1}^{N}\omega_{j}-\frac{1+q}{L}\Bigg(N(L-N)+\Big(\sum_{j=1}^{N}\omega_{j}\Big)\Big(\sum_{j=1}^{N}\omega_{j}^{-1}\Big)\Bigg)\;,
\end{equation}
which generalizes (\ref{lambda ginf TASEP}) to ASEP, and leads to the same degree $d_{+}$ for the characteristic polynomial as for TASEP. From the expansions above, analytic continuation along large loops for $g$ multiplies all the $\omega_{j}$ by the same factor $\rme^{2\rmi\pi/L}$, which amounts to shifting the corresponding sets $J$ for TASEP by one modulo $L$, and we observe that the points with $g=\infty$ have the same ramification indices as for TASEP.

The Gallavotti-Cohen automorphism $\varphi$ on $\R_{q}$ acts on Bethe roots as $y_{j}\to(qy_{j})^{-1}$, i.e. $Y(\varphi(p))=(qY(p))^{-1}$. The automorphism $\varphi$ sends the points with $g=\infty$ to points with $g=0$, and Bethe roots $y_{j}=1$ to Bethe roots $y_{j}=1/q$. The points of $p\in\R_{q}$ with $g(p)=0$ thus have the same ramification indices as the ones with $g(p)=\infty$, which is a major difference with the situation for TASEP. In fact, since the Gallavotti-Cohen symmetry exchanges the domains $|g|<q^{L/2}$ and $|g|>q^{L/2}$, we observe that the TASEP limit is singular since the ``half'' of the Riemann surface with $|g|<q^{L/2}$ degenerate into points with $g=0$: roughly speaking, the Riemann surface $\R_{q}$ for ASEP can be viewed as two copies of the Riemann surface for TASEP glued together.

Poles of the eigenvalue $\lambda$ may only happen at points $p\in\R_{q}$ where some coefficients of $M(g(p))$ are divergent, i.e. only when $g(p)=0$ and $g(p)=\infty$. From the Bethe ansatz perspective, the expression (\ref{lambda[y] ASEP}) for the eigenvalue implies that poles of $\lambda$ require that some Bethe roots are equal to $1$ or $1/q$. This is indeed the case for $g=0$ and $g=\infty$, as discussed above. However, unlike for TASEP, numerics reveal that there does exist other points $p_{0}\in\R_{q}$ with some Bethe root $y_{j}=1$. Assuming $g(p_{0})\neq0,\infty$, the Bethe equations then imply that there is another Bethe root $y_{k}=1/q$, and vice versa, and since $\lambda$ can not have a pole at $p_{0}$, one must have $1-y_{j}\simeq1-qy_{k}$ near $p_{0}$. Numerics seem to indicate that those points $p_{0}$ are not ramified for $g$, at least for generic values of $q$. For the system with $N=2$ particles, we observe in particular that there are two such points $p_{0}$, corresponding to $g(p_{0})=\pm(\rmi\sqrt{q})^{L}$.

\begin{table}
	\begin{center}
	\begin{tabular}{|lcc|}\hline
	Point $p\in\R_{q}$ & Bethe roots & Ramification for $g$\\\hline
	$g=1$, $\lambda=0$ & all $y_{j}=0$ & non-ramified\\[2mm]
	$g=q^{L}$, $\lambda=0$ & all $y_{j}=\infty$ & non-ramified\\[3mm]
	\hspace{-2mm}\begin{tabular}{r}Some points with $g=q^{k}$\\$k=1,\ldots,L-1$\end{tabular} & \begin{tabular}{c}$\max(N-k,0)$ vanishing\\$\max(k+N-L,0)$ divergent\end{tabular} & non-ramified\\[5mm]
	$g=\infty$ & all $y_{j}=1$ & ramified\\[2mm]
	$g=0$ & all $y_{j}=1/q$ & ramified\\[2mm]
	Points at which some Bethe roots \hspace{-14mm} & $y_{j}=1$ and $y_{k}=1/q$\hspace{5mm} & non-ramified\\[2mm]
	Bethe roots $y_{j}=y_{k}=y_{*}$, with \hspace{-6mm} & $LV(y_{*})+\sum_{k=1}^{N}X(y_{*},y_{k})=0$ & non-ramified\\[2mm]
	Non-trivial zeroes of $\det(K_{Y})$ \hspace{-100mm} & unremarkable & ramified twice\\\hline
	\end{tabular}
	\end{center}
	\caption{Special points on the Riemann surface $\R_{q}$ for ASEP. All the points ramified for $g$ and the points with vanishing or divergent Bethe roots are shown.}
	\label{table points ASEP}
\end{table}

We now turn to special points of $\R_{q}$ where some Bethe roots vanish (related by the Gallavotti-Cohen symmetry to points where some Bethe roots diverge), and call $S$ the set of indices $j$ such that $y_{j}=0$, with cardinal $|S|=n$. Then, for $j\in S$, the Bethe equation (\ref{Bethe equations ASEP}) leads to $\prod_{k\in S}(y_{j}-qy_{k})/(y_{k}-qy_{j})\simeq(-1)^{n-1}g/q^{N-n}$. Taking the product over all $j\in S$ then implies $g=\rme^{2\rmi\pi r/n}\,q^{N-n}$ for some integer $r$. A more sophisticated argument using Baxter's equation (\ref{Baxter equation ASEP}) shows that $r=0$, i.e. $g=q^{N-n}$. Indeed, when $Q(0)\neq0$, Baxter's equation implies $T(0)=g+q^{N}$. On the other hand, when $n$ Bethe roots vanish, writing $Q(z)=z^{n}\tilde{Q}(z)$ with $\tilde{Q}(0)\neq0$, factoring out $z^{n}$ from Baxter's equation and setting $z=0$ leads to $T(0)=q^{n}g+q^{N-n}$. Continuity of $T(0)$ then implies $g+q^{N}=q^{n}g+q^{N-n}$, which is indeed equivalent to $g=q^{N-n}$. In the context of the XXZ spin chain with twisted boundaries, these special values of $g$ correspond to the so called root of unity case, where peculiar symmetry algebras appear \cite{K2004.1}.

By the Gallavotti-Cohen symmetry, the discussion above for vanishing $y_{j}$ shows that a number $n$ of Bethe roots may diverge only at $g=q^{L-N+n}$. From numerics up to $L=9$, we observe that the number of solutions of the Bethe equations with vanishing or divergent Bethe roots when $g\to q^{k}$, $k=0,\ldots,L$ is equal to $\min({{L}\choose{k}},{{L}\choose{N}})$. In particular, for the special case $k=0$, only the stationary eigenstate with $g=1$ and eigenvalue $\lambda=0$ has its Bethe roots equal to zero. Additionally, when $L-N<k<N$, the same solutions of the Bethe equations appear to have both vanishing and divergent Bethe roots. Thus, any solution with vanishing or divergent Bethe roots at $g=q^{k}$ always has $\max(N-k,0)$ vanishing and $\max(k+N-L,0)$ divergent Bethe roots. Numerics also indicates that the points on $\R$ corresponding to those solutions are not ramified for $g$, at least for generic values of $q$. When $q\to0$, all those points with $1\leq k\leq L$ merge together with the points with $g=0$, which are ramified for $g$, and the limit $q\to0$ is thus highly singular around those points.

We have seen above that several Bethe roots may coincide when $g\in\{0,\infty\}$ and when $g\in\{q^{k},k=0,\ldots,L\}$. We consider now other situations with $n\geq2$ coinciding Bethe roots $y_{j}=y_{*}\not\in\{0,1,q^{-1},\infty\}$. From Baxter's equation (\ref{Baxter equation ASEP}), assuming $Q(z)=(z-y_{*})^{n}\,\tilde{Q}(z)$ gives an apparent pole at $z=y_{*}$ for $T(z)$, which must cancel since $T$ is a polynomial. This leads to the constraint $LV(y_{*})+\sum_{k=1}^{N}X(y_{*},y_{k})=0$, where
\begin{equation}
\label{VX}
V(y)=\frac{1}{1-y}-\frac{q}{1-qy}
\qquad\text{and}\qquad
X(y,z)=\frac{1}{y-qz}-\frac{1}{y-z/q}\;.
\end{equation}
Again, numerics indicate that those points are not ramified for $g$.

We finally consider ramification points for $g$, and assume $g$ different from $0$ and $\infty$, which were already treated above. As seen in the previous section, ramification points for $g$ on $\R_{q}$ may only happen at branch points $g_{*}$ of some $y_{j}$. These branch points can be located by interpreting the Bethe equations as a system of ordinary differential equations. Taking the logarithmic derivative of (\ref{Bethe equations ASEP}) with respect to $g$, one has
\begin{equation}
\label{dyj/dg ASEP}
g\,\frac{\rmd y_{j}}{\rmd g}=\sum_{k=1}^{N}(K_{Y}^{-1})_{j,k}\;,
\end{equation}
where the matrix elements of the $N\times N$ matrix $K_{Y}$ with $Y=\{y_{1},\ldots,y_{N}\}$ are
\begin{equation}
\label{KY}
(K_{Y})_{i,j}=\delta_{i,j}\Big(LV(y_{j})+\sum_{k=1}^{N}X(y_{j},y_{k})\Big)-X(y_{j},y_{i})\;,
\end{equation}
with $V$ and $X$ given in (\ref{VX}). The solution $Y(g)$ of the system of ordinary differential equations (\ref{dyj/dg ASEP}) is locally analytic in the variable $g$ as long as $(K_{Y}^{-1})_{j,k}$ is finite. By Cramer's rule, $(K_{Y}^{-1})_{j,k}=A_{j,k}/\det(K_{Y})$, with $A_{j,k}$ polynomial in the matrix elements of $K_{Y}$. Thus, ramification points for $y_{j}$ may only happen at points of $\R_{q}$ where either some $(K_{Y})_{i,j}$ diverges or $\det(K_{Y})$ vanishes.

The matrix elements $(K_{Y})_{i,j}$ may only diverge when there exist two Bethe roots $y_{j}$ and $y_{k}$ with $y_{j}-qy_{k}=0$ (from $X(y_{j},y_{k})$), or when there exists a Bethe root $y_{j}$ equal to $1$ or $1/q$ (from $V(y_{j})$). The case with $y_{j}\in\{1,1/q\}$ is a special case of $y_{j}-qy_{k}=0$ since Bethe roots equal to $1$ and $1/q$ have to come together, as explained above. Conversely, assuming $y_{j}-qy_{k}=0$ with all Bethe roots finite and non-zero, and $y_{j}-q^{-1}y_{\ell}\neq0$ for any $\ell$ (which can always be achieved by a suitable choice for $j$ and $k$), the Bethe equation (\ref{Bethe equations ASEP}) implies $y_{j}=1$ and thus $y_{k}=1/q$. Thus, the branch points for $g$ with $g\neq0,\infty$ corresponding to divergent matrix elements $(K_{Y})_{i,j}$ are either the points with $g\in q^{\lb0,L\rb}$ where some Bethe roots vanish or diverge, or the points where Bethe roots equal to $1$ and $1/q$ appear together. As said above, numerics indicate that both of those points are not ramification points for $g$, which means that cancellations between $A_{j,k}$ and $\det(K_{Y})$ must happen.

The only ramification points for $g$ with $g\neq0,\infty$ are thus zeroes of $\det(K_{Y})$. We observe that the points with divergent Bethe roots or coinciding Bethe roots verify $\det(K_{Y})=0$, but as seen above from numerics, such points are not actually ramification points for $g$, and cancellations must happen again between $A_{j,k}$ and $\det(K_{Y})$. The other, non-trivial zeroes of $\det(K_{Y})$ are ramification points for $g$, with ramification index $2$ for generic values of $q$, which is confirmed by numerics. This is consistent with the fact that $\det(K_{Y})$ coincides with the celebrated Gaudin determinant \cite{GC2014.1} giving the normalization $\langle\psi|\psi\rangle$ of Bethe eigenvectors, and thus appears as the denominator of the function $\mathcal{N}$ on $\R$ (see next section for explicit expressions for some initial conditions), whose poles must be ramified for $g$, see section~\ref{section pole structure}.
\end{subsubsection}

\begin{subsubsection}{Function \texorpdfstring{$\mathcal{N}$}{N}}\hfill\\
For stationary initial condition, one has from \cite{P2016.2}
\begin{equation}
\mathcal{N}_{\text{stat}}=\frac{g^{N}\prod_{j=0}^{N-1}(1-q^{j}/g)^{2}}{{{L}\choose{N}}\,(1-q)^{N}}\,\frac{V_{q}^{2}}{(\prod_{j=1}^{N}y_{j}^{2})\det(K_{Y})}\;,
\end{equation}
with $K_{Y}$ given by (\ref{KY}) and
\begin{equation}
\label{Vq2}
V_{q}^{2}=\prod_{j=1}^{N}\prod_{k=j+1}^{N}\frac{(y_{j}-y_{k})^{2}}{(y_{j}-qy_{k})(qy_{j}-y_{k})}\;.
\end{equation}
For the Gallavotti-Cohen symmetric initial condition (\ref{PGC}), similar calculations as in \cite{P2016.2} lead instead to
\begin{equation}
\label{NGC}
\mathcal{N}_{\text{GC}}=\frac{\prod_{j=0}^{N-1}((g-q^{j})(1-q^{L-j}/g))}{{{L}\choose{N}}_{q}\,(1-q)^{N}}\,\frac{V_{q}^{2}}{(\prod_{j=1}^{N}y_{j})\det(K_{Y})}\;.
\end{equation}
The expressions $\prod_{j=1}^{N}y_{j}$, $V_{q}^{2}$ and $\det(K_{Y})$ are symmetric functions of the Bethe roots, and are thus all meromorphic on $\R$. The same is then true for $\mathcal{N}_{\text{stat}}$ and $\mathcal{N}_{\text{GC}}$, which is expected from the discussion in section~\ref{section eigenvectors on R}. As explained in the previous section, the non-trivial zeroes of the Gaudin determinant $\det(K_{Y})$ are ramification points for $g$ with ramification index $2$. They are also poles of the meromorphic functions $\mathcal{N}_{\text{stat}}$ and $\mathcal{N}_{\text{GC}}$. We argue that all the other poles and zeroes of $\mathcal{N}_{\text{stat}}$ and $\mathcal{N}_{\text{GC}}$, which we study in the rest of this section, correspond to simple values of $g$. We focus for convenience on $\mathcal{N}_{\text{GC}}$, which is invariant under the Gallavotti-Cohen automorphism $g\to q^{L}/g$, $y_{j}\to(qy_{j})^{-1}$, and from which the case of $\mathcal{N}_{\text{stat}}$ can be deduced easily.

When $g\to\infty$, all the Bethe roots converge to $1$, and one finds $\det(K_{Y})\simeq L^{N}\prod_{j=1}^{N}(1-y_{j})^{-1}$. Using (\ref{yj ginfinity ASEP}) then leads to $\det(K_{Y})\sim g^{N/L}$, $V_{q}^{2}\sim g^{-\frac{N(N-1)}{L}}$ and finally $\mathcal{N}_{\text{GC}}\sim g^{\frac{N(L-N)}{L}}$ when $g\to\infty$. The function $\mathcal{N}_{\text{GC}}$ has thus $|\Omega|N(L-N)/L$ poles (counted with multiplicity) at the points of $\R$ with $g=\infty$, and then also $|\Omega|N(L-N)/L$ poles with $g=0$ by the Gallavotti-Cohen symmetry (the fact that the points on $\R$ with $g=0$ and $g=\infty$ are ramified for $g$ does not change the counting, since the sum of the ramification indices is necessarily equal to $|\Omega|$).

When $g\to q^{k}$, $k=0,\ldots,L$, we observe that each eigenstate with no vanishing or divergent Bethe roots corresponds to a simple zero for $\mathcal{N}_{\text{GC}}$. On the other hand, cancellations happen for the eigenstates with vanishing or divergent Bethe roots, whose number is conjectured in the previous section, and numerics up to $L=9$ with $q$ generic indicate that those points of $\R_{q}$ are neither poles nor zeroes of $\mathcal{N}$. This gives a total of $\max({{L}\choose{N}}-{{L}\choose{k}},0)$ zeroes for each value of $k$.

We conjecture that $\mathcal{N}_{\text{GC}}$ has no other poles or zeroes than the ones mentioned above. In particular, cancellations happen between $V_{q}^{2}$ and $\det(K_{Y})$ at the non-trivial points of $\R_{q}$ with coinciding Bethe roots and at the points for which Bethe roots equal to $1$ and $1/q$ appear, which is confirmed by numerics. Then, the fact that the meromorphic function $\mathcal{N}_{\text{GC}}$ must have as many zeroes as poles counted with multiplicity implies that the number of non-trivial zeroes of the Gaudin determinant (i.e. the number of ramification points for $g$ with $g\not\in\{0,\infty\}$) must be equal to $-\frac{2N(L-N)}{L}{{L}\choose{N}}+\sum_{k=0}^{L}\max({{L}\choose{N}}-{{L}\choose{k}},0)$, which reduces after some simplifications to $2\sum_{k=0}^{\min(N,L-N)-1}(\frac{2k+1}{L}{{L}\choose{N}}-{{L}\choose{k}})$. This expression was checked up to $L=7$ against a numerical computation of the branch points for $g$ by solving $P(\lambda,g)=P^{(1,0)}(\lambda,g)=0$ for generic values of $q$, with $P$ the characteristic polynomial. In the next section, we use this result for the number of ramification points to compute the genus of $\R_{q}$ when $L$ and $N$ are co-prime.

The functions $\mathcal{N}_{\text{stat}}$ and $\mathcal{N}_{\text{GC}}$ can in principle be written in exponential form, by computing explicitly $\rmd\log\mathcal{N}$ using (\ref{dyj/dg ASEP}). However, in the absence of a good analogue for the function $\kappa$ appearing in (\ref{dlogkappaN stat}) and (\ref{dlogkappaN dw}) for TASEP, we do not expect that $\rmd\log\mathcal{N}$ alone can be simplified in a meaningful way, which currently prevents us from taking the KPZ scaling limit for the probability of the current $Q_{t}$.
\end{subsubsection}

\begin{subsubsection}{Genus}\hfill\\
While the local ramification data of $\R_{q}$ follows rather directly from the Bethe equations, see the previous sections, questions about the global connectivity of $\R_{q}$ are more difficult in the absence of a nice parametrization as for TASEP.

In particular, while the number of connected components follows directly from the algebra of the generators (\ref{aj}) for TASEP, we did not find a way to compute it directly from Bethe ansatz. Resorting to the factorization of the characteristic polynomial instead, a study up to $L=9$ for generic values of $q$ seems to indicate that $\R_{q}$ has a single connected component if and only if $L$ and $N$ are co-prime, like for TASEP, but the number of connected components for ASEP and TASEP differ in general.

For the genus, the Riemann-Hurwitz formula supplemented with the number of ramification points with $\det(K_{Y})=0$ conjectured in the previous section implies for $L$ and $N$ co-prime
\begin{equation}
\label{g ASEP}
\mathrm{g}=\sum_{k=1}^{\min(N,L-N)-1}\Bigg(\frac{2k+1}{L}{{L}\choose{N}}-{{L}\choose{k}}\Bigg)\;.
\end{equation}
At large $L$, $N$ with fixed density $\rho=N/L$, this expression for the genus grows as $\mathrm{g}\simeq\frac{\sqrt{L}\,\min(\rho,1-\rho)^{2}}{\sqrt{2\pi\rho(1-\rho)}}\,\rme^{-L(\rho\log\rho+(1-\rho)\log(1-\rho))}$, which is again much smaller than for generic transition rates, see below (\ref{g generic bidirectional}).

The difference between the genus with generic transition rates, given by (\ref{g generic bidirectional}), and the genus for ASEP (\ref{g ASEP}) must be equal to the number of singular points on the algebraic curve $\mathcal{A}$, if all of them are nodal and the number of connected components of $\R_{q}$ is equal to one. For $L=5$, $N=2$, a numerical solution of $P(\lambda,g)=P^{(1,0)}(\lambda,g)=P^{(0,1)}(\lambda,g)=0$ with $P$ the characteristic polynomial of $M(g)$ gives for instance $16$ singular points, which are all nodal. This is indeed compatible with the presence of $17$ points in the interior of the Newton polygon, see figure~\ref{fig Newton polygon ASEP}, and a genus equal to $1$ from (\ref{g ASEP}).
\end{subsubsection}

\end{subsection}

\end{section}

\begin{section}{Conclusion}
We have studied in this paper the statistics of integer counting processes $Q_{t}$ from the point of view of the Riemann surface $\R$ associated to the algebraic curve built from the characteristic equation of the generator of $Q_{t}$. We have in particular showed that the probability of $Q_{t}$ can be written as an integral (\ref{Prob[int R]}) over a closed contour on $\R$.

While the formalism most easily leads to explicit formulas when the genus of $\R$ is equal to zero, with an example given in section~\ref{section simple example}, an alternative representation of $\R$ in terms of Bethe ansatz for the integrable examples of TASEP and ASEP also allows for a somewhat explicit expression for the integrand of the probability of $Q_{t}$.

A major open question that would be worth elucidating in the future is what can be done exactly with this formalism beyond the simple examples considered in this paper. On the side of integrable models, TASEP with open boundary conditions, whose Bethe ansatz equations \cite{CN2018.1} have a similar nature as the ones for TASEP with periodic boundaries, is a good candidate \cite{GP2020.1}. Regarding non-integrable models, it would be interesting to study some examples with small genus.

Additionally, we would like to understand whether more sophisticated tools from algebraic geometry can be useful in the context of integer counting processes. In particular, the Riemann-Roch theorem, which relates the dimensions of meromorphic functions and differentials with prescribed pole structure, might shed some light on the space of meromorphic differentials $\rmd\log\mathcal{N}$ corresponding to admissible initial conditions. On the integrable side, the formalism of vector bundles appears to be the right setting to study the monodromy of the vector of Bethe roots $(y_{1},\ldots,y_{N})$ on $\R$, and tools from K-theory \cite{MS2014.2,PSZ2020.1} might lead to some progress toward more explicit formulas for ASEP.
\end{section}

\vspace{10mm}

\end{document}